\documentclass[journal]{vgtc}                
\ifpdf
  \pdfoutput=1\relax                   
  \usepackage{graphicx}                
  \DeclareGraphicsExtensions{.pdf,.png,.jpg,.jpeg} 
\else
  \usepackage{graphicx}                
  \DeclareGraphicsExtensions{.eps}     
\fi%

\graphicspath{{figures/}{pictures/}{images/}{./}} 

\usepackage{microtype}                 
\PassOptionsToPackage{warn}{textcomp}  
\usepackage{textcomp}                  
\usepackage{mathptmx}                  
\usepackage{times}                     
\usepackage{cite}                      
\usepackage{tabu}                      
\usepackage{booktabs}                  

\usepackage[usenames]{color}
\definecolor{darkblue}{rgb}{0,0,0.75}
\definecolor{darkgreen}{rgb}{0,0.5,0}
\definecolor{darkred}{rgb}{0.8,0,0}
\definecolor{darkgray}{rgb}{0.3,0.3,0.3}

\usepackage{psfrag}
\usepackage{amssymb}
\usepackage{amsmath}
\usepackage{bbold}

\onlineid{1647}

\vgtccategory{Research}
\vgtcpapertype{please specify}

\title{Copula-based statistical dependence visualizations}

\author{Arturo Erdely and Manuel Rubio-S\'{a}nchez}
\authorfooter{
\item
 Arturo Erdely is with Universidad Nacional Aut\'{o}noma de M\'{e}xico. E-mail: aerdely@acatlan.unam.mx.
\item
 Manuel Rubio-S\'{a}nchez is with Universidad Rey Juan Carlos. E-mail: manuel.rubio@urjc.es.
}

\shortauthortitle{Copula-based statistical dependence visualizations}

\abstract{A frequent task in exploratory data analysis consists in examining pairwise dependencies between data variables. Popular approaches include visualizing correlation or scatter plot matrices. However, both methods can be misleading. The former is primarily limited because it reports a single value for a pair of random variables. Furthermore, scatter plots can fail to convey the dependency structure between variables properly. In this paper we discuss these shortcomings and present alternative and richer visualizations based on copula functions, which fully determine the dependency between continuous random variables. Since copulas seldom appear in the data visualization literature we first review essential theory, and propose alternative scatter plots and several heatmaps for assessing the statistical association between two continuous random variables. These visualizations not only allow users to detect independence, but also increasing and/or decreasing trends in the data through a color coding, which can also be applied in other methods such as parallel coordinates.} 


\CCScatlist{
\CCScat{Human-centered computing}{}{Visualization}{Visualization techniques}
\CCScat{Human-centered computing}{}{Visualization}{Visualization theory, concepts and paradigms}
}

\vgtcinsertpkg

\begin{document}



\maketitle

\section{Introduction}

Exploratory data analysis (EDA)~\cite{Tukey77} is one of the first stages of a data analysis process where the goal is to obtain an overview of the data, and often involves statistical graphs or data visualization methods and tools. A frequent task in EDA consists in assessing dependencies between the data variables by combining information from visualizations and numerical measures of association. For example, these dependence relationships can be exploited when performing feature or model selection \cite{Hall00,Guyon03,May11,Krause14} by revealing redundant variables, or the strength of the associations with a target variable.


The most popular method for visualizing two quantitative data variables is the ubiquitous scatter plot. Since it displays the data ``as it is'' (i.e., no information is lost through its visual encoding) it is one of the most useful and popular statistical graphs \cite{friendly05}. Scatter plots may seem to be the most adequate and straightforward technique to display the relationship between two variables. However, when it comes to revealing \textit{statistical dependence} we will explain, through the theory of copula functions \cite{Nelsen1999}, that scatter plots can actually be misleading. Consequently, scatter plot matrices (SPLOM) \cite{Carr87} also suffer from this drawback.

Statisticians and data analysts also often rely on single measures of association between two quantitative variables such as Pearson's correlation coefficient, or Spearman's and Kendall's rank correlation coefficients. These scalar values are used in correlation matrices, also known as corrgrams \cite{Friendly02}, and frequently complement the scatter plots in SPLOM. While these popular correlations may provide useful summaries, they do not constitute formal \textit{dependence measures} and can also be misleading when quantifying dependence. Again, we will use the theory of copulas to justify this claim, showing the connection between these correlation measures and copulas.

The paper has two main contributions. Firstly, we study and question the appropriateness of using scatter plots as a visualization tool to estimate the dependence between continuous random variables. Specifically, we will show that the so called \textit{pseudo-observations}, which constitute the most popular approach for visualizing copulas, are better suited for detecting independence or weak dependencies. In particular, we will show that by choosing different marginal distributions a bivariate distribution represented as a scatter plot can display patterns that can mislead users when interpreting independence or weak dependencies.

Secondly, we propose several heatmaps for visualizing information related to an empirical copula. The first two allow users to approximate association measures such as Spearman's rank orrelation coefficient and Schweizer-Wolff's dependence measure \cite{Schweizer1981}. More importantly, we introduce a normalized measure of differences between an empirical copula and a copula that represents independence. With this measure we construct a heatmap that highlights deviations from independence, i.e., increasing and/or decreasing trends in data. Moreover, these heatmaps can be used to assign colors to data pairs that can be exploited in other visualizations such as scatter plots or parallel coordinate plots.

The paper is structured as follows. Section~\ref{sec:relatedwork} reviews related work, while Sec.~\ref{sec:background} summarizes essential elements of copula theory. In Sec.~\ref{sec:copulavisualizations} we describe methods for visualizing information (in particular, associations between continuous random variables) related to copulas. In Sec.~\ref{sec:experiments} we present the results of a user study regarding pseudo-observations. Finally, Sec.~\ref{sec:discussion} includes the main conclusions and a discussion.

\section{Related work}
\label{sec:relatedwork}

Statistical dependence is usually analyzed and quantified through computational methods. To the best of our knowledge, the most common approach for visualizing relationships between continuous random variables consists of the standard scatter plot, or scatter plot matrices when working with more than two variables. In this latter case it is also common to use correlation matrices or \textit{corrgrams} \cite{Friendly02}, which, for instance, show the values of correlation measures through a color coding. These types of visualizations can be extended in a number of ways, by including ellipses, fitted curves, polygons, winglets, glyphs, graphs, etc. (see \cite{Friendly06,Elmqvist08b,Chan13,Yates14,Nguyen16,Nguyen17,Lu20}). There are a few techniques that are able to indicate independence for categorical data, such as mosaic plots (see \cite{Friendly01}). However, the study in this paper will focus exclusively on continuous data, where we will use basic scatter plots and heatmaps.

Several works have carried out in-depth studies related to the perception of Pearson's correlation coefficients for Gaussian random variables \cite{Doherty2007,Rensik2010,Harrison2014,Sher2017}. In this paper we will focus on a more general concept of statistical dependence, which is more complex but also more powerful and realistic. Moreover, we will focus on other measures of association such as Spearman's concordance or Schweizer-Wolff's dependence measure.

Lastly, copula theory has not been fully exploited in the visualization literature. The work in \cite{Hazarika18} uses copulas to model dependence between random variables, which is arguably the most common application of copula theory. However, it is not focused on the same goal as our paper, which is visualizing and interpreting dependence in terms of copulas and their transformations.


\section{Background}
\label{sec:background}

For the reader's convenience, in this section we provide a summary of copula theory and results for quick reference.

\subsection{Copula functions}

Modeling and quantifying dependence between random variables is central to statistical science. Yet surprisingly, little attention has been paid to a formal definition of what should be understood as a \textit{dependence measure} and what should not, as mentioned in \cite{Kotz2001}, where the authors point out that the first book devoted to dependence concepts was published as late as 1997 \cite{Joe1997}.

A key concept regarding statistical dependence is \textit{copula functions}, which are defined as follows:

\textbf{Definition 1} A bivariate \textit{copula} is a function $C:[0,1]^2\rightarrow[0,1]$ that satisfies the following:
\begin{enumerate}
    \item $C(u,0)\,=\,0\,=\,C(0,v)\,;$
    \item $C(u,1)\,=\,u\,$ and $\,C(1,v)=v\,;$
    \item For all $u_1\leq u_2\,$ and $\,v_1\leq v_2\,:$ $$C(u_2,v_2)-C(u_2,v_1)-C(u_1,v_2)+C(u_1,v_1)\geq 0.$$
\end{enumerate}

Copula functions were introduced and coined in 1959 by Sklar \cite{Sklar1959}, who proved that there exists a functional link between the marginal distribution functions of a random vector and its joint distribution function precisely through a copula function. If $(X,Y)$ is a random vector with joint probability distribution function $F_{X,Y}(x,y)=P(X\leq x,Y\leq y)$ and marginal continuous distribution functions $F_X(x)=F_{X,Y}(x,+\infty)$ and $F_Y(y)=F_{X,Y}(+\infty,y)$ then by Sklar's Theorem \cite{Sklar1959} there exists a unique copula function $C_{X,Y}:[0,1]^2\rightarrow[0,1]$ such that
\begin{equation}\label{eq:SklarThm}
  F_{X,Y}(x,y) = C_{X,Y}(F_X(x),F_Y(y)).
\end{equation}
It is important to note that since the marginal distributions have no information about how each random variable may interact with others, all the information regarding their dependence is in the underlying copula function. Thus, it is possible to express dependence measures through copulas, with no involvement of the marginal distributions.

\begin{figure*}
    \centering

  \begin{minipage}{0.27\textwidth}
   \centering
    \begin{psfrags}

      \psfrag{u}[cc][cc]{\footnotesize $u$}
      \psfrag{v}[cc][cc]{\footnotesize $v$}
      \psfrag{P(u,v)}[cc][cc]{\footnotesize $\Pi(u,v) = uv$}

     \includegraphics[clip=true,width=\textwidth]{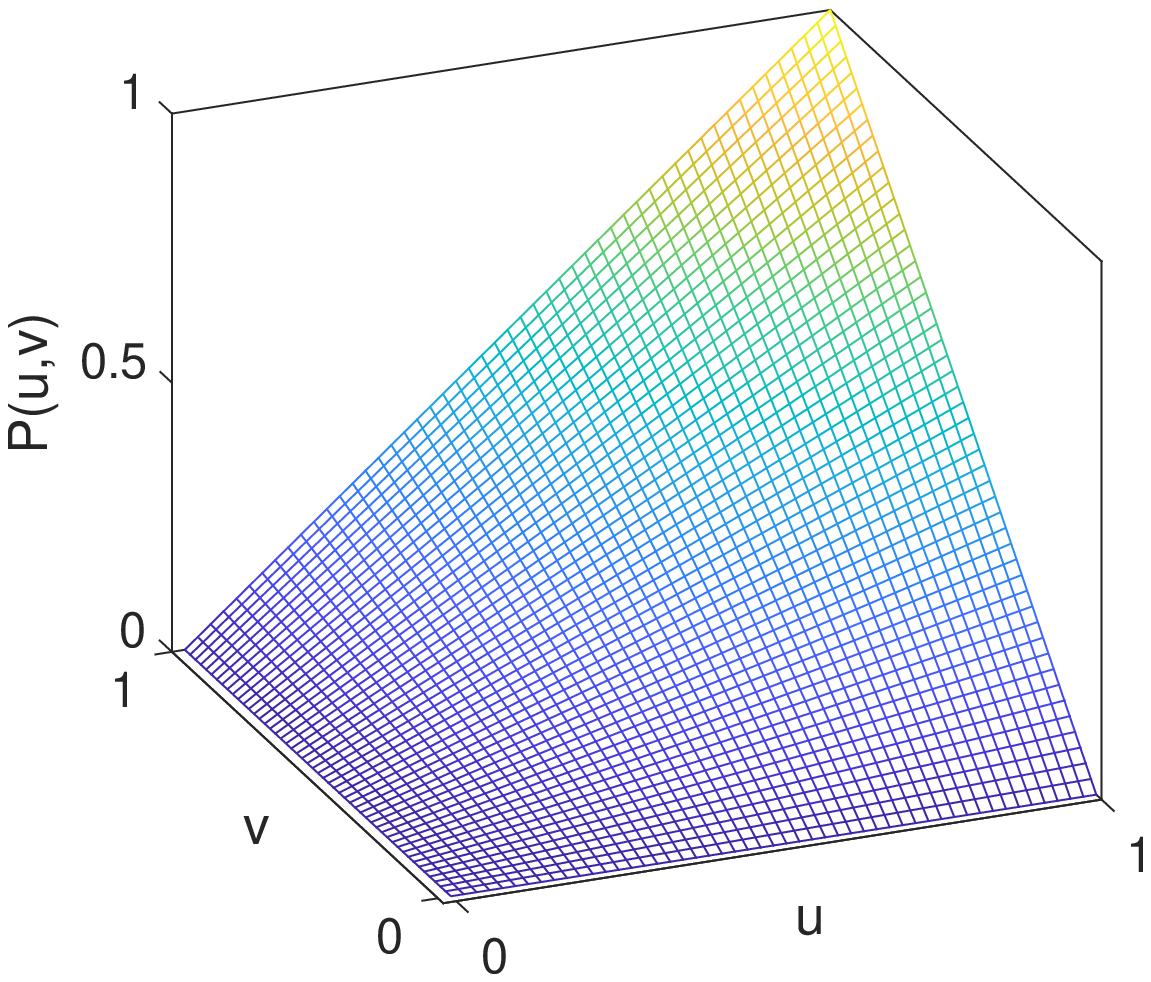}\\ \smallskip

    \end{psfrags}

   \hspace{0.38cm} \footnotesize (a) Product copula
  \end{minipage}
  \hfill
  \begin{minipage}{0.27\textwidth}
   \centering
    \begin{psfrags}

      \psfrag{u}[cc][cc]{\footnotesize $u$}
      \psfrag{v}[cc][cc]{\footnotesize $v$}
      \psfrag{W(u,v)}[cc][cc]{\footnotesize $W(u,v) = \max(u+v-1,0)$}

     \includegraphics[clip=true,width=\textwidth]{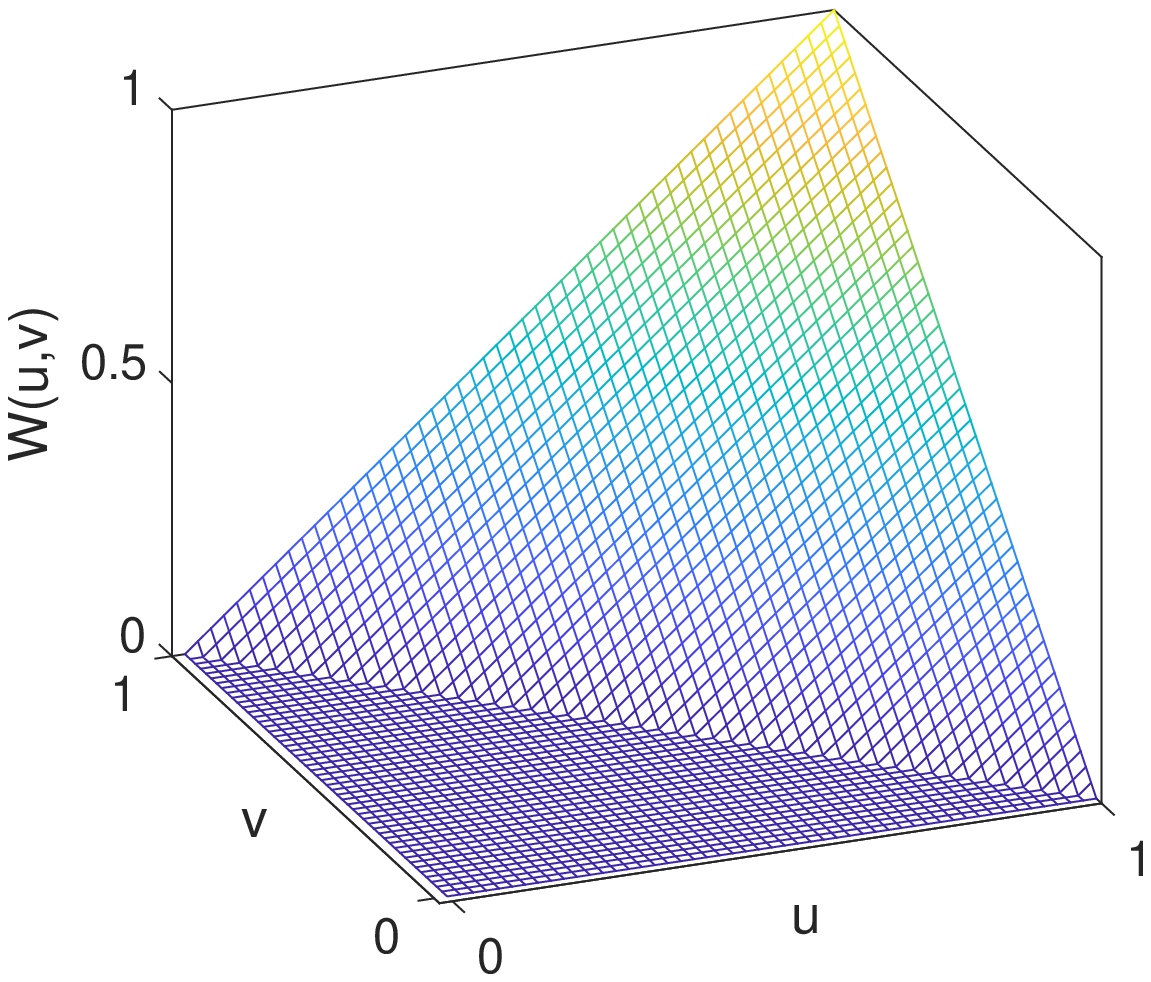}\\ \smallskip

    \end{psfrags}

   \hspace{0.38cm} \footnotesize (b) Lower bound copula
  \end{minipage}
  \hfill
  \begin{minipage}{0.27\textwidth}
   \centering
    \begin{psfrags}

      \psfrag{u}[cc][cc]{\footnotesize $u$}
      \psfrag{v}[cc][cc]{\footnotesize $v$}
      \psfrag{M(u,v)}[cc][cc]{\footnotesize $M(u,v) = \min(u,v)$}

     \includegraphics[clip=true,width=\textwidth]{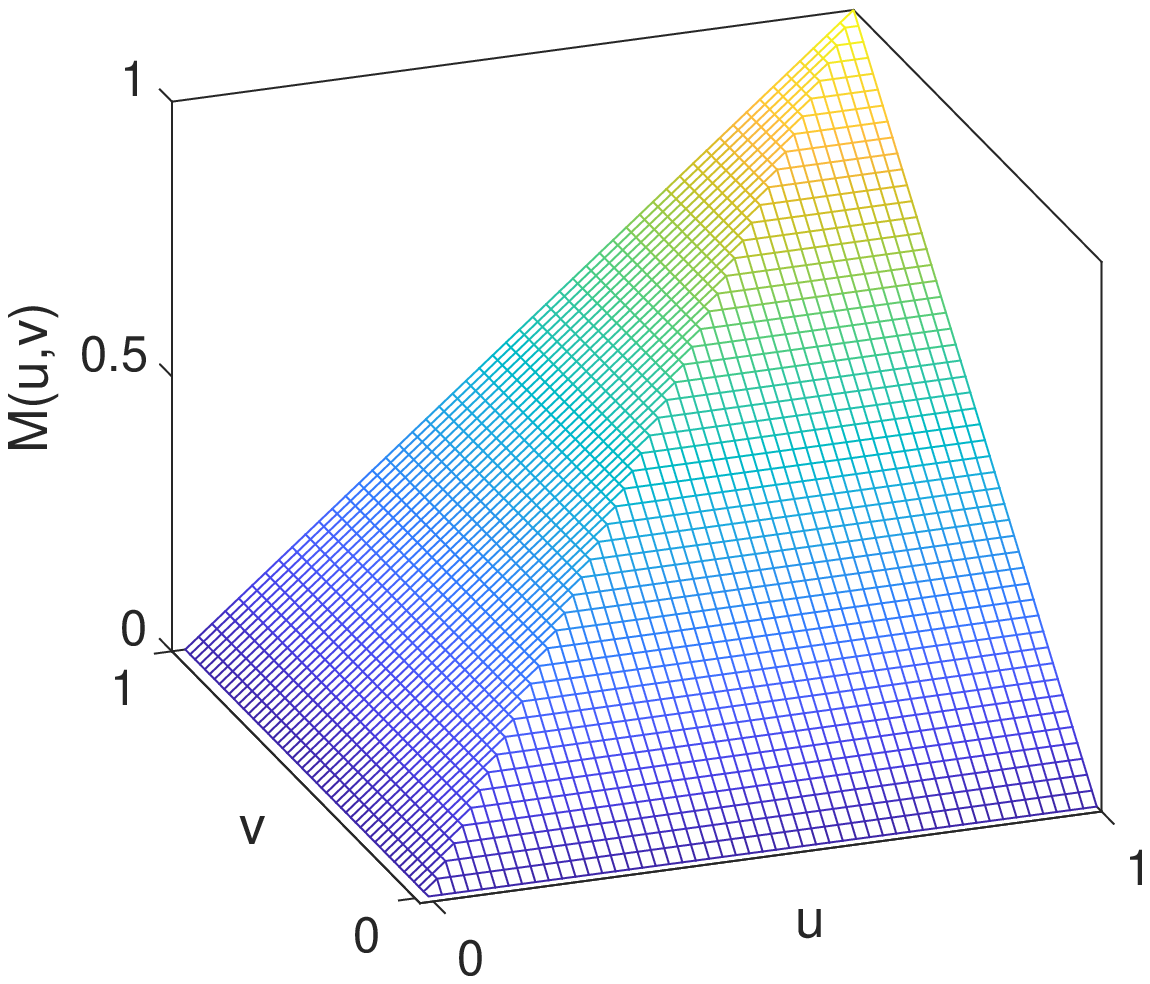}\\ \smallskip

    \end{psfrags}

   \hspace{0.38cm} \footnotesize (c) Upper bound copula
  \end{minipage}
  \hfill
  \begin{minipage}{0.05\textwidth}
   \centering

     \includegraphics[clip=true,height=4.2cm]{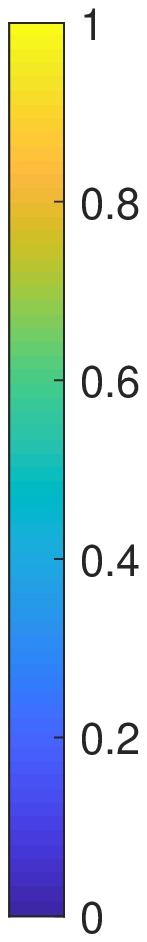}\\ \smallskip

     \footnotesize \mbox{}
  \end{minipage}

  \caption{Copula functions $\Pi$, $W$, and $M$, in (a), (b), and (c), respectively.}
\label{fig:copulasbasicas}
\end{figure*}
Furthermore, since $X$ and $Y$ are independent continuous random variables if and only if their joint distribution function is the product of the marginal distributions, i.e., $F_{X,Y}=F_X(x)F_Y(y)$, it follows that the unique underlying copula for independence is
\begin{equation}\label{eq:copulaPi}
    \Pi(u,v)=uv.
\end{equation}
We will also call this function the \textit{product copula}, which we show in Figure~\ref{fig:copulasbasicas}(a).

Sklar's Theorem also shows that for any copula function $C$ and any univariate continuous distribution functions $G$ and $H$ the function defined by
\begin{equation}\label{eq:SklarThm2}
    K(x,y) = C(G(x),H(y))
\end{equation}
is in fact a joint distribution with marginals $K(x,+\infty)=G(x)$ and $K(+\infty,y)=H(y).$ Therefore, (\ref{eq:SklarThm2}) provides a flexible way to build multivariate probability models with any given copula and any desired marginal distributions. An immediate corollary of (\ref{eq:SklarThm}) or (\ref{eq:SklarThm2}) is that the underlying copula function can be expressed as follows:
\begin{equation}\label{eq:SklarThm3}
    C(u,v) = K(G^{-1}(u),H^{-1}(v)),
\end{equation}
and therefore from any given joint distribution $K$ we can obtain its underlying copula function.

Two important copulas arise as a consequence of the \textit{Fréchet-Hoeffding bounds} for joint distribution functions (see \cite{Frechet1951} and \cite{Hoeffding1949}), combined with Sklar's Theorem. In particular, any copula $C$ is bounded by:
\begin{equation}\label{eq:FHbounds}
    W(u,v) \leq C(u,v) \leq M(u,v),
\end{equation}
where $W(u,v)=\max\{u+v-1,0\}$ and $M(u,v)=\min\{u,v\}$ are also copulas, ant therefore such bounds are best possible. Figure~\ref{fig:copulasbasicas}(b) and (c) show $W$ and $M$, which we will denote as the lower and upper bound copulas.

\subsection{Measures of association}
\label{sec:measuresassociation}

One of the early attempts to provide a formal definition for a dependence measure was the work by \cite{Renyi1959}, published in the same year as \textit{Sklar's Theorem}. Therefore, it did not take into consideration copula functions and the proposal was too restrictive. In 1981 \cite{Schweizer1981} published copula-based dependence measures for the first time. The first book devoted to copula functions was published in 1999 by \cite{Nelsen1999} and includes a formal definition of what should be understood as a \textit{dependence measure} and a \textit{concordance measure}, which we describe below.

\subsubsection{Dependence measures}

\textbf{Definition 2} A numeric measure $\delta$ of association between two continuous random variables $X$ and $Y$ whose copula is $C$ is a \textit{dependence measure} if it satisfies the following properties (where we write $\delta_{X,Y}$ or $\delta_C$ if convenient):
\begin{enumerate}
    \item $\delta_{X,Y}=\delta_{Y,X}\,;$
    \item $0\leq\delta_{X,Y}\leq 1\,;$
    \item $\delta_{X,Y}=0$ if and only $X$ and $Y$ are independent;
    \item $\delta_{X,Y}=1$ if and only if each of $X$ and $Y$ is almost surely a strictly monotone function the other;
    \item if $\alpha$ and $\beta$ are almost surely strictly monotone functions on the supports of the random variables $X$ and $Y,$ respectively, then $\delta_{\alpha(X),\beta(Y)}=\delta_{X,Y}\,;$
    \item if $\{(X_n,Y_n): n = 1,2,\ldots\}$ is a sequence of vectors of continuous random variables with copulas $C_n,$ and if $\{C_n:n=1,2,\ldots\}$ converges pointwise to $C,$ then $\lim_{n\rightarrow\infty}\delta_{C_n}=\delta_C\,.$
\end{enumerate}

As mentioned by \cite{Schweizer1981} and \cite{Nelsen1999}, any suitably normalized measure of distance between the surfaces $z=C(u,v)$ and $z=\Pi(u,v),$ that is, any $L_p$-distance, yields a symmetric nonparametric measure of dependence for $p\in[1,\infty).$ In particular, the $L_1$-distance is known as \textit{Schweizer-Wolff dependence measure}
\begin{equation}\label{eq:SchweizerWolff}
    \sigma_C = 12\int_0^1\!\!\!\int_0^1 |C(u,v) - \Pi(u,v)|\,dudv,
\end{equation}
which is based on the works of \cite{Renyi1959} and \cite{Schweizer1981}, and is formally defined in \cite{Nelsen1999}. We will use this measure throughout the rest of the paper to quantify the dependence between two continuous random variables. More generally, for $1\leq p<\infty$ any $L_p$-distance between $C$ and $\Pi$ given by
\begin{equation}\label{eq:Lp}
    \delta_C(p) := \left(k_p\int_0^1\!\!\!\int_0^1|C(u,v)-\Pi(u,v)|^p dudv\right)^{1/p}
\end{equation}
also satisfies Definition 2 for a dependence measure. The constant $k_p$ is chosen so that the quantity in (\ref{eq:Lp}) is equal to $1$ whenever $C=W$ or $C=M$. It can be shown that the $L_{\infty}$-distance satisfies all but property 4 in Definition 2 for a dependence measure \cite{Nelsen1999}
\begin{equation}\label{eq:Linf}
    \Lambda_C = 4\sup |C(u,v) - \Pi(u,v)|.
\end{equation}

Regarding property 5 it is worth mentioning that for the particular case when $\alpha$ and $\beta$ are strictly increasing the underlying copula for $(X,Y)$ is the same as for $(\alpha(X), \beta(Y))$. Thus, the dependence measure does not change given a fixed copula $C$:
\begin{equation}\label{eq:scaleCopula}
    C_{\alpha(X),\beta(Y)} = C_{X,Y} \qquad \mbox{if}\,\,\alpha\uparrow\,,\,\beta\uparrow .
\end{equation}

\subsubsection{Concordance measures}

\textbf{Definition 3} \cite{Scarsini1984} A numeric measure $\kappa$ of association between two continuous random variables $X$ and $Y$ whose copula is $C$ is a \textit{concordance measure} if it satisfies the following properties (where we write $\kappa_{X,Y}$ or $\kappa_C$ if convenient):
\begin{enumerate}
    \item $-1\leq\kappa_{X,Y}\leq 1\,,\,$ $\kappa_{X,X}=1\,,\,$ $\kappa_{X,-X}=-1\,;$
    \item $\kappa_{X,Y}=\kappa_{Y,X}\,;$
    \item if $X$ and $Y$ are independent then $\kappa_{X,Y} = 0\,;$
    \item $\kappa_{-X,Y}=\kappa_{X,-Y}=-\kappa_{X,Y}\,;$
    \item if $C_1\leq C_2$ then $\kappa_{C_1}\leq\kappa_{C_2}\,;$
    \item if $\{(X_n,Y_n): n = 1,2,\ldots\}$ is a sequence of vectors of continuous random variables with copulas $C_n,$ and if $\{C_n:n=1,2,\ldots\}$ converges pointwise to $C,$ then $\lim_{n\rightarrow\infty}\kappa_{C_n}=\kappa_C\,.$
\end{enumerate}

It is important to emphasize that a zero value for a concordance measure does not imply independence, in contrast to dependence measures. If $\kappa$ is a concordance measure for continuous random variables $X$ and $Y$ it has been proved (see \cite{Nelsen1999}) that the following properties hold:
\begin{itemize}
    \item[a)] if $Y=g(X)$ (almost surely) with $g$ a strictly increasing function then $\kappa_{X,Y}=1\,;$
    \item[b)] if $Y=g(X)$ (almost surely) with $g$ a strictly decreasing function then $\kappa_{X,Y}=-1\,;$
    \item[c)] if $\alpha$ and $\beta$ are almost surely strictly increasing functions on the supports of the random variables $X$ and $Y,$ respectively, then $\kappa_{\alpha(X),\beta(Y)}=\kappa_{X,Y}\,.$
\end{itemize}

The concordance measure by Spearman \cite{Spearman1904} was published quite before Sklar's Theorem, but it has been shown that it can be expressed just in terms of the underlying copula \cite{Nelsen1999}:
\begin{equation}\label{eq:Spearman}
    \rho_C \,=\, 12\int_0^1\!\!\!\int_0^1 (C(u,v) - \Pi(u,v))\,dudv
\end{equation}

Spearman's $\rho_C$ and Schweizer's $\sigma_C$ have an interpretation in terms of the following type of dependence as discussed in \cite{Lehmann1966}:

\textbf{Definition 4} Random variables $X$ and $Y$ are \textit{positively quadrant dependent} (PQD) if for all $(x,y)\in\mathbb{R}^2$:
\begin{equation}\label{eq:PQD1}
    P(X\leq x, Y\leq y) \geq P(X\leq x)P(Y\leq y)
\end{equation}
or equivalently:
\begin{equation}\label{eq:PQD2}
    P(X>x, Y>y) \geq P(X>x)P(Y>y)
\end{equation}
As clearly explained by \cite{Nelsen1999} $X$ and $Y$ are PQD if the probability that they are simultaneously small (or simultaneously large) is greater or equal than in the case of independence. By reversing the sense of the inequalities (\ref{eq:PQD1}) and (\ref{eq:PQD2}) we get \textit{negatively quadrant dependent} (NQD) random variables, with the following interpretation: the probability that they are simultaneously small (or simultaneously large) is less or equal than in the case of independence. These types of dependence are specially relevant in regression analysis. Applying Sklar's Theorem to (\ref{eq:PQD1}) we obtain that PQD and NQD are equivalent to $C(u,v)\geq uv$ and $C(u,v)\leq uv$, respectively.

The pair $(\rho_C,\sigma_C)$ provides valuable information about dependence. Note that the integrand in (\ref{eq:SchweizerWolff}) is the absolute value of the integrand in (\ref{eq:Spearman}), which implies that $-\sigma_C\leq\rho_C\leq\sigma_C.$  Therefore, if $\rho_C=\sigma_C$ then $X$ and $Y$ are PQD, and if $\rho=-\sigma_C$ then $X$ and $Y$ are NQD. If $-\sigma_C<\rho_C<\sigma_C$ this implies that $C>\Pi$ and $C<\Pi$ in disjoint subsets of its domain. Thus, $\rho_C$ is an average value of quadrant dependence, which could be zero. If $\rho_C=0$ with $C\neq\Pi$ (i.e., the random variables are not independent) then certainly $\sigma_C>0$.

A very popular measure of association is
Pearson's correlation coefficient, given by $r_{X,Y}=\mbox{Cov}(X,Y)/\sqrt{\mbox{V}(X)\mbox{V}(Y)},$ where Cov and V denote covariance and variance, respectively. However, it is neither a dependence measure nor a concordance measure, and is prone to a number of pitfalls as discussed extensively in \cite{Embrechts2002}. Firstly, $r_{X,Y} = 0$ does not imply independence. Thus, it is not a dependence measure according to property 3 in Definition 2. Secondly, copula-based measures always exist, since by Sklar's Theorem the underlying copula always exists. However, $r_{X,Y}$ depends on the existence of second moments of the random variables. Additionally, the covariance can be expressed as: $$\mbox{Cov}(X,Y)=\int_0^1\!\!\!\int_0^1(C(u,v)-\Pi(u,v))\,dF_X^{-1}(u)dF_Y^{-1}(v).$$
Although it involves the underlying copula function, it also depends on information from the marginal distributions, which have no information about the dependence between the random variables. Therefore, for a fixed copula (i.e., for a specific dependence structure), $r_{X,Y}$ may provide different values just by changing the marginal distributions.

\subsection{Empirical copula and measures of association}

An empirical copula is a consistent nonparametric estimation of the true underlying copula (see \cite{Deheuvels1979}), based on an observed random sample $\{(x_1,y_1),\ldots,(x_n,y_n)\}$ from a random vector $(X,Y),$ and which is a function $C_n:\{0,1/n,\ldots,(n-1)/n,1\}^2\rightarrow[0,1]$:
\begin{equation}\label{eq:copem}
    C_n\left(\frac{i}{n},\frac{j}{n}\right) = \frac{1}{n}\sum_{k\,=\,1}^n \mathbb{1}_{\{x_k\,\leq\,x_{(i)}\,,\,y_k\,\leq\,y_{(j)}\}},
\end{equation}
where $x_{(i)}$ represents the order statistic with rank $i\in\{1,\ldots,n\}$, $C_n(i/n,0) = 0 = C_n(0,j/n)$, and $\mathbb{1}_A$ is the indicator function of condition $A$. In terms of such empirical copula it is possible to obtain an empirical version of Spearman's concordance measures (\ref{eq:Spearman}) \cite{Nelsen1999}:
\begin{equation}\label{eq:rhon}
    \rho_n \,=\, \frac{12}{n^2-1}\sum_{i=1}^n\sum_{j=1}^n\left[C_n\left(\frac{i}{n},\frac{j}{n}\right)-\frac{ij}{n^2}\right],
\end{equation}
and adapting (\ref{eq:rhon}) we obtain an empirical estimation of Schweizer's dependence measure (\ref{eq:SchweizerWolff}):
\begin{equation}\label{eq:sigman}
    \sigma_n = \frac{12}{n^2-1}\sum_{i=1}^n\sum_{j=1}^n\left|C_n\left(\frac{i}{n},\frac{j}{n}\right)-\frac{ij}{n^2}\right|.
\end{equation}

\section{Copula Visualizations}
\label{sec:copulavisualizations}

Copula functions can be shown through 3D surfaces such as the ones in Figure~\ref{fig:copulasbasicas}, or as 2D contour plots. However, these visualizations of a single copula are generally not useful for extracting information regarding dependence, mainly because in practice it is necessary to compare a copula with $\Pi$, $W$ and/or $M$. Thus, other visualizations are needed in order to exploit the information in copulas. In this section we describe two main types of copula visualizations. The first approach is often used in the literature and consists of a transformation of a regular scatter plot. The second is a novel approach that is based on heatmaps that show differences between the empirical copula and $\Pi$. These heatmaps can also be used to color other types of visualizations (e.g., scatter plots or parallel coordinates).

\subsection{Pseudo-observations: marginal-free scatter plots}
\label{sec:pseudoobservations}

\begin{figure*}[ht!]
    \centering

  \begin{minipage}{0.27\textwidth}
   \centering

     \includegraphics[clip=true,width=\textwidth]{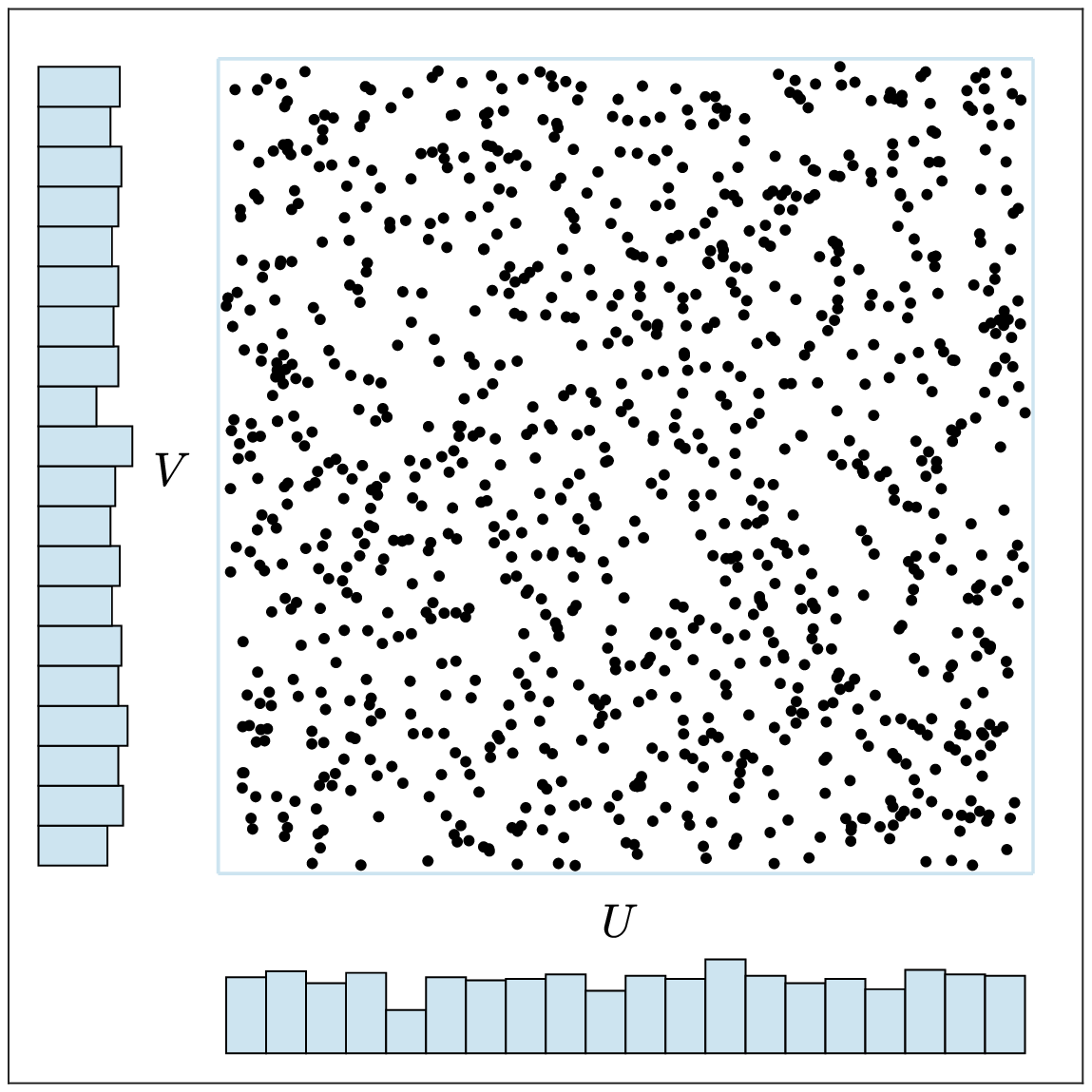}\\

   \hspace{0cm} \footnotesize (a) 
  \end{minipage}
  \hfill
  \begin{minipage}{0.27\textwidth}
   \centering

     \includegraphics[clip=true,width=\textwidth]{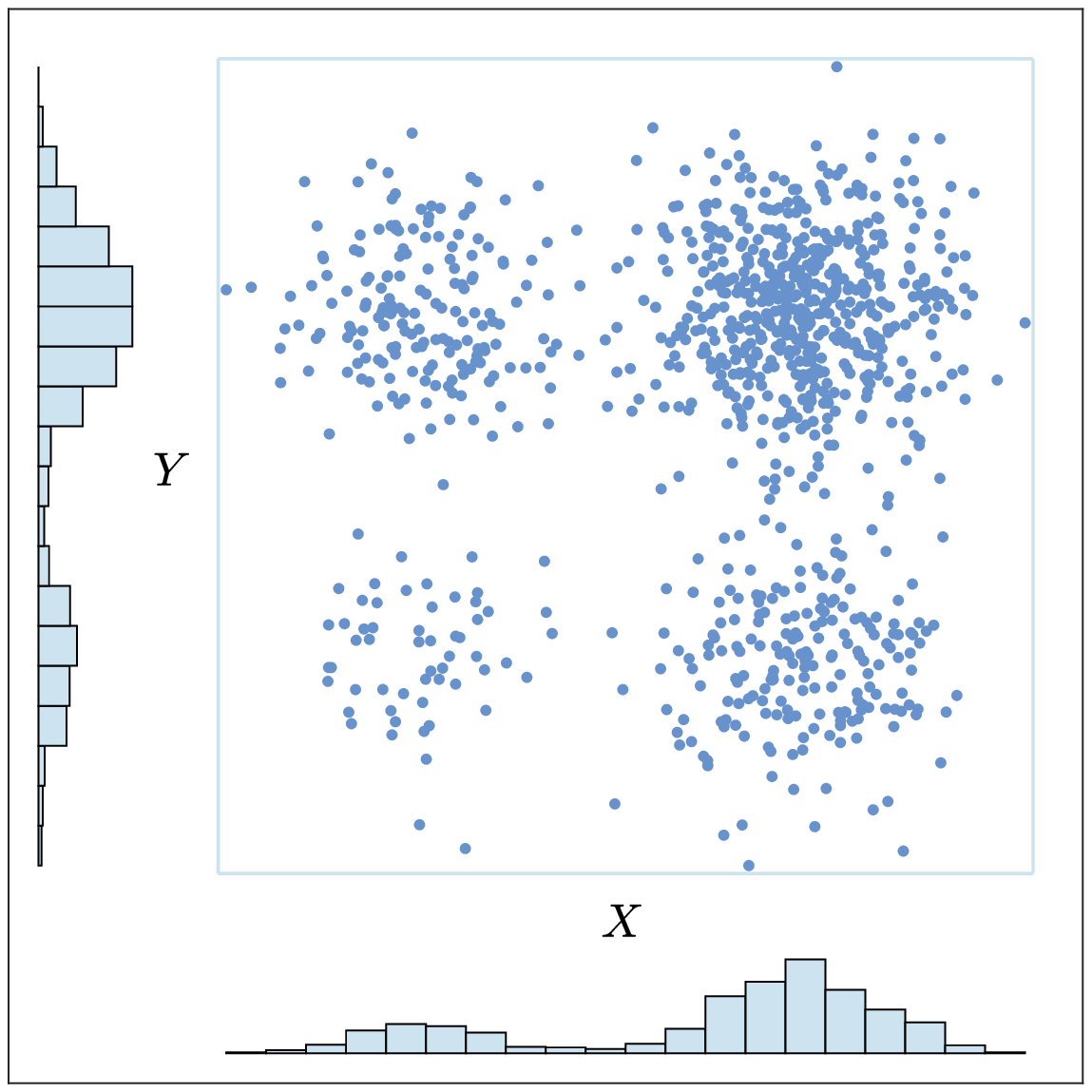}\\

   \hspace{0cm} \footnotesize (b) 
  \end{minipage}
  \hfill
  \begin{minipage}{0.27\textwidth}
   \centering

     \includegraphics[clip=true,width=\textwidth]{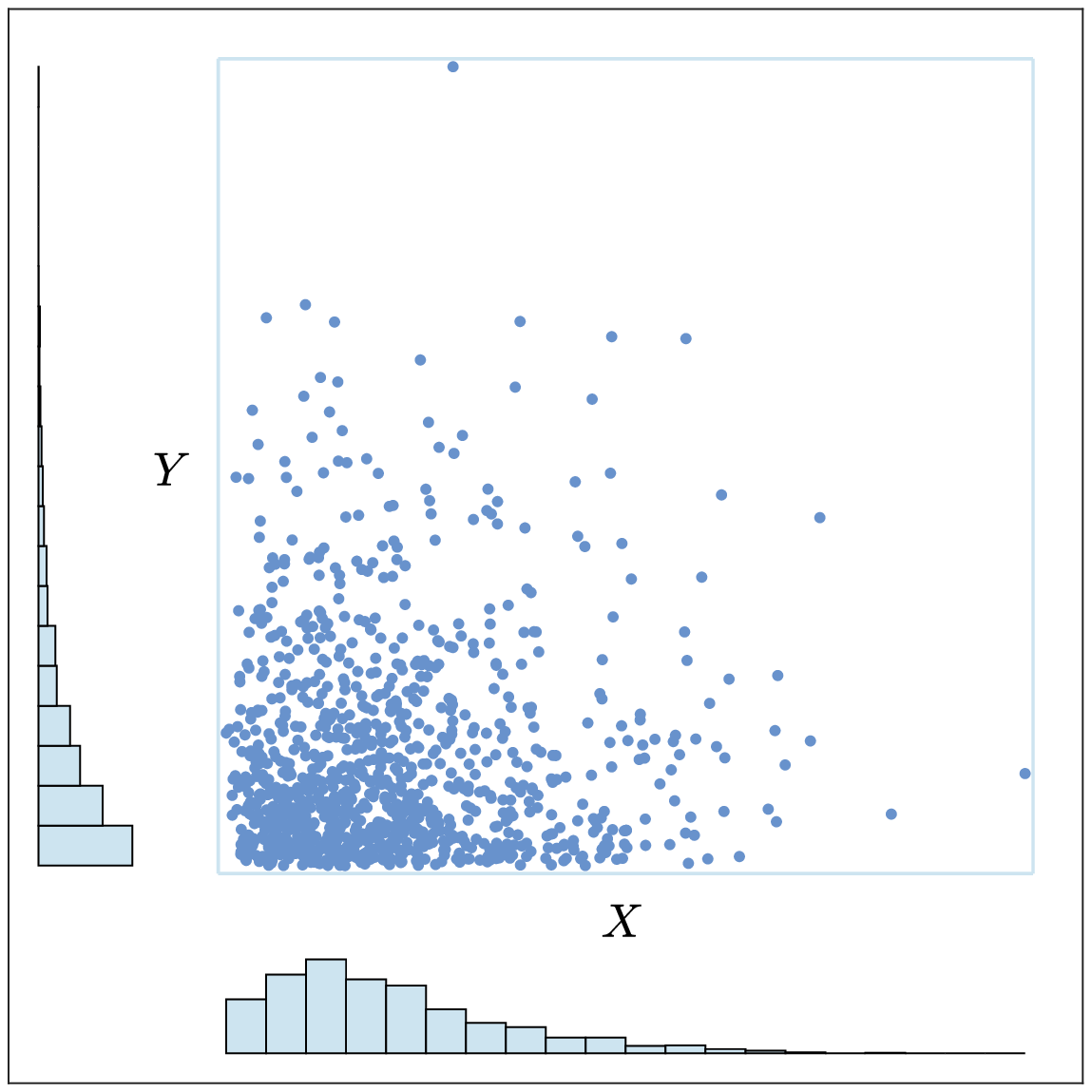}\\

   \hspace{0cm} \footnotesize (c) 
  \end{minipage}

  \caption{Three plots with the same dependence structure. The visualization in (a) is a scatter plot of two independent and uniformly-distributed random variables $U$ and $V$ over an open interval $(0,1)$. Since the marginals are uniform it is also a plot of pseudo-observations. By applying $X := F_{X}^{-1}(U)$ and $Y := F_{Y}^{-1}(V)$ for different marginals ($F_{X}$ and $F_{Y}$) we can produce very different bivariate distributions, as shown in (b) and (c). Nevertheless, the dependency between $X$ and $Y$ (in this case, independence) is identical as that of the pseudo-observations. Moreover, the pseudo-observations of (b) and (c) are precisely the ones shown in (a). Thus, the pseudo-observations allow users to examine the dependence structure while avoiding ``noise'' introduced by the marginal distributions. In other words, the pseudo-observations are marginal-free scatter plots. In this case the independence between $U$ and $V$ (and $X$ and $Y$) is more apparent in (a) since the bivariate distribution is uniform.}
\label{fig:pseudoobservations}
\end{figure*}

The most common approach for visualizing information related to a copula is through a scatter plot of ``observed values'' of the copula, known as \textit{pseudo-observations}. Let $(X,Y)$ be a random vector of continuous random variables with underlying copula $C_{X,Y}$ and marginal (continuous) distribution functions $F_X$ and $F_Y.$ If we define $U:=F_X(X)$ and $V:=F_Y(Y)$ it is well known from elementary probability theory that $U$ and $V$ are continuous uniform random variables with support in the open interval $(0,1)$. By (\ref{eq:scaleCopula}) we have $C_{X,Y}=C_{U,V}$, and therefore the observed random samples $\{(x_1,y_1),\ldots,(x_n,y_n)\}$ and $\{(u_1,v_1),\ldots,(u_n,v_n)\}$, where $u_k=F_X(x_k)$ and $v_k=F_Y(y_k)$, have the same underlying copula. Moreover, since $F_U(u)=u$ and $F_V(v)=v$ then by Sklar's Theorem:
\begin{equation}\label{eq:CopulaObs}
    F_{U,V}(u,v) = C_{U,V}(F_U(u),F_V(v)) = C_{U,V}(u,v) = C_{X,Y}(u,v).
\end{equation}
Thus, $\{(u_1,v_1),\ldots,(u_n,v_n)\}$ may be considered as marginal-free observations from copula $C_{X,Y}.$ If the marginals $F_X$ and $F_Y$ are unknown, we may replace them by their empirical estimations \cite{Wasserman2006}:
\begin{equation}\label{eq_edf}
    G_n(x)=\frac{1}{n}\sum_{k=1}^n\mathbb{1}_{\{x_k\,\leq\,x\}} \quad,\quad H_n(y)=\frac{1}{n}\sum_{k=1}^n\mathbb{1}_{\{y_k\,\leq\, y\}}.
\end{equation}
The observed values $\{(\hat{u}_1,\hat{v}_1),\ldots,(\hat{u}_n,\hat{v}_n)\},$ given by $\hat{u}_k=G_n(x_k)$ and $\hat{v}_k=H_n(y_k),$ are estimations of the observed values of the copula, i.e., the pseudo-observations \cite{Fermanian2013}, which are also marginal-free.

Assessing dependence visually through a scatter plot of pseudo-observations has an advantage over a regular scatter plot of the data. Specifically, while both use dependence information (the copula), the latter mixes it with information that has nothing to do with the dependence (the marginals). Thus, it introduces ``noise'' that can produce patterns that might be misleading. 

Figure~\ref{fig:pseudoobservations} illustrates three bivariate distributions that we have generated with the same underlying copula, and therefore the same dependence structure (in this case, independence). However, by choosing different marginals we obtain quite different scatter plots. In particular, we first generated data of two independent uniform distributions, $U$ and $V$, as shown in (a). Since the marginals are uniform the scatter plot can also be considered to be a plot of pseudo-observations. Note that the bivariate uniform distribution indicates that the variables are independent. Subsequently, in (b) and (c) we generated new data by choosing different marginals $F_{X}$ and $F_{Y}$, and setting $X := F_{X}^{-1}(U)$ and $Y := F_{Y}^{-1}(V)$. Note that the three data sets share the same copula as an immediate consequence of (\ref{eq:SklarThm2}), (\ref{eq:SklarThm3}), and (\ref{eq:scaleCopula}). Thus, the data variables in (b) and (c) are also independent. However, users generally interpret that there exists some degree of dependence between the variables due to the patterns that appear in the scatter plots. In Sec.~\ref{sec:experiments} we will show results of a study regarding how users estimate dependence through pseudo-observations and regular scatter plots.

\subsection{Copula-based heatmaps}

Since Spearman's concordance measure (\ref{eq:Spearman}) is the average of quadrant dependence $C(u,v)-\Pi(u,v)$ over $[0,1]^2,$ and Schweizer's dependence measure (\ref{eq:SchweizerWolff}) is an average of the absolute difference $|C(u,v)-\Pi(u,v)|$, we propose heatmaps as straightforward visualizations that show the following empirical differences:
\begin{equation}\label{eq:heatmaprho}
    \mathcal{H}_{\rho} = \left\{12\left[C_n(\frac{i}{n},\frac{j}{n})-\frac{ij}{n^2}\right]\,:\,i,j\in\{1,\ldots,n-1\}\right\},
\end{equation}
\begin{equation}\label{eq:heatmapsigma}
    \mathcal{H}_{\sigma} = \left\{12\left|C_n(\frac{i}{n},\frac{j}{n})-\frac{ij}{n^2}\right|\,:\,i,j\in\{1,\ldots,n-1\}\right\},
\end{equation}
where the values of $\mathcal{H}_{\rho}$ are in a $[-3,3]$ scale, and the values of $\mathcal{H}_{\sigma}$ are in a $[0,3]$ scale, as an immediate consequence of (\ref{eq:Linf}). It is important to note that the mean of the values in $\mathcal{H}_{\rho}$ and $\mathcal{H}_{\sigma}$ are nonparametric estimates of $\rho_{C}$ and $\sigma_{C}$, respectively. Thus, with an appropriate coloring scale it is possible to approximate $\rho_{C}$ and $\sigma_{C}$ by estimating the mean of the colors of the heatmaps.

We also propose using heatmaps to visualize normalized differences between $\Pi$ and the empirical copula as a proportion with respect to the the distance to the closest Fr\'echet-Hoeffding bound:   
\[
    D_{n}\left(\frac{i}{n},\frac{j}{n}\right) = 
\left\{
\begin{array}{ll}
  \displaystyle \frac{C_{n}(\frac{i}{n},\frac{j}{n})-\Pi(\frac{i}{n},\frac{j}{n})}{M(\frac{i}{n},\frac{j}{n})-\Pi(\frac{i}{n},\frac{j}{n})} & \text{if  } C_{n}(\frac{i}{n},\frac{j}{n}) \geq \Pi(\frac{i}{n},\frac{j}{n}), \\[10pt]
  { } \\ 
  \displaystyle -\frac{\Pi(\frac{i}{n},\frac{j}{n})-C_n(\frac{i}{n},\frac{j}{n})}{\Pi(\frac{i}{n},\frac{j}{n})-W(\frac{i}{n},\frac{j}{n})} & \text{if  } C(\frac{i}{n},\frac{j}{n}) < \Pi(\frac{i}{n},\frac{j}{n}),
\end{array}  
\right.
\]
for $i,j\in\{1,\ldots,n-1\}.$ The corresponding heatmap can be defined as:
\begin{equation}\label{eq:Dn}
    \mathcal{H} = \left\{ D_{n}(\frac{i}{n},\frac{j}{n}):\,i,j\in\{1,\ldots,n-1\} \right\}.
\end{equation}

\begin{figure*}[ht!]
    \centering

  \begin{minipage}{0.25\textwidth}
   \centering
     Scatter plot \\ \smallskip

     \hspace{-0.3cm} \includegraphics[clip=true,height=3.3cm]{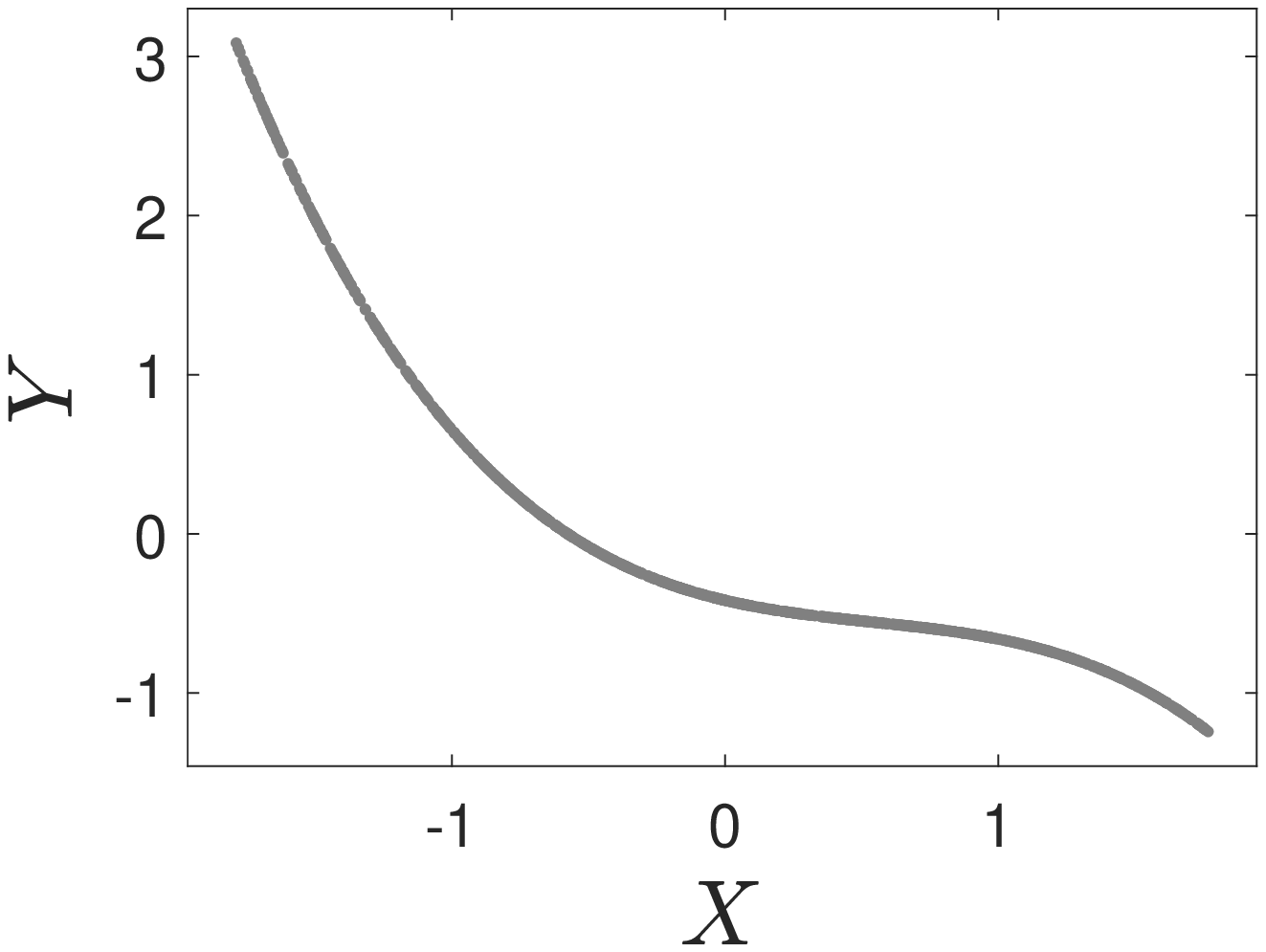}\\

   \hspace{0.27cm} \footnotesize (a) 
  \end{minipage}
  \hfill
  \begin{minipage}{0.23\textwidth}
   \centering
     $\mathcal{H}_{\rho}$ \\ \smallskip

     \includegraphics[clip=true,height=3.3cm]{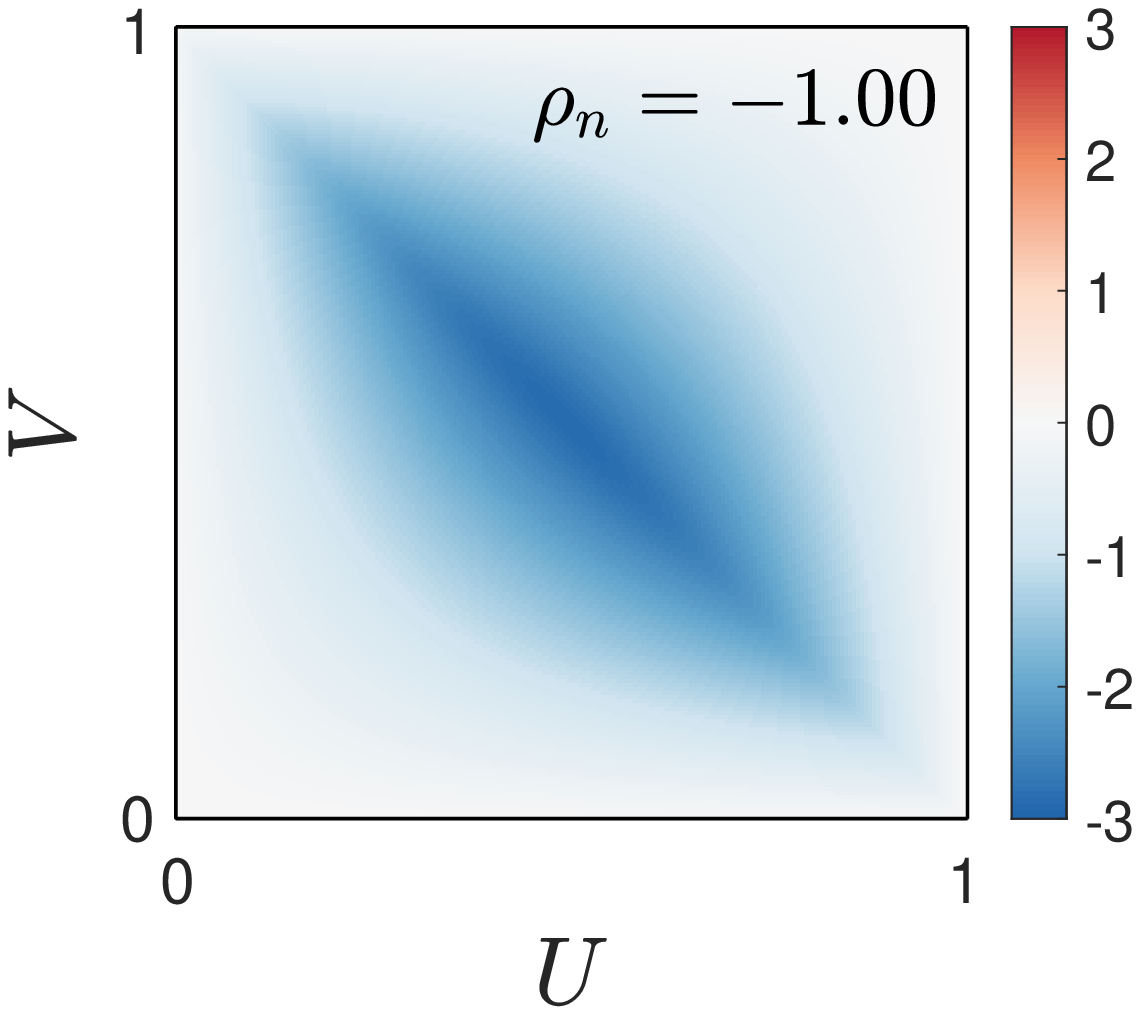}\\

   \hspace{-0.13cm} \footnotesize (b) 
  \end{minipage}
  \hfill
  \begin{minipage}{0.23\textwidth}
   \centering
        $\mathcal{H}_{\sigma}$ \\ \smallskip

     \includegraphics[clip=true,height=3.3cm]{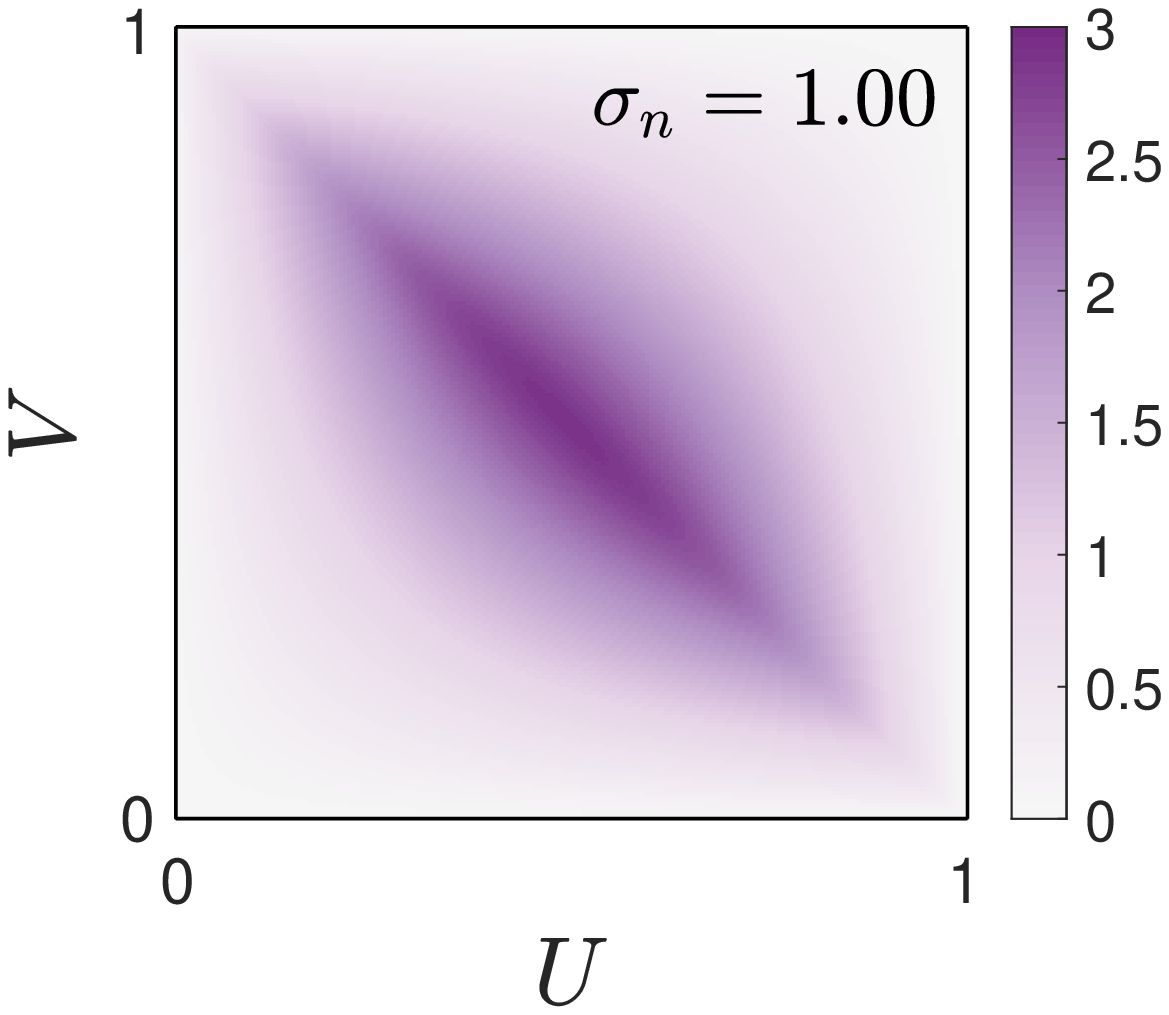}\\

   \hspace{-0.25cm} \footnotesize (c) 
  \end{minipage}
  \hfill
  \begin{minipage}{0.23\textwidth}
   \centering
     $\mathcal{H}$\\ \smallskip

     \includegraphics[clip=true,height=3.3cm]{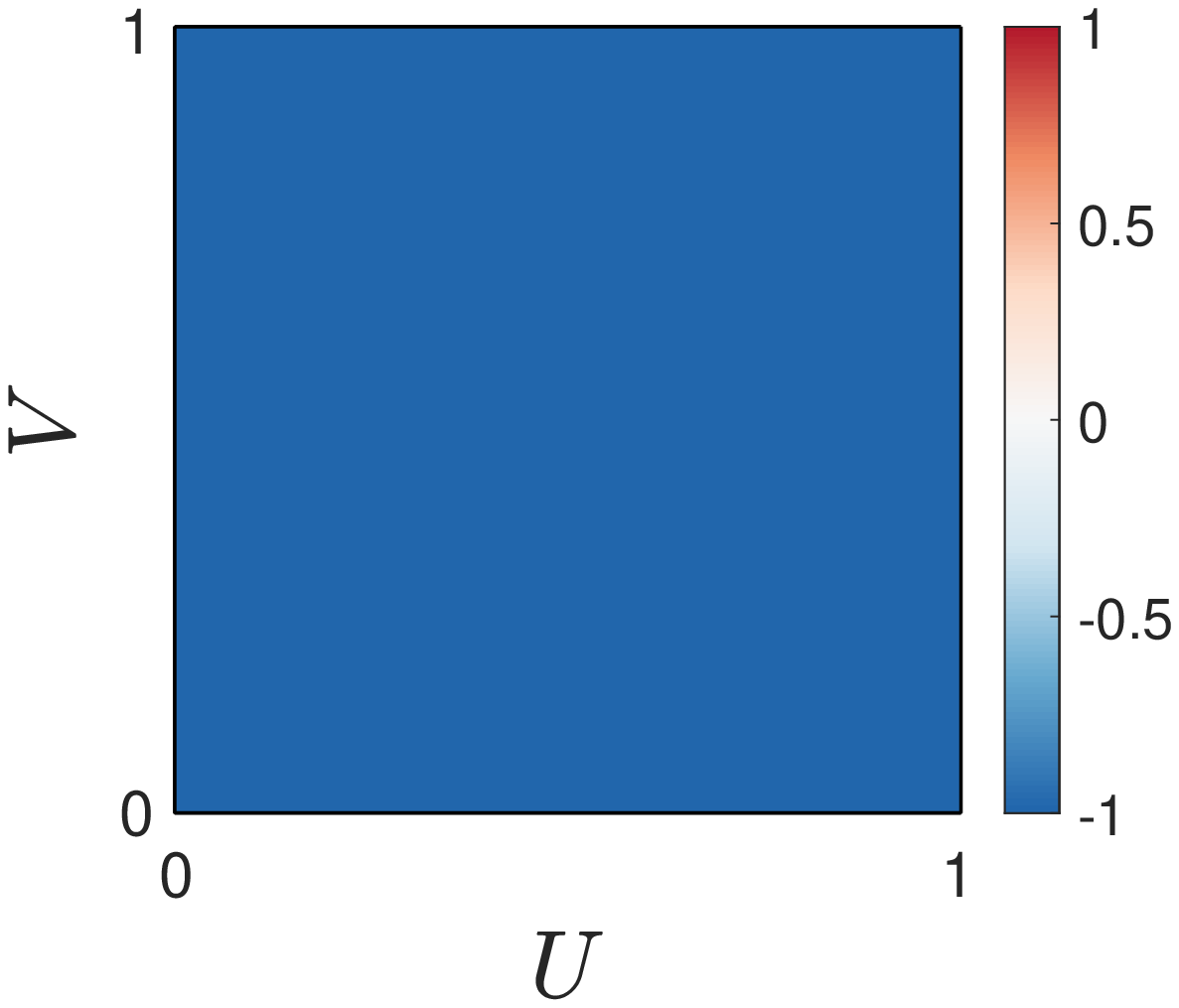}\\

   \hspace{-0.32cm} \footnotesize (d) 
  \end{minipage}
  
  \mbox{} \medskip

  \begin{minipage}{0.25\textwidth}
   \centering

     \hspace{-0.3cm} \includegraphics[clip=true,height=3.3cm]{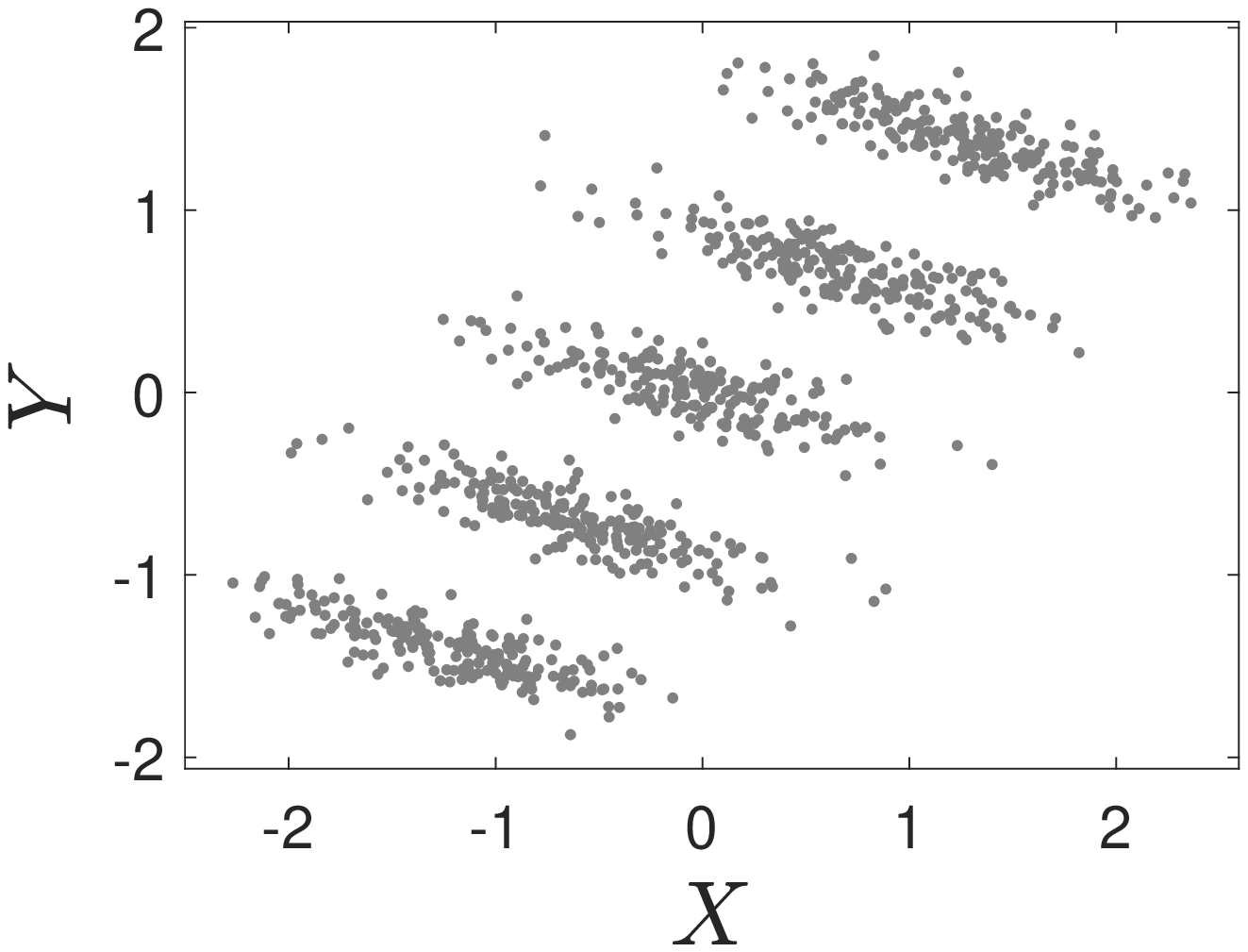}\\

   \hspace{0.27cm}  \footnotesize (e) 
  \end{minipage}
  \hfill
  \begin{minipage}{0.23\textwidth}
   \centering

     \includegraphics[clip=true,height=3.3cm]{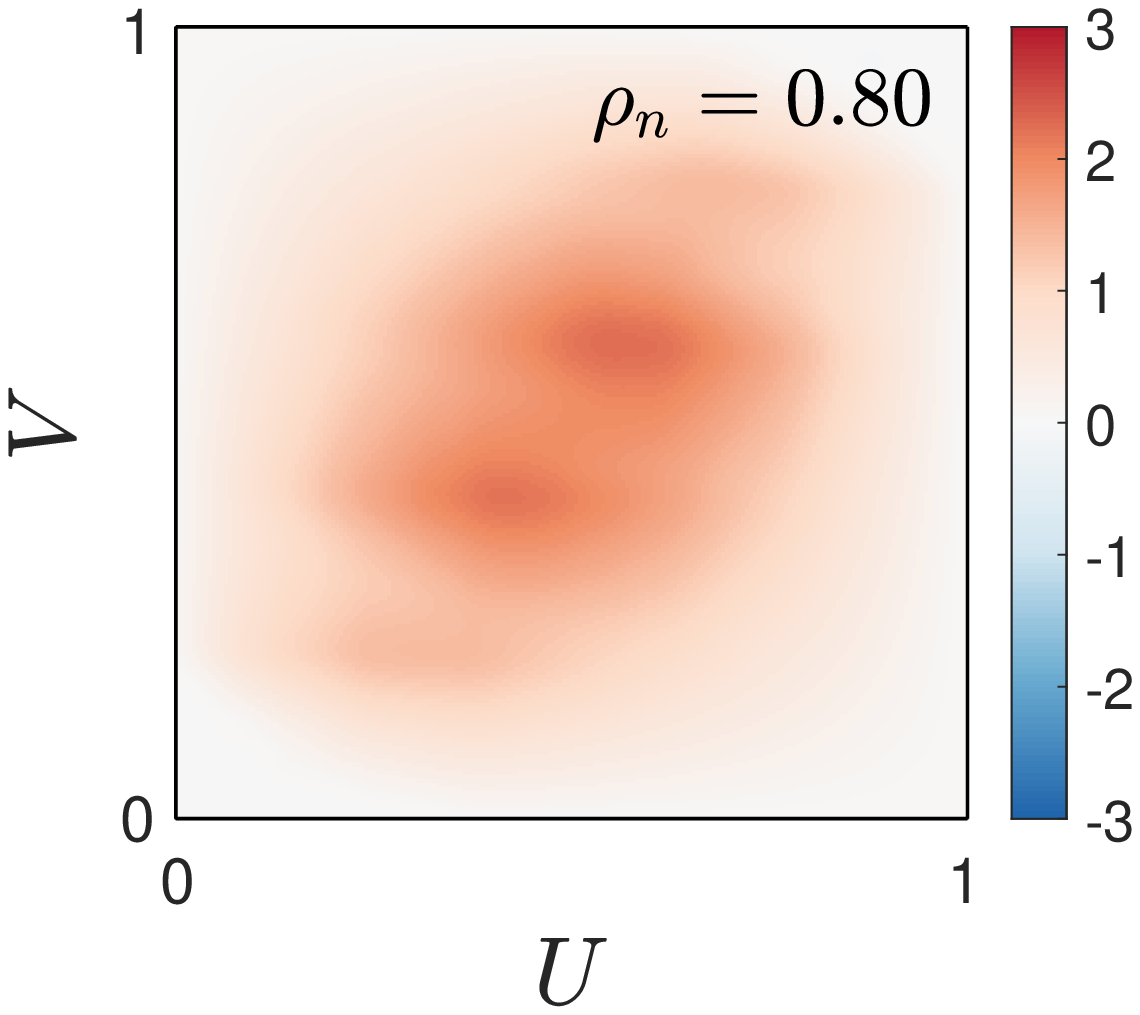}\\

   \hspace{-0.13cm} \footnotesize (f) 
  \end{minipage}
  \hfill
  \begin{minipage}{0.23\textwidth}
   \centering

     \includegraphics[clip=true,height=3.3cm]{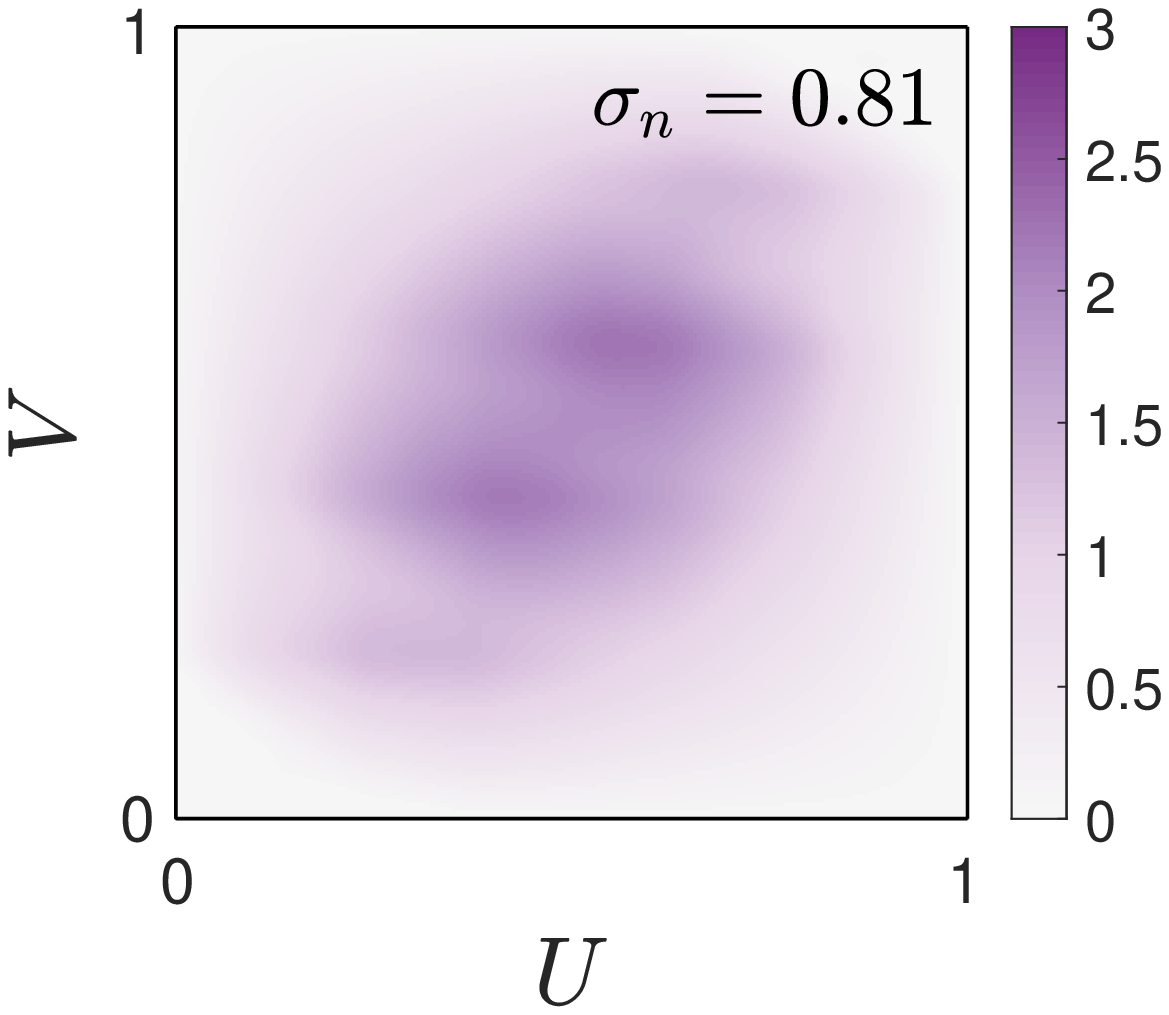}\\

   \hspace{-0.25cm} \footnotesize (g) 
  \end{minipage}
  \hfill
  \begin{minipage}{0.23\textwidth}
   \centering

     \includegraphics[clip=true,height=3.3cm]{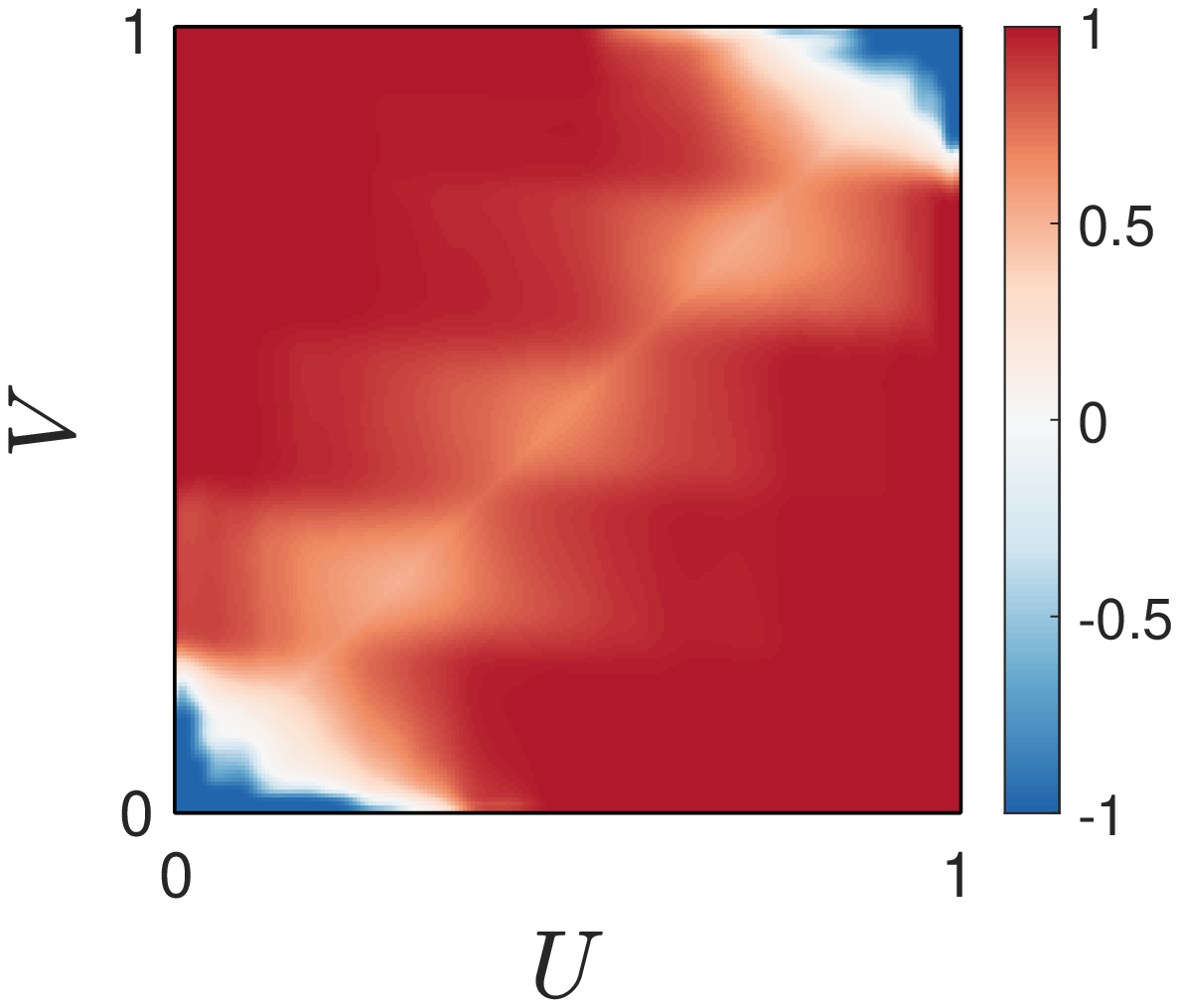}\\

   \hspace{-0.32cm} \footnotesize (h) 
  \end{minipage} 
  
    \mbox{} \medskip
 
  \begin{minipage}{0.25\textwidth}
   \centering

     \hspace{-0.3cm} \includegraphics[clip=true,height=3.3cm]{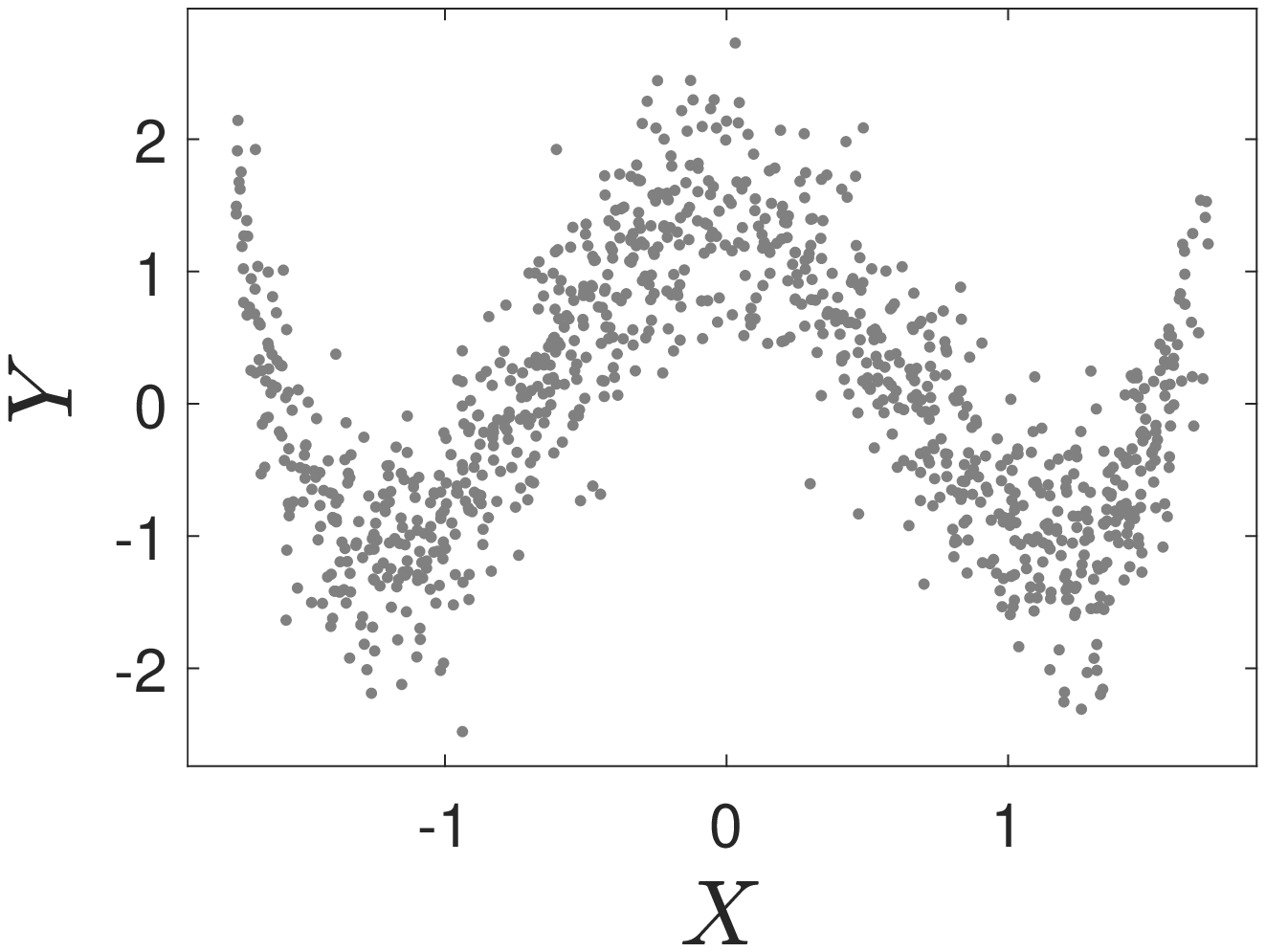}\\

   \hspace{0.27cm}  \footnotesize (i) 
  \end{minipage}
  \hfill
  \begin{minipage}{0.23\textwidth}
   \centering

     \includegraphics[clip=true,height=3.3cm]{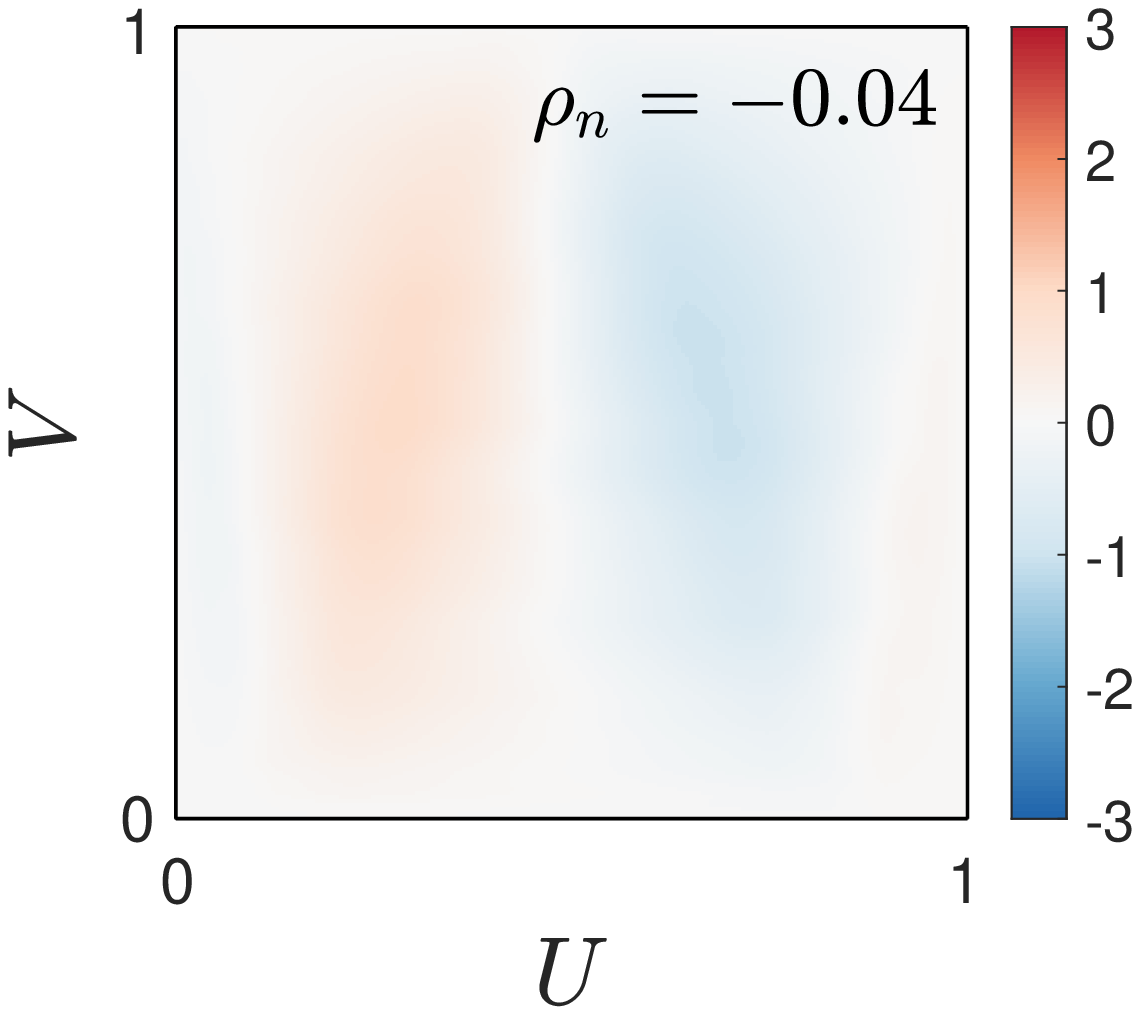}\\

   \hspace{-0.13cm} \footnotesize (j) 
  \end{minipage}
  \hfill
  \begin{minipage}{0.23\textwidth}
   \centering

     \includegraphics[clip=true,height=3.3cm]{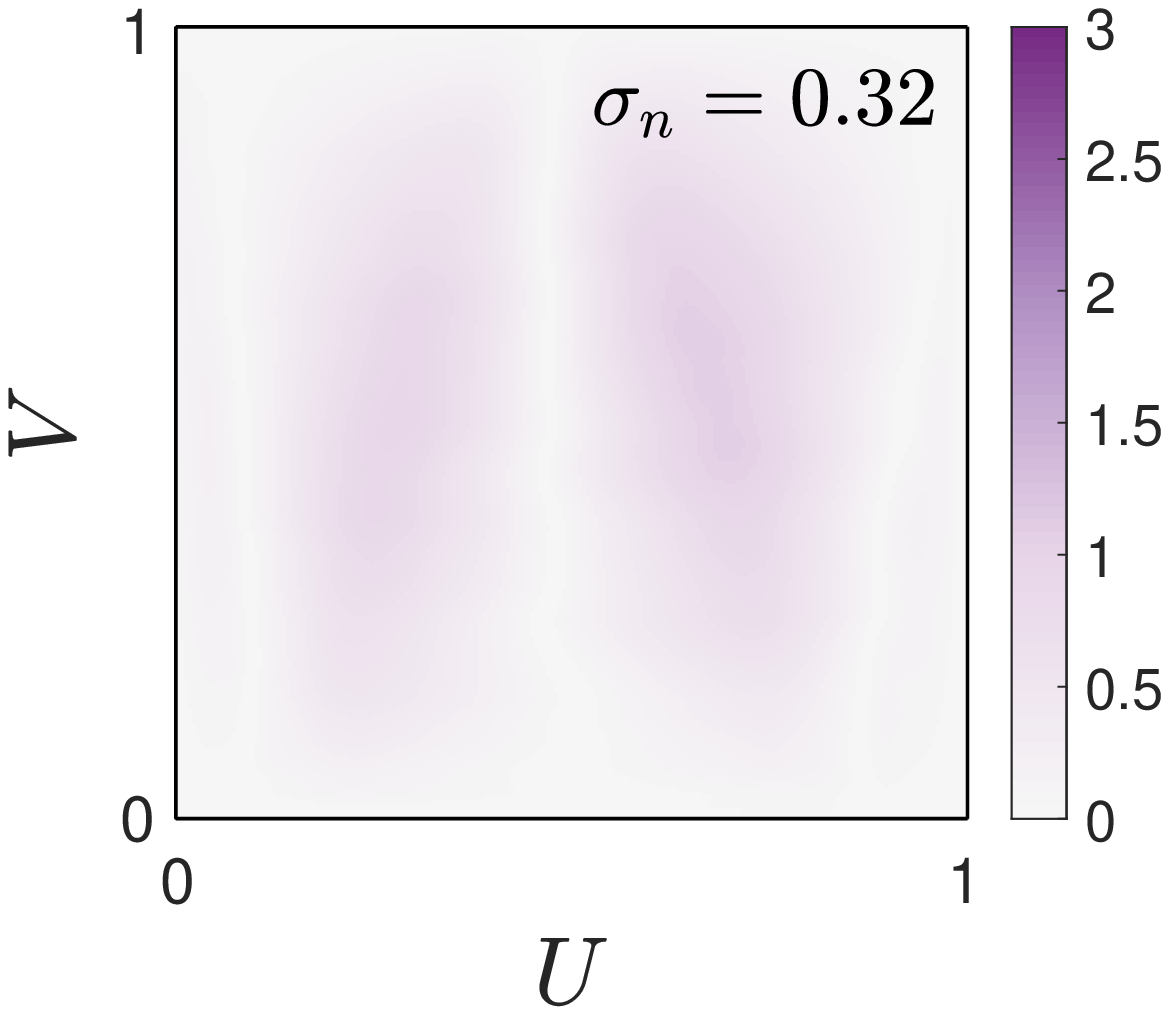}\\

   \hspace{-0.25cm} \footnotesize (k) 
  \end{minipage}
  \hfill
  \begin{minipage}{0.23\textwidth}
   \centering

     \includegraphics[clip=true,height=3.3cm]{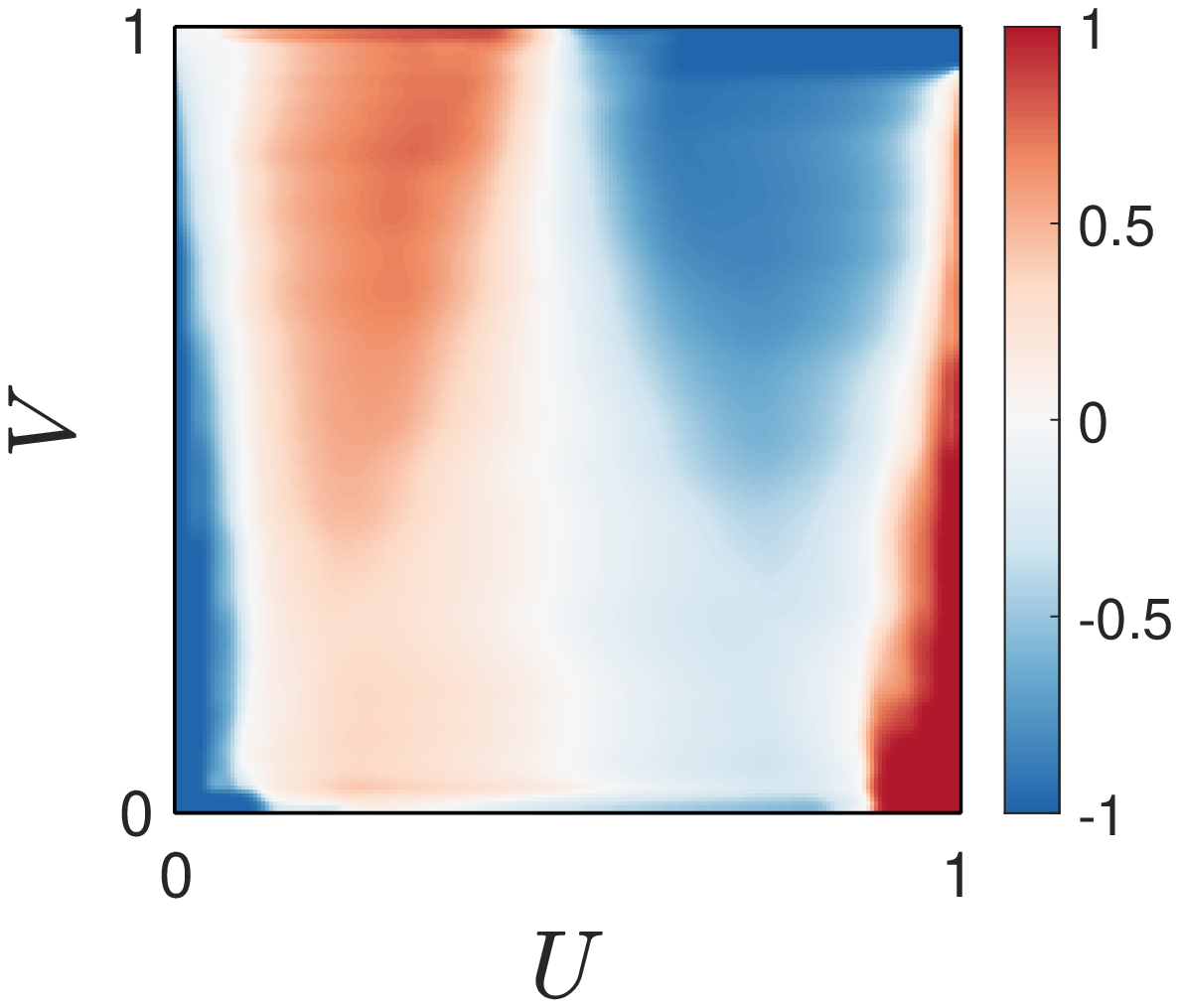}\\

   \hspace{-0.32cm} \footnotesize (l) 
  \end{minipage}  

  \mbox{} \medskip

  \begin{minipage}{0.25\textwidth}
   \centering

     \hspace{-0.3cm} \includegraphics[clip=true,height=3.3cm]{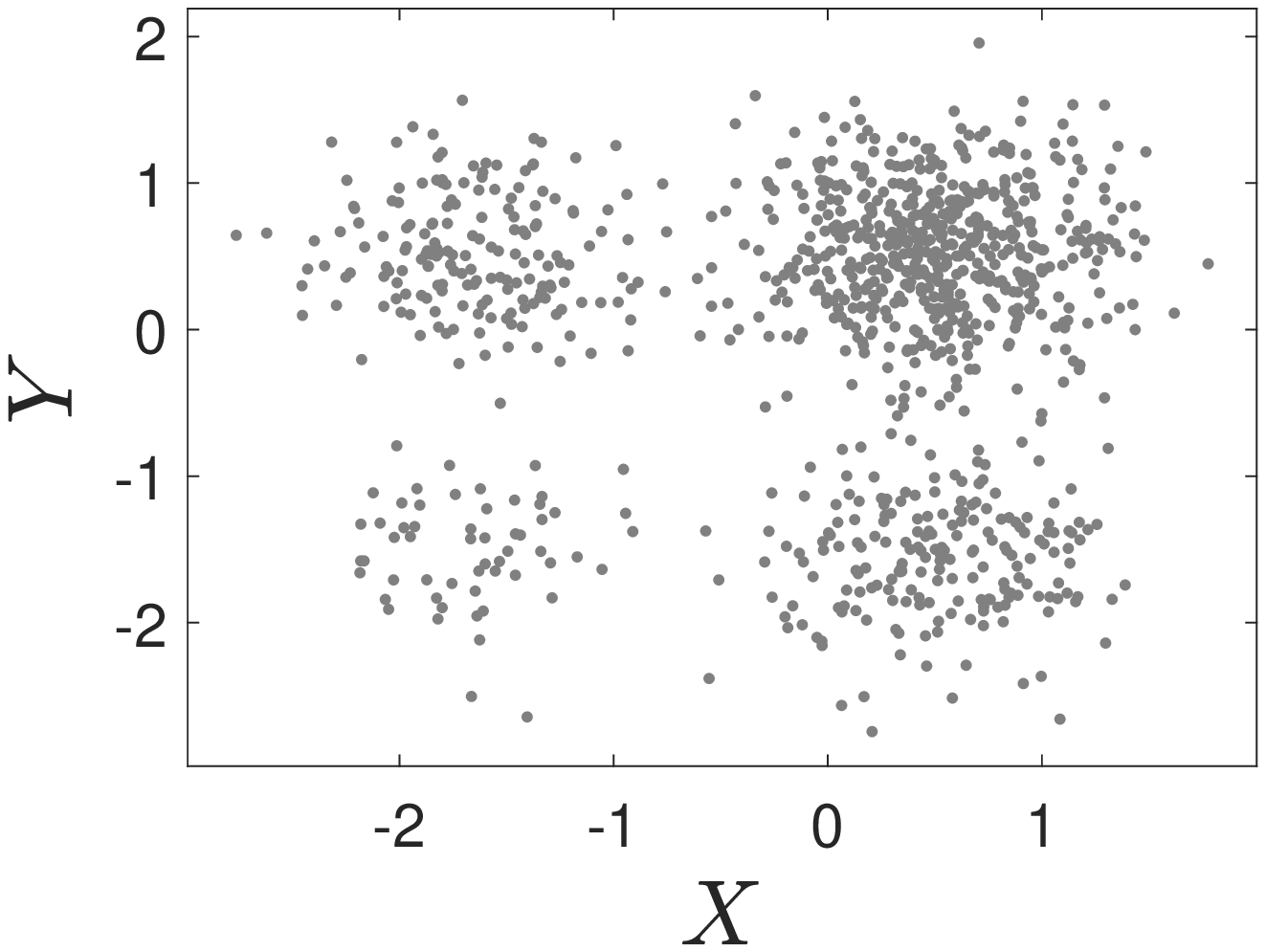}\\

   \hspace{0.27cm}  \footnotesize (m) 
  \end{minipage}
  \hfill
  \begin{minipage}{0.23\textwidth}
   \centering

     \includegraphics[clip=true,height=3.3cm]{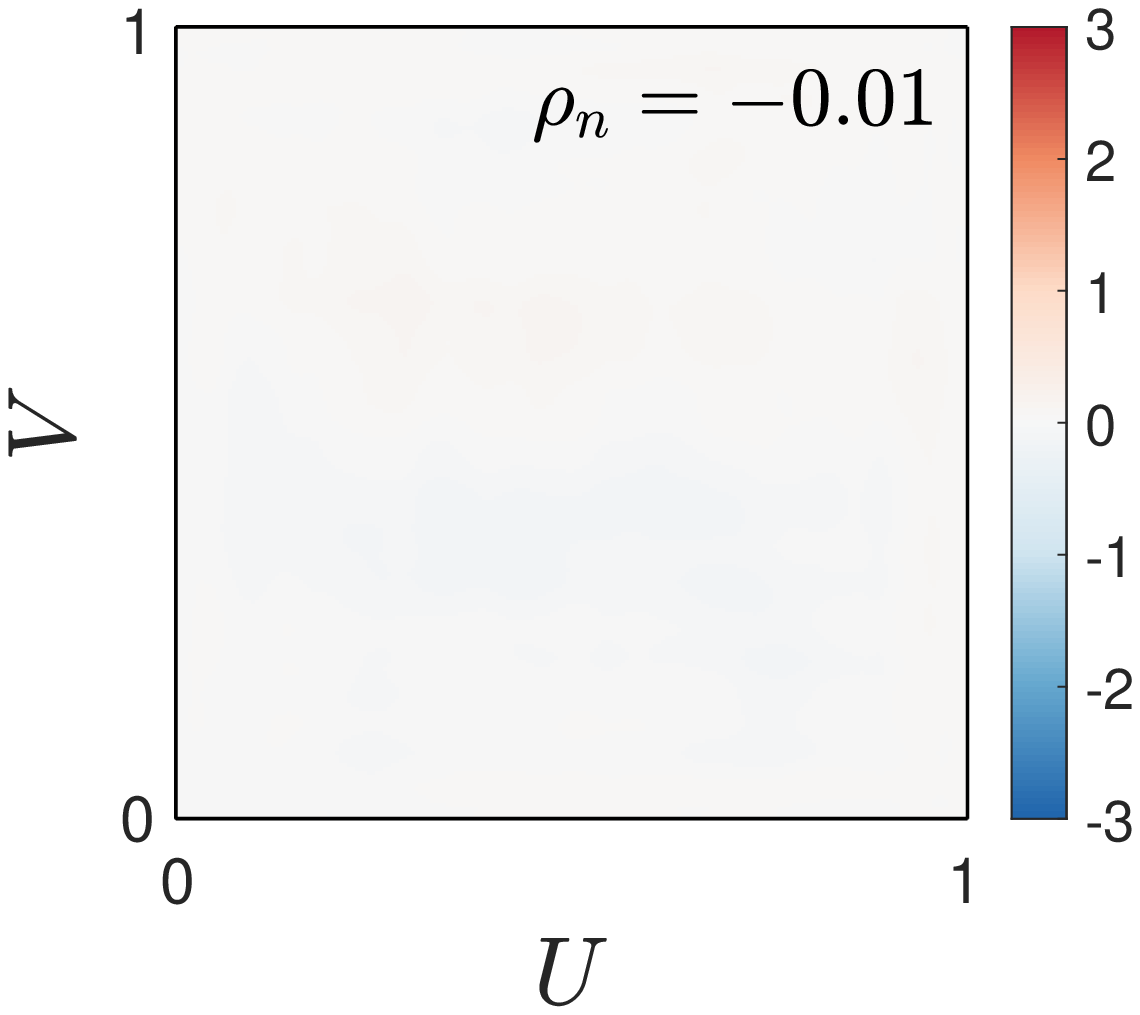}\\
     
   \hspace{-0.13cm} \footnotesize (n) 
  \end{minipage}
  \hfill
  \begin{minipage}{0.23\textwidth}
   \centering

     \includegraphics[clip=true,height=3.3cm]{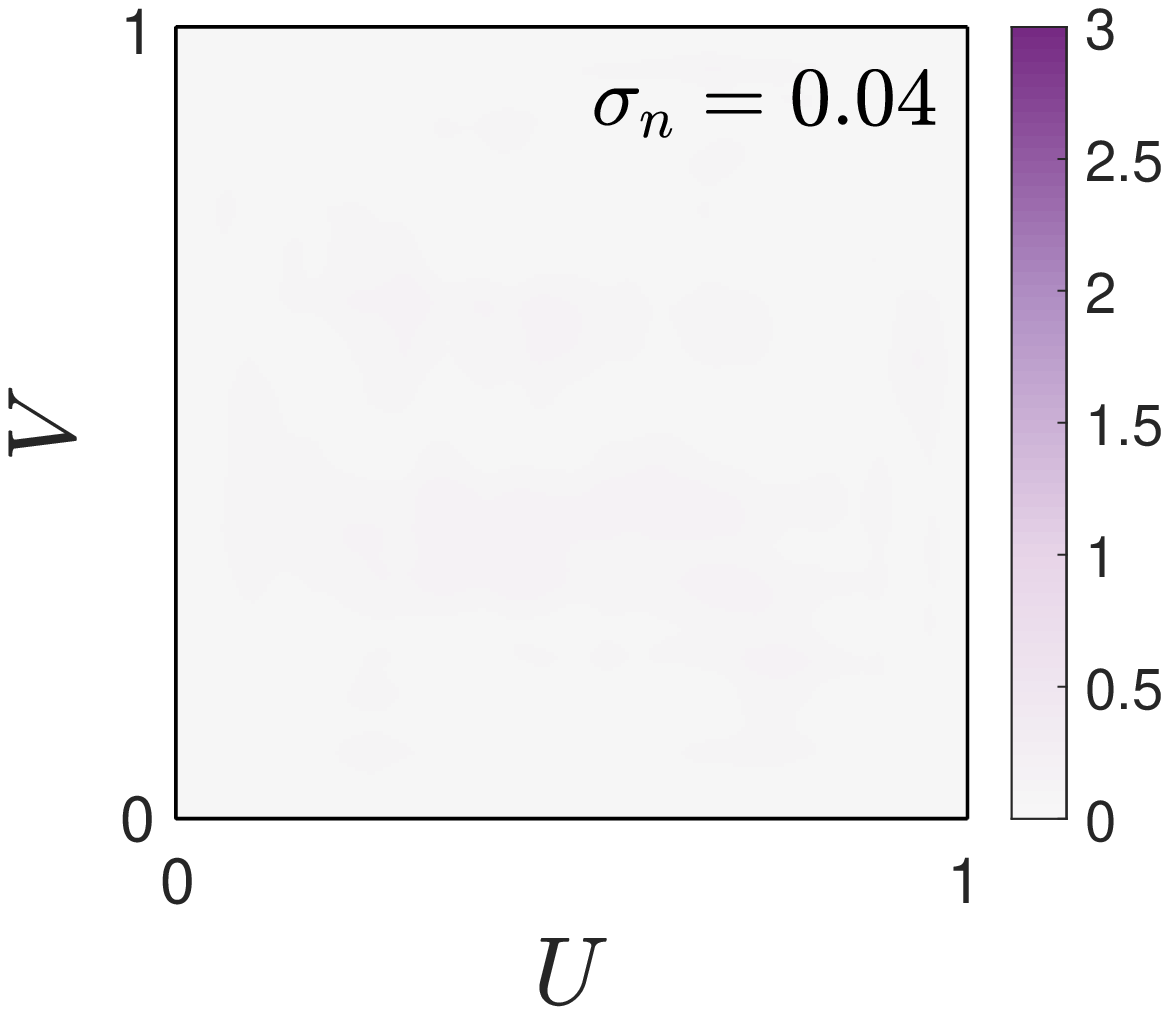}\\
     
   \hspace{-0.25cm} \footnotesize (o) 
  \end{minipage}
  \hfill
  \begin{minipage}{0.23\textwidth}
   \centering

     \includegraphics[clip=true,height=3.3cm]{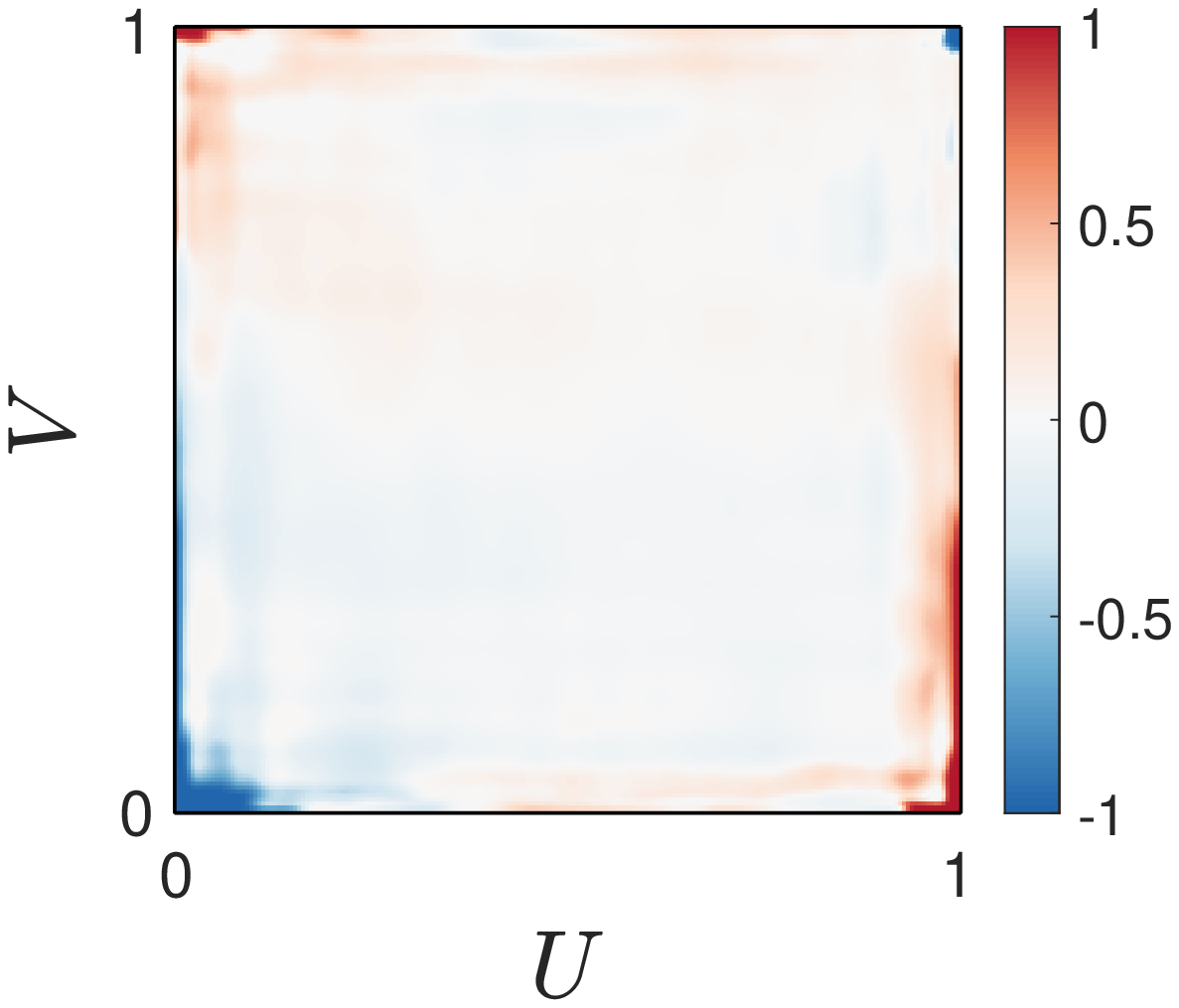}\\
     
   \hspace{-0.32cm} \footnotesize (p) 
  \end{minipage}

  \caption{Copula heatmaps $\mathcal{H}_{\rho}$, $\mathcal{H}_{\sigma}$, $\mathcal{H}$ for four distributions (one in each row) with different dependence structures. The average colors (i.e., values) of $\mathcal{H}_{\rho}$ and $\mathcal{H}_{\sigma}$ are estimates of $\rho_{C}$ and $\sigma_{C}$. However, it is difficult to estimate these values visually, in part because the scale of the colors does not match the scale of the association measures. Furthermore, the large number of values close to 0 can also hamper the estimation, regardless of the color palette used in the visualization. Alternatively, the colors in $\mathcal{H}$ stand out more and allow users to detect ascending (red) or descending (blue) trends (i.e., deviations from independence) in the data more clearly.} 
\label{fig:heatmaps}
\end{figure*}

\begin{figure*}[ht!]
    \centering

  \begin{minipage}{0.23\textwidth}
   \centering
     $\mathcal{H}$ + pseudo-observations \\ \medskip

     \hspace{-0.35cm} \includegraphics[clip=true,height=3.3cm]{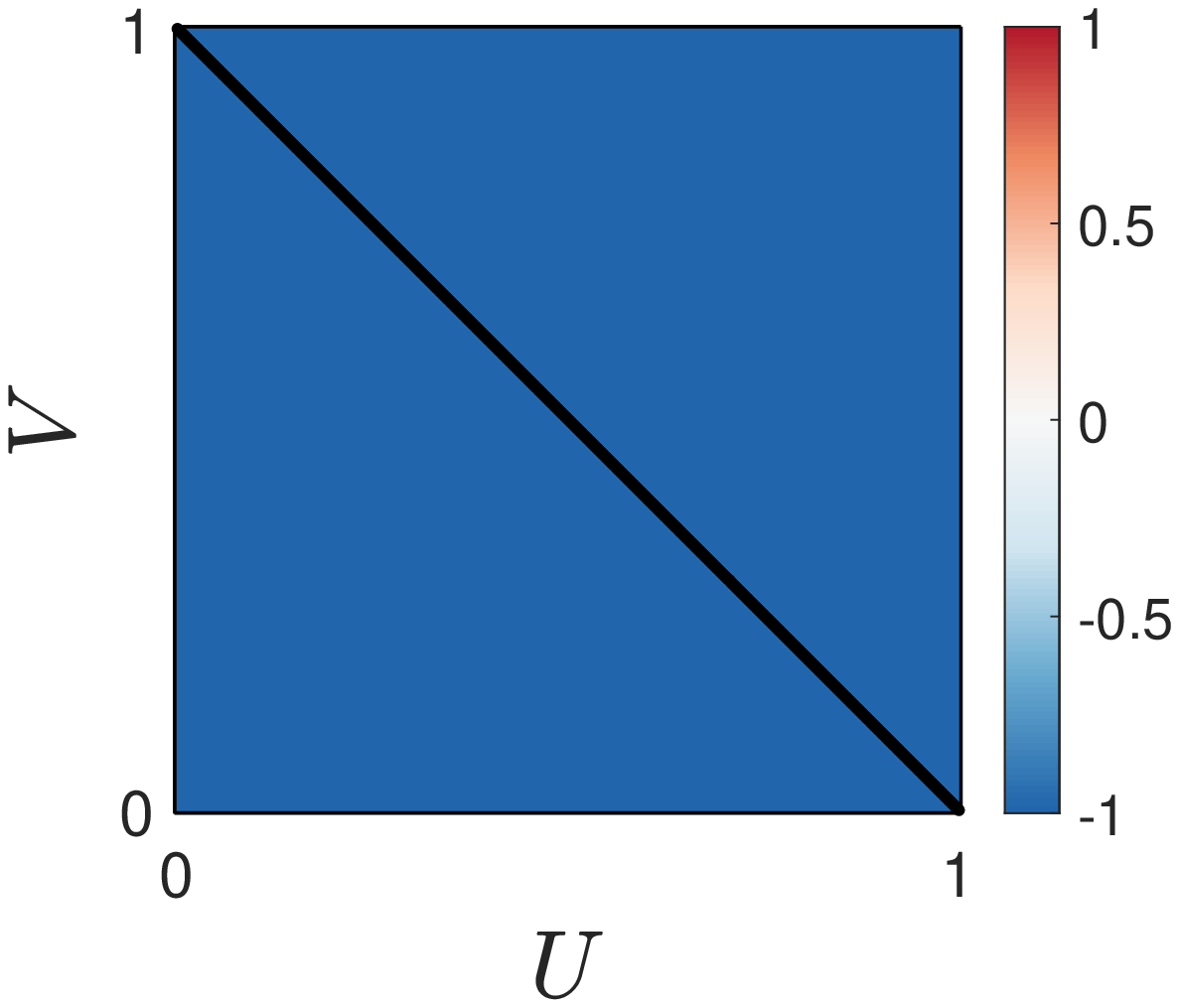}\\

   \hspace{-0.56cm} \footnotesize (a) 
  \end{minipage}
  \hfill
  \begin{minipage}{0.23\textwidth}
   \centering
     Colored pseudo-observations \\ \medskip

     \includegraphics[clip=true,height=3.3cm]{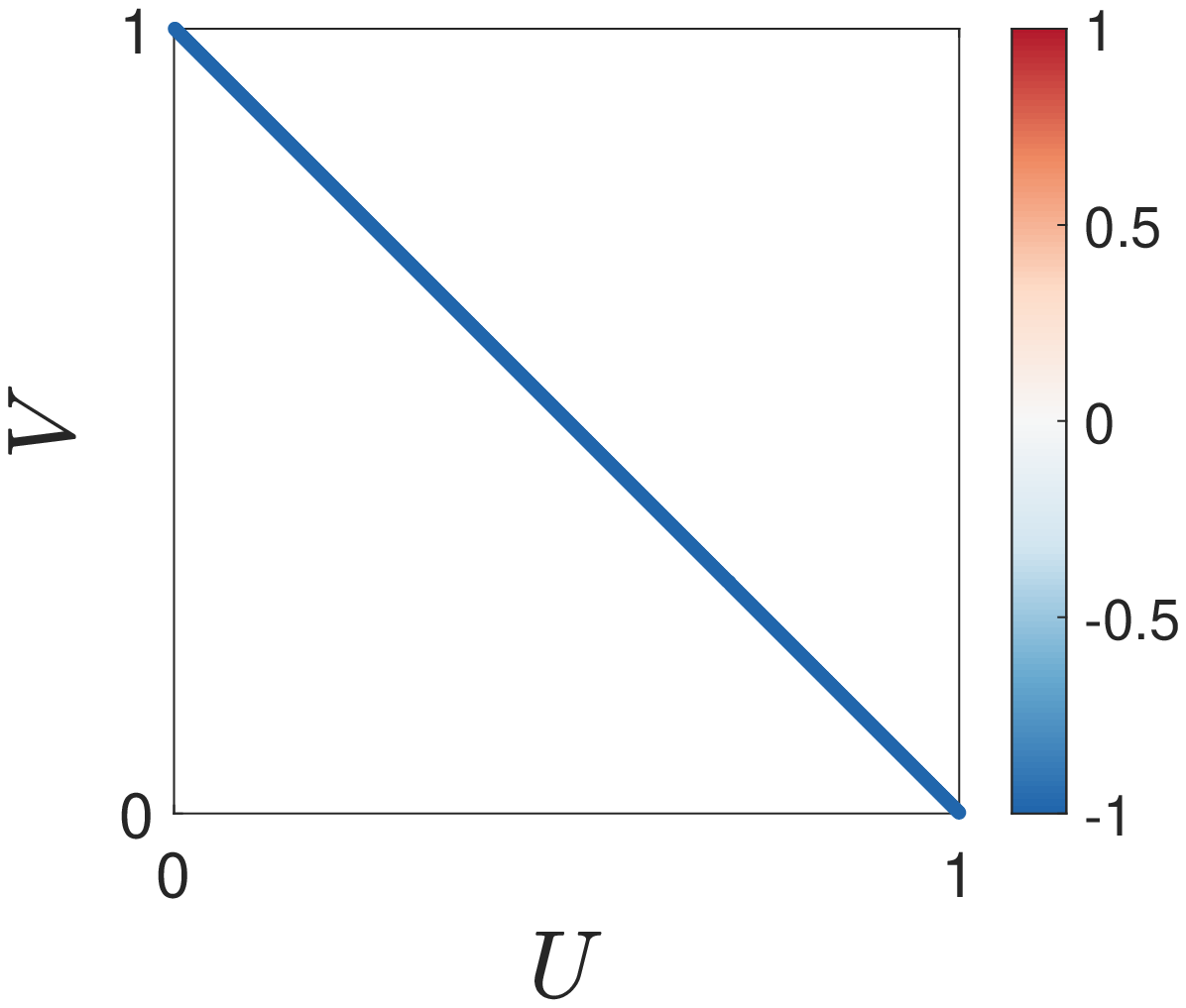}\\

   \hspace{-0.31cm} \footnotesize (b) 
  \end{minipage}
  \hfill
  \begin{minipage}{0.27\textwidth}
   \centering
        Colored scatter plots \\ \medskip

     \includegraphics[clip=true,height=3.3cm]{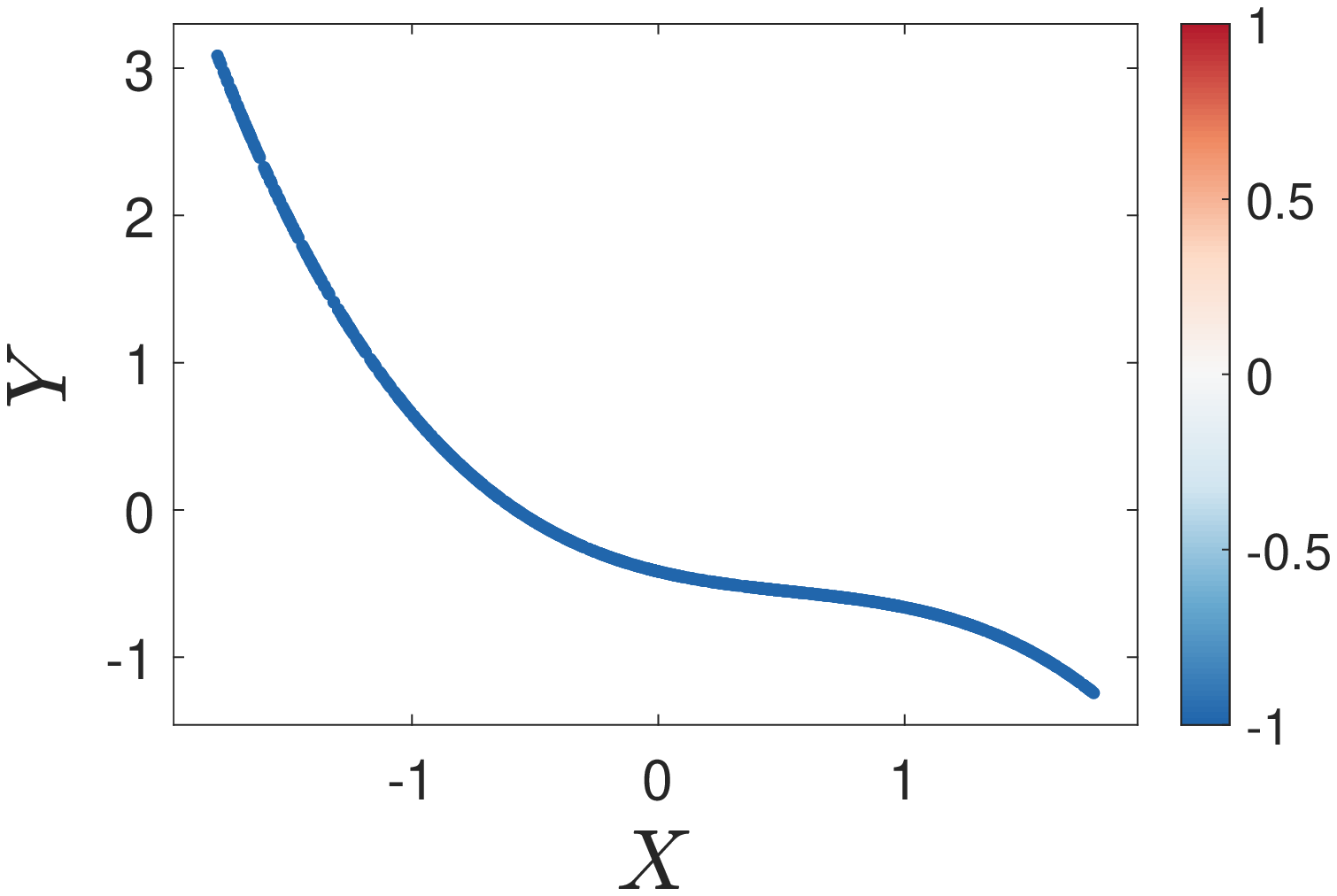}\\

   \hspace{-0.18cm} \footnotesize (c) 
  \end{minipage}
  \hfill
  \begin{minipage}{0.20\textwidth}
   \centering
     Colored parallel coordinates\\ \medskip

     \includegraphics[clip=true,height=2.9cm]{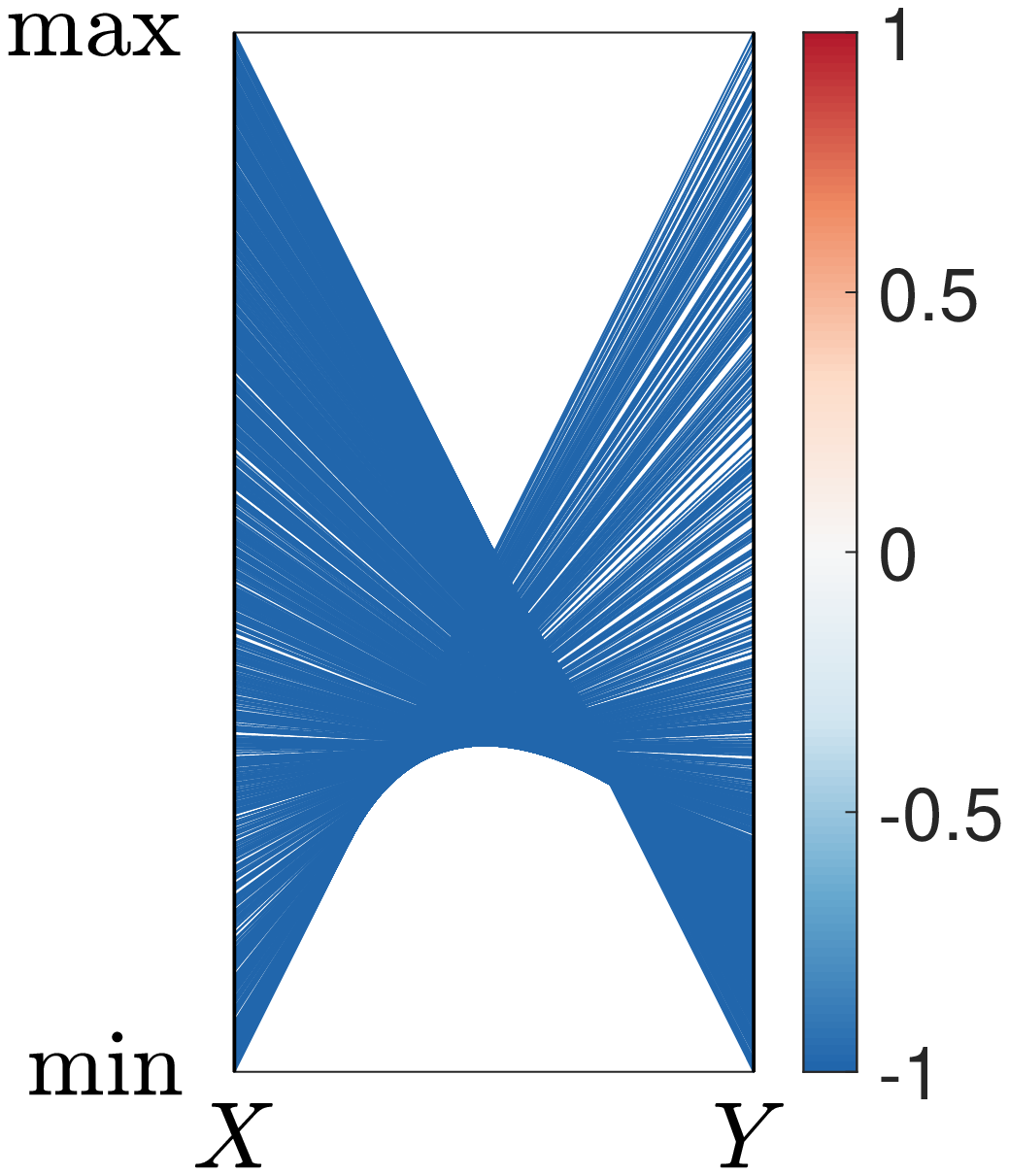}\\ \vspace{0.4cm}

   \hspace{-0.20cm} \footnotesize (d) 
  \end{minipage}
  
  \mbox{} \bigskip
  
  \begin{minipage}{0.23\textwidth}
   \centering

     \hspace{-0.3cm} \includegraphics[clip=true,height=3.3cm]{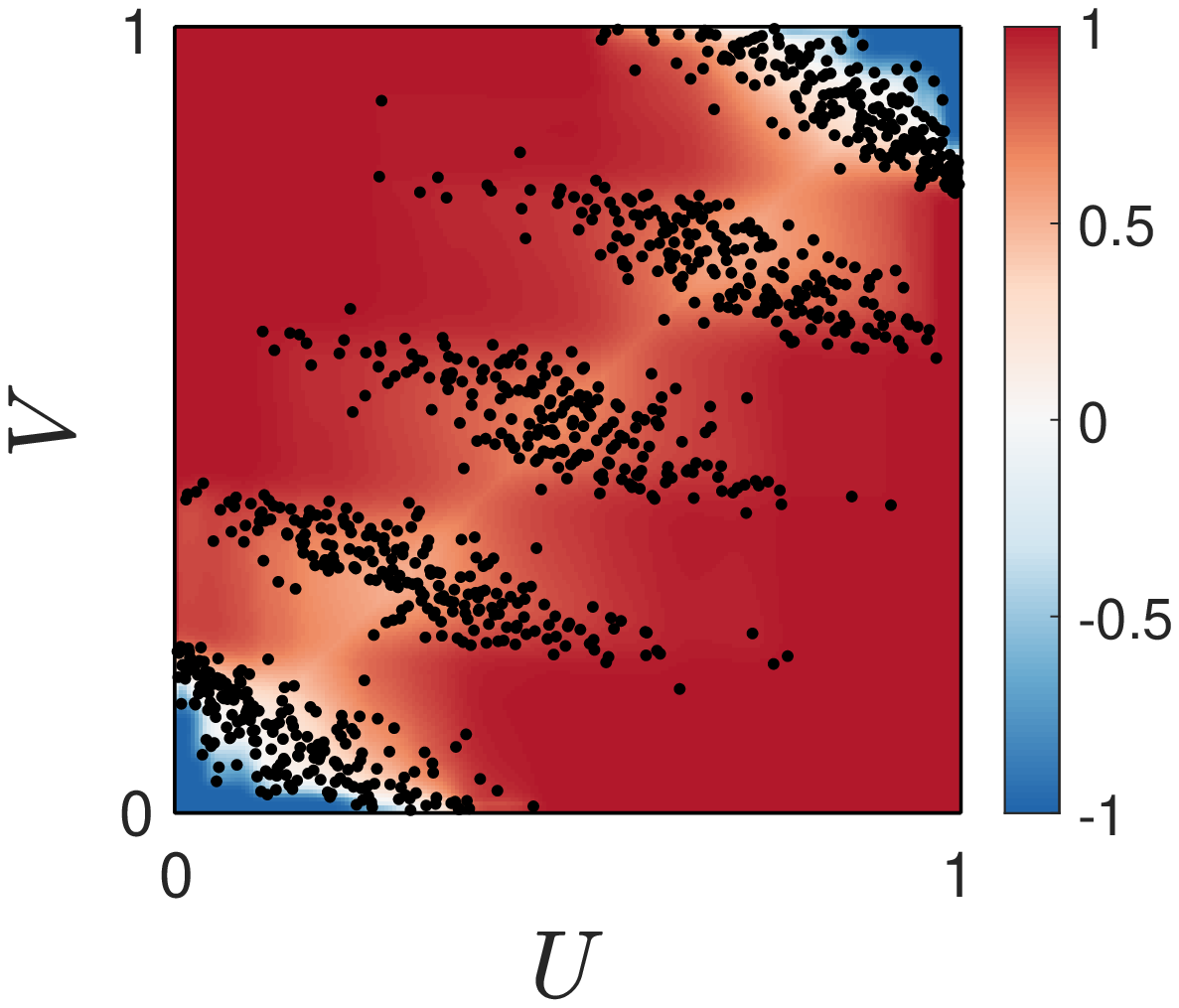}\\

   \hspace{-0.56cm}  \footnotesize (e) 
  \end{minipage}
  \hfill
  \begin{minipage}{0.23\textwidth}
   \centering

     \includegraphics[clip=true,height=3.3cm]{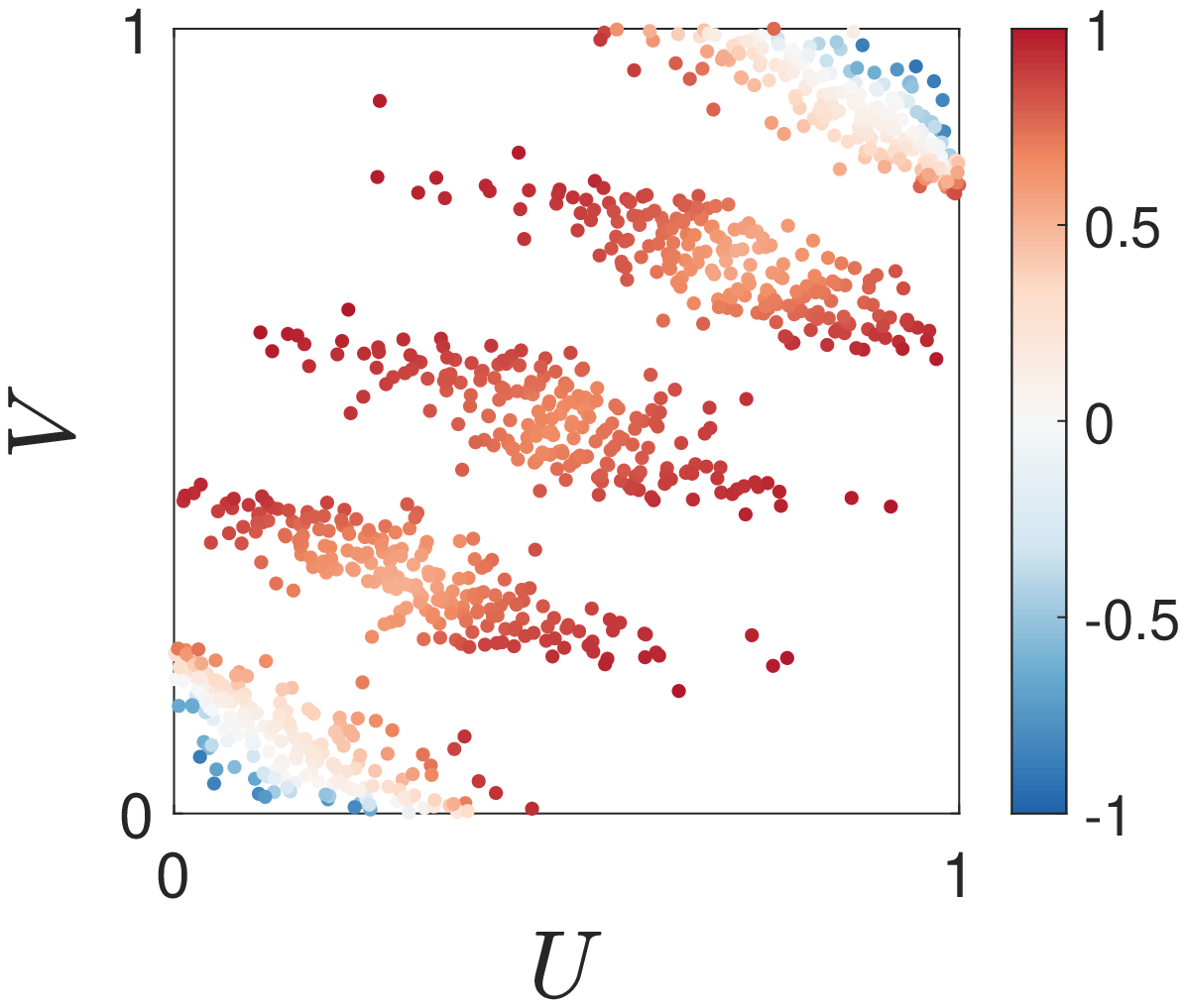}\\

   \hspace{-0.31cm} \footnotesize (f) 
  \end{minipage}
  \hfill
  \begin{minipage}{0.27\textwidth}
   \centering

     \includegraphics[clip=true,height=3.3cm]{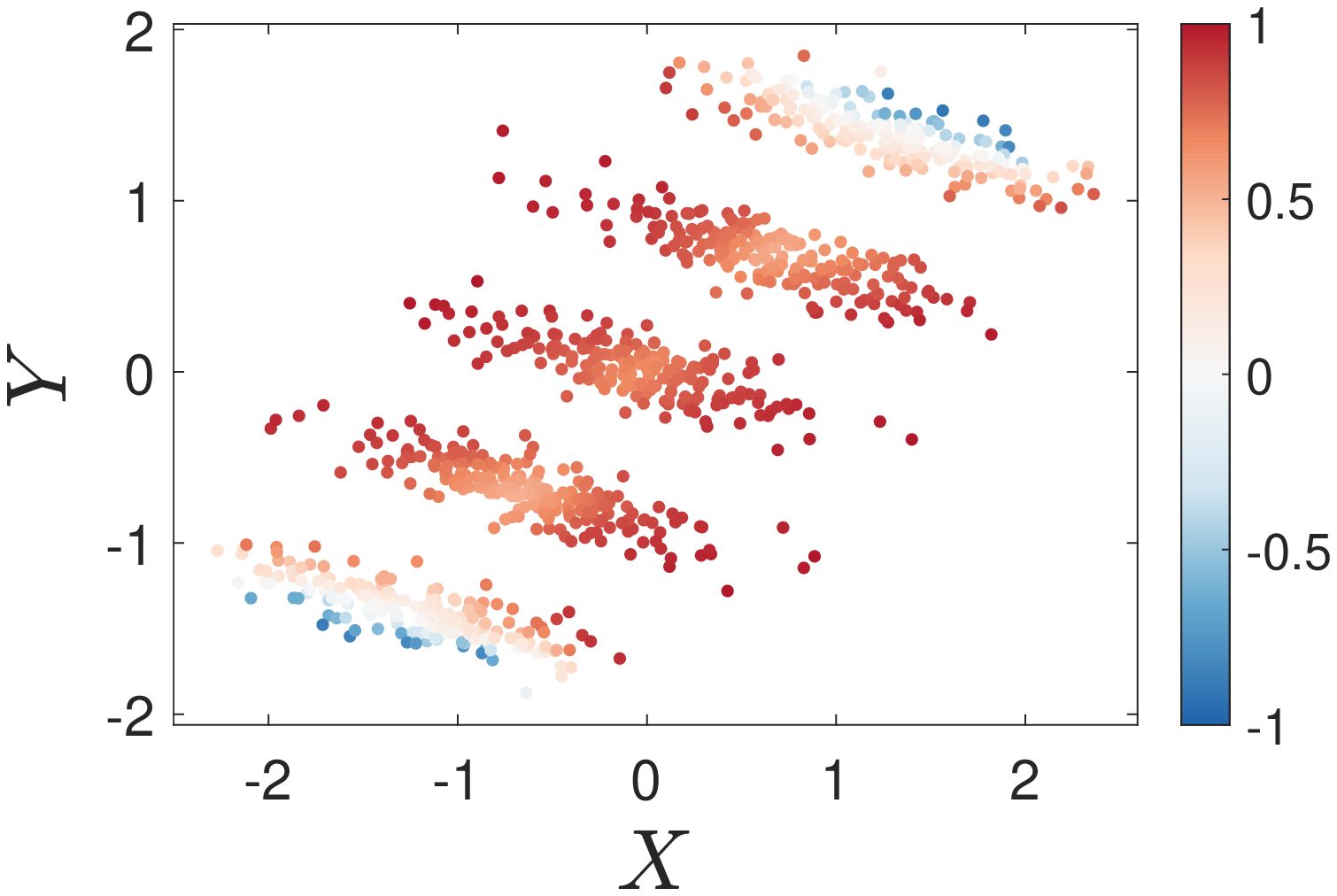}\\

   \hspace{-0.18cm} \footnotesize (g) 
  \end{minipage}
  \hfill
  \begin{minipage}{0.20\textwidth}
   \centering

     \includegraphics[clip=true,height=2.9cm]{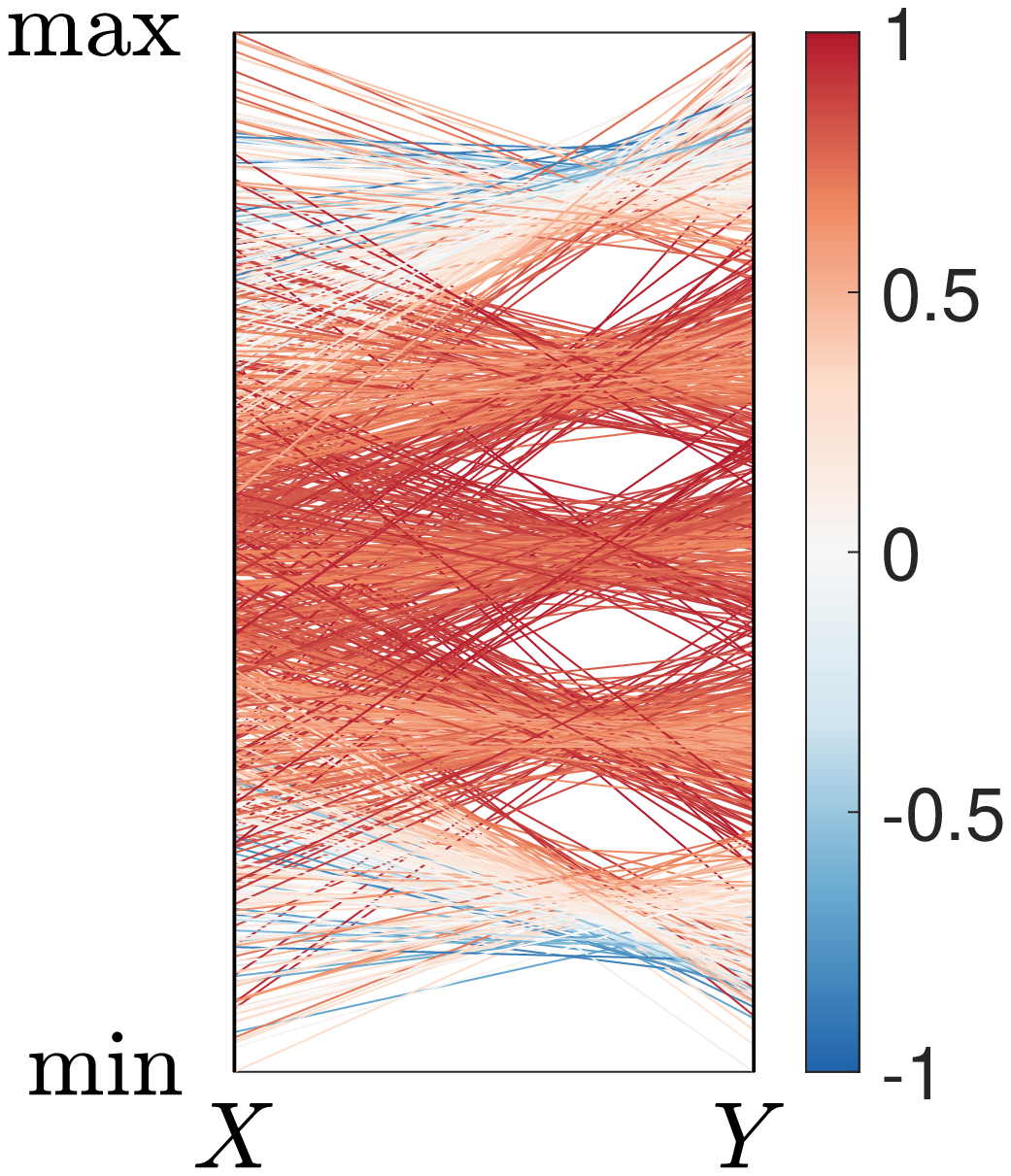}\\ \vspace{0.4cm}

   \hspace{-0.20cm} \footnotesize (h) 
  \end{minipage} 
  
    \mbox{} \bigskip

  \begin{minipage}{0.23\textwidth}
   \centering

     \hspace{-0.3cm} \includegraphics[clip=true,height=3.3cm]{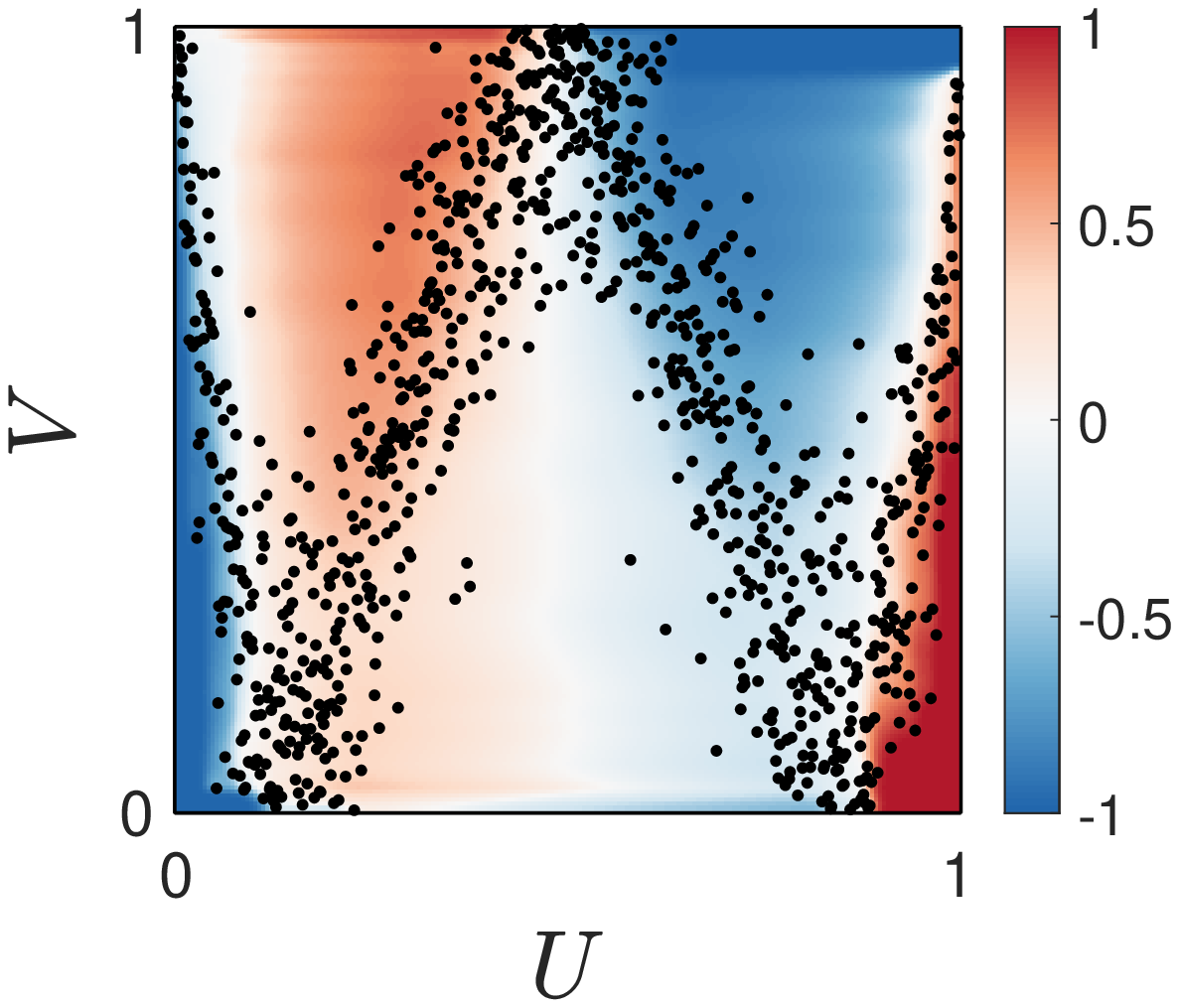}\\

   \hspace{-0.56cm}  \footnotesize (i) 
  \end{minipage}
  \hfill
  \begin{minipage}{0.23\textwidth}
   \centering

     \includegraphics[clip=true,height=3.3cm]{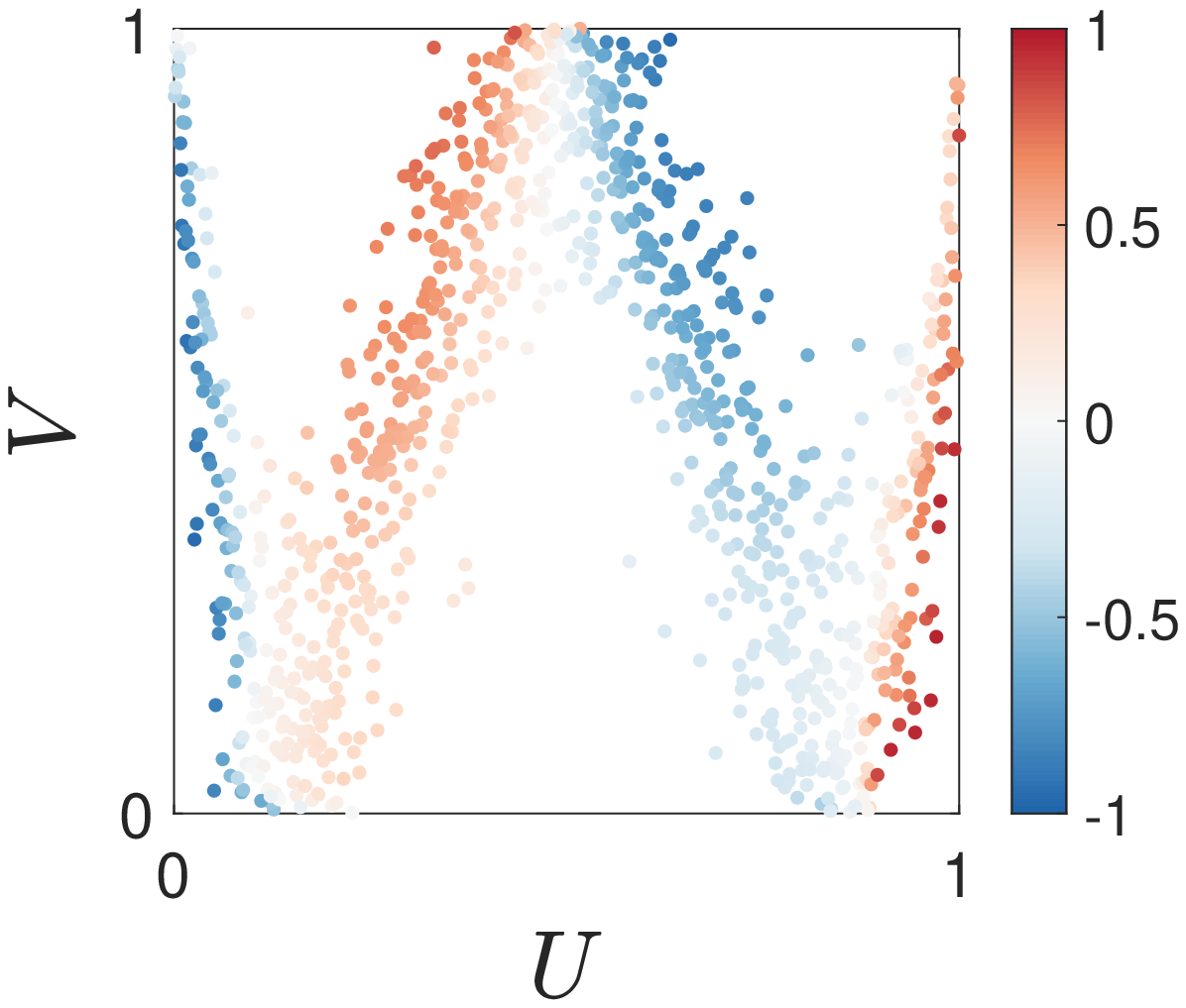}\\

   \hspace{-0.31cm} \footnotesize (j) 
  \end{minipage}
  \hfill
  \begin{minipage}{0.27\textwidth}
   \centering

     \includegraphics[clip=true,height=3.3cm]{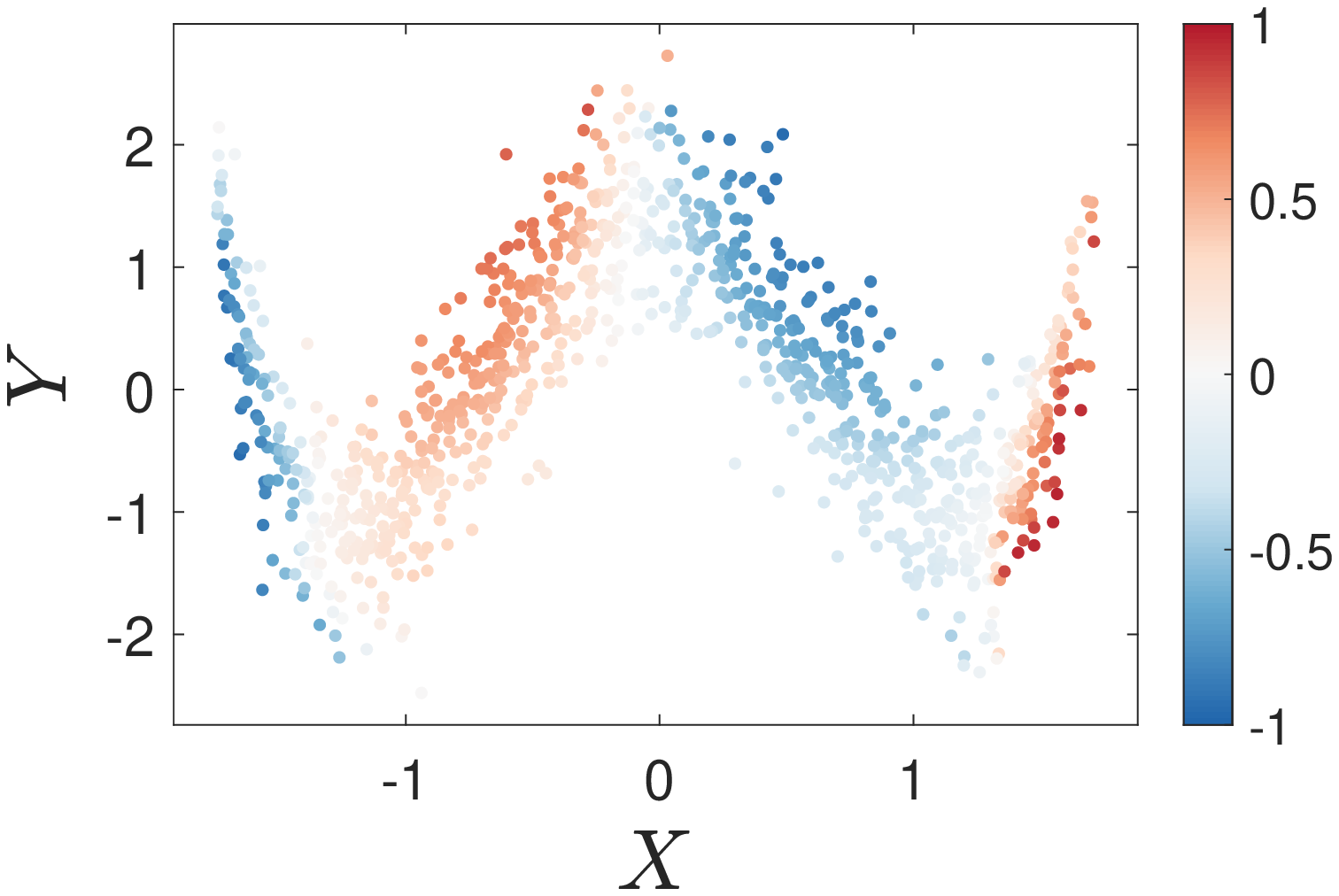}\\

   \hspace{-0.18cm} \footnotesize (k) 
  \end{minipage}
  \hfill
  \begin{minipage}{0.20\textwidth}
   \centering

     \includegraphics[clip=true,height=2.9cm]{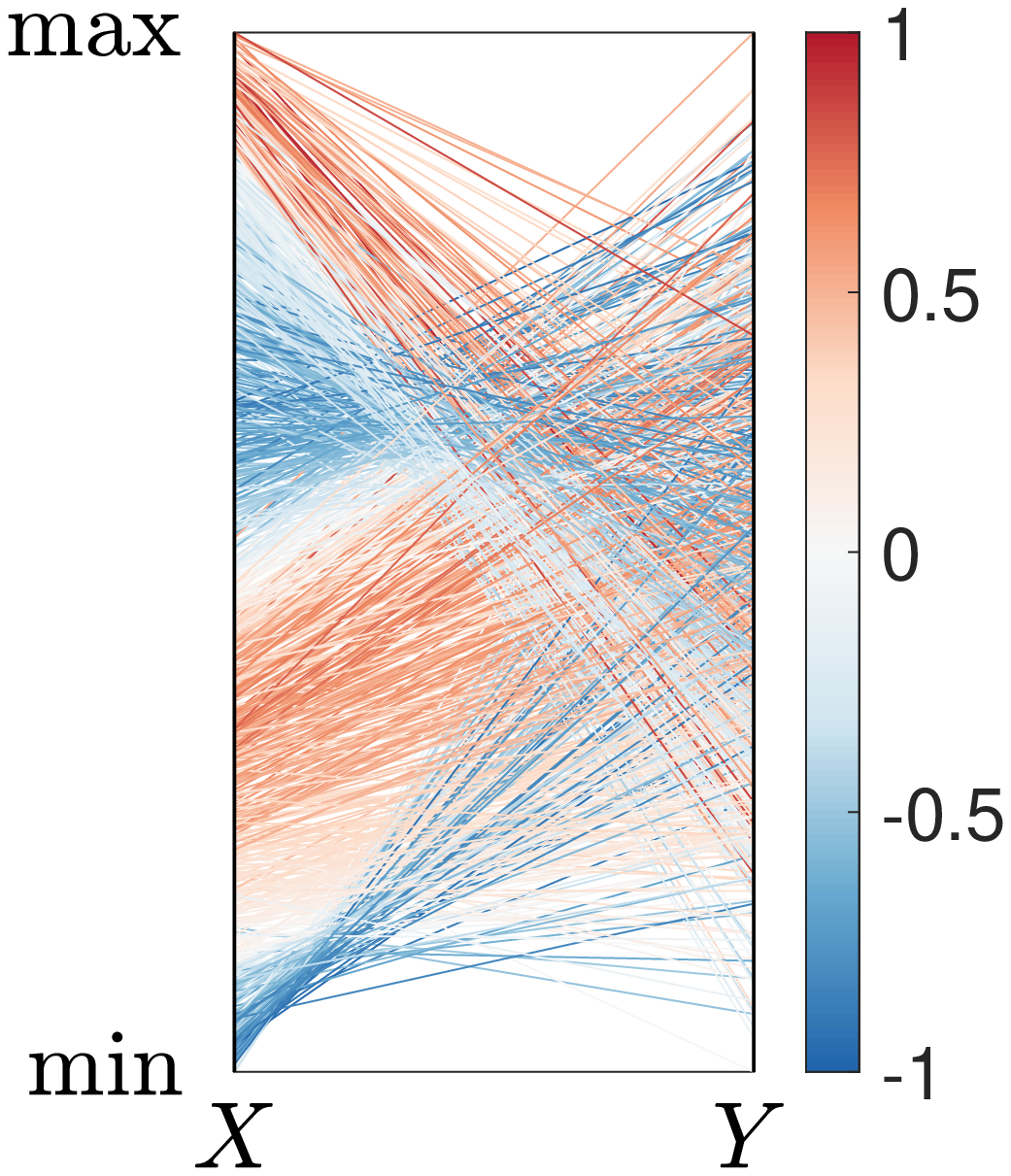}\\ \vspace{0.4cm}

   \hspace{-0.20cm} \footnotesize (l) 
  \end{minipage}  

      \mbox{} \bigskip

  \begin{minipage}{0.23\textwidth}
   \centering

     \hspace{-0.3cm} \includegraphics[clip=true,height=3.3cm]{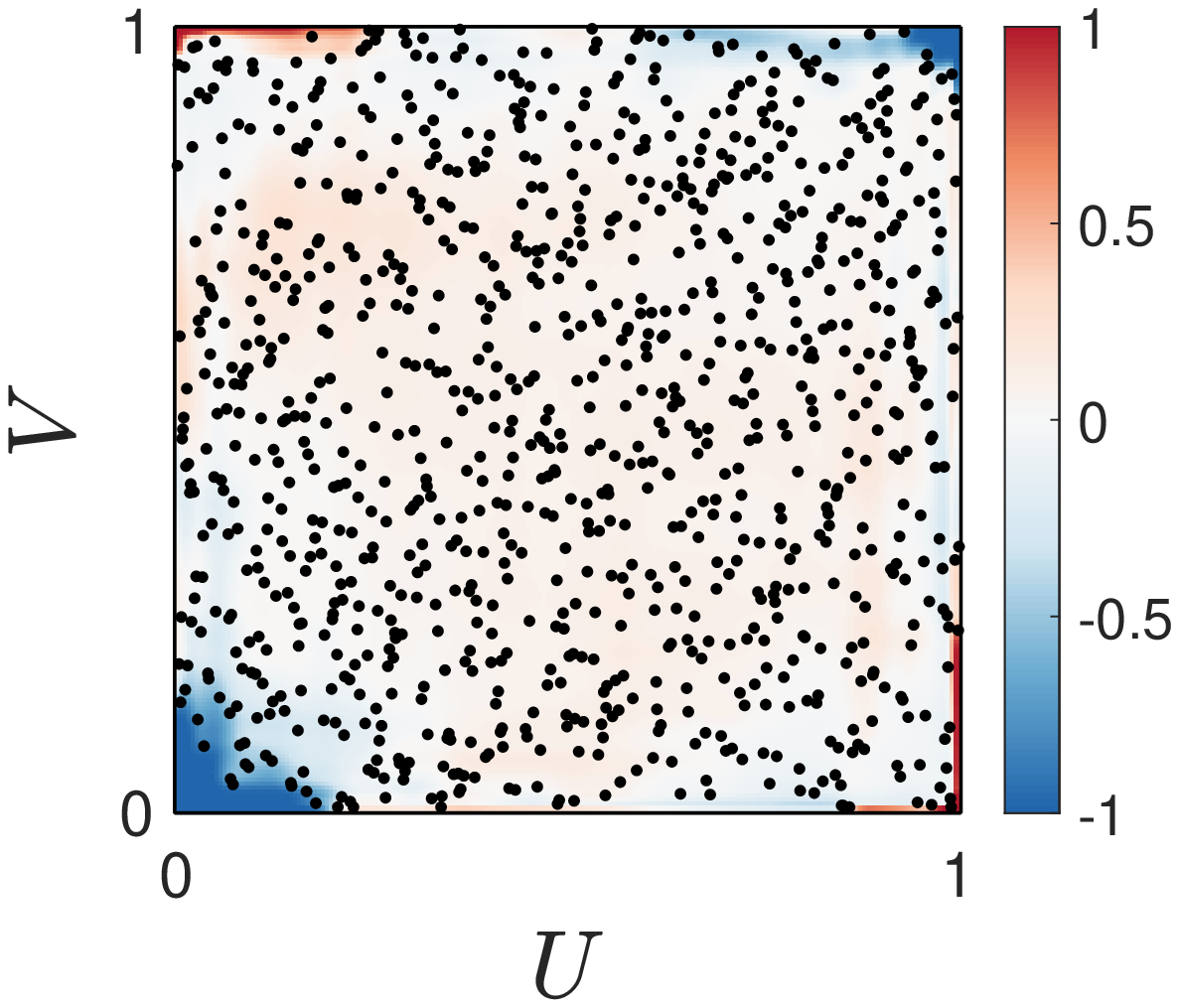}\\

   \hspace{-0.56cm}  \footnotesize (m) 
  \end{minipage}
  \hfill
  \begin{minipage}{0.23\textwidth}
   \centering

     \includegraphics[clip=true,height=3.3cm]{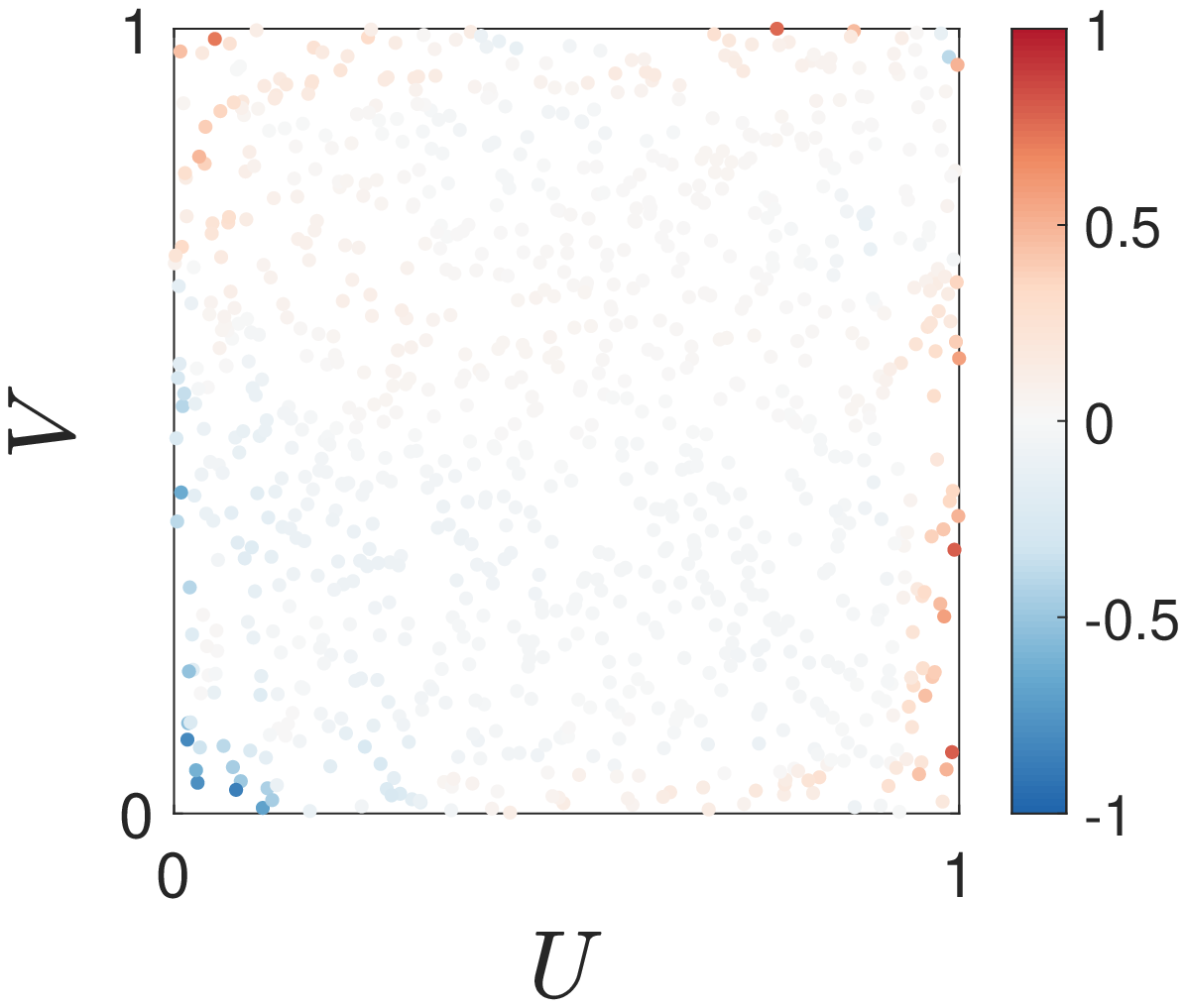}\\
     
   \hspace{-0.31cm} \footnotesize (n) 
  \end{minipage}
  \hfill
  \begin{minipage}{0.27\textwidth}
   \centering

     \includegraphics[clip=true,height=3.3cm]{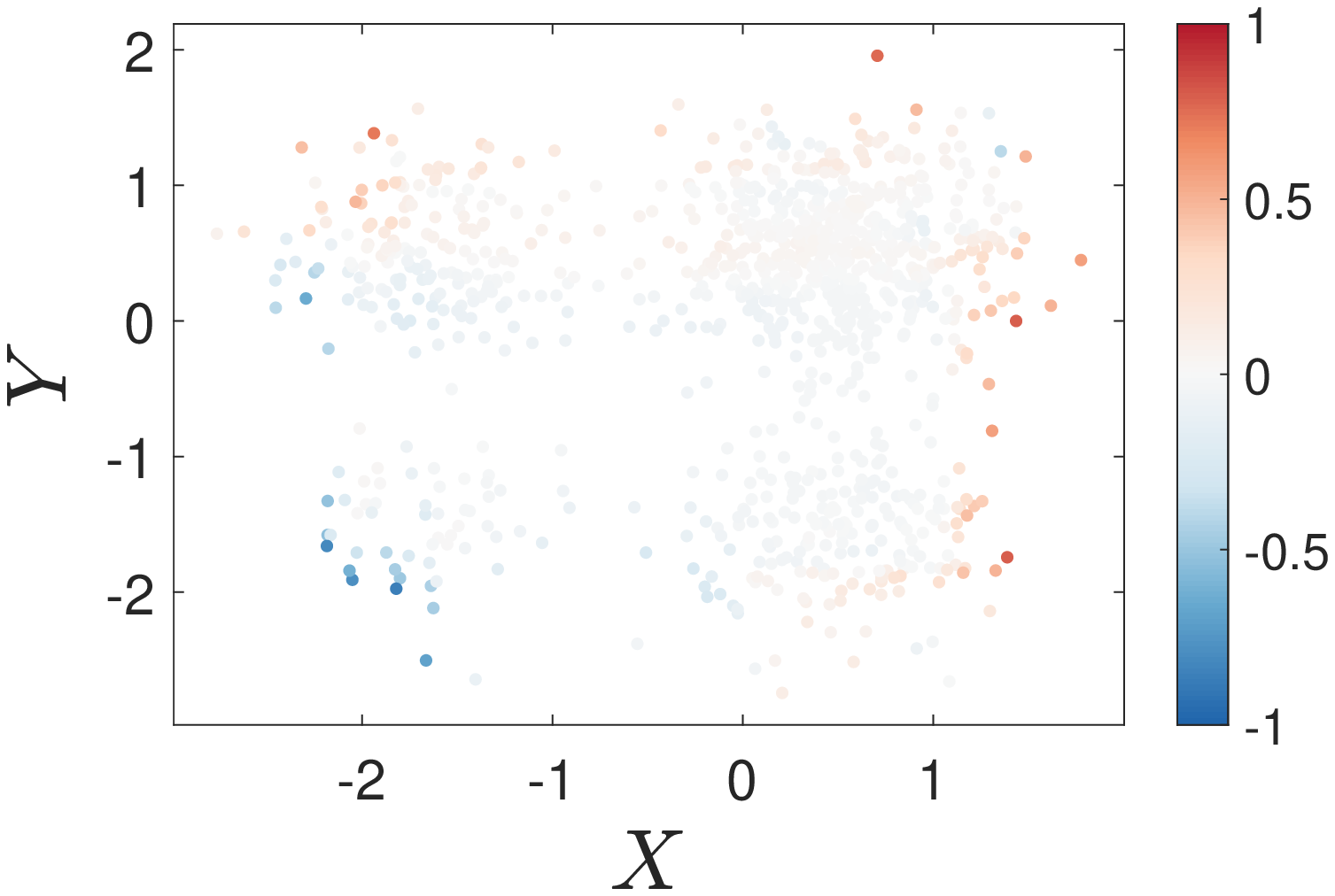}\\
     
   \hspace{-0.18cm} \footnotesize (o) 
  \end{minipage}
  \hfill
  \begin{minipage}{0.20\textwidth}
   \centering

     \includegraphics[clip=true,height=2.9cm]{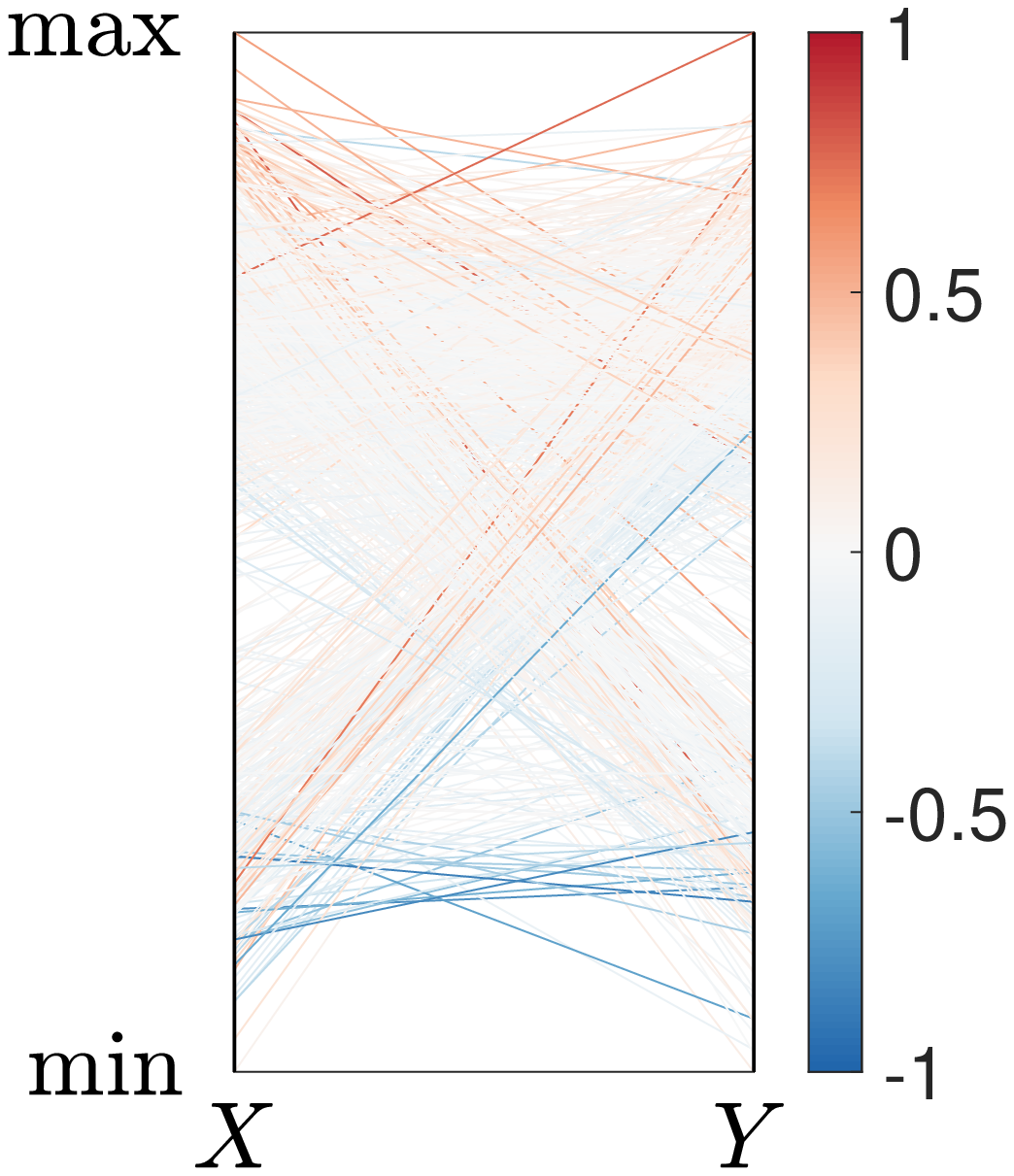}\\ \vspace{0.4cm}

   \hspace{-0.20cm} \footnotesize (p) 
  \end{minipage}

  \caption{Colored visualizations based on $\mathcal{H}$ and the coordinates of pseudo-observations. In the first column we show $\mathcal{H}$ together with the pseudo-observations of the data in Fig.~\ref{fig:heatmaps} simply to illustrate how we assign colors in $\mathcal{H}$ to individual data pairs. These pairs can then be shown with the assigned colors in other visualizations. In this figure we have colored pseudo-observations, scatter plots, and parallel coordinates in columns 2, 3 and 4, respectively. In this example red indicates increasing patterns, blue is associated with decreasing trends, and white suggests independence. It is important to note that the colors must be interpreted globally. For example, although each cluster in the third data set (row) shows a decreasing pattern, overall the relationship between the variables is increasing.}
\label{fig:coloredvisualizations}
\end{figure*}

Figure~\ref{fig:heatmaps} shows these heatmaps for four data sets whose scatter plot is shown in the first column. The second and third columns show the corresponding heatmaps $\mathcal{H}_{\rho}$ and $\mathcal{H}_{\sigma}$, respectively. In (a) we have chosen a data set for which the relationship between the variables $X$ and $Y$ is strictly decreasing, which implies that $\rho_{C} = -1$ and $\sigma_{C} = 1$. Equivalently, since the copula for this data set is $W$, the differences between it and the product copula $\Pi$ are maximal, where the greater values are located along the $V = 1-U$ diagonal. The corresponding heatmaps for $\mathcal{H}_{\rho}$ and $\mathcal{H}_{\sigma}$ are shown in (b) and (c), respectively. It is worth mentioning that the average of the colors (i.e., heatmap values) is $-1$ for $\mathcal{H}_{\rho}$ and $1$ for $\mathcal{H}_{\sigma}$. However, it is cumbersome to estimate these means visually for two main reasons. On the one hand, the heatmap is not uniform. On the other hand, the colors can range from [-3,3] in $\mathcal{H}_{\rho}$ and from [0,3] in $\mathcal{H}_{\sigma}$, whereas the values to be estimated are in $[-1,1]$ for $\rho_{C}$ and in $[0,1]$ for $\sigma_{C}$. Note that the colors close to the borders of the heatmaps will be very pale (for the particular chosen palette) since all of the copulas are identical at those borders (i.e., the differences between the copulas will be close to 0 near the borders). Alternatively, in (d) we use $\mathcal{H}$, which leads to a uniform heatmap for this data set.

The data set in (e) contains an overall increasing trend, but it is formed by five Gaussian clusters that individually would present a negative correlation (it could serve as an example of the well-known Simpson's paradox). In this case the relationship between the variables is also strong ($\rho_{n}\approx 0.80$, $\sigma_{n} = 0.81$) and increasing. Thus, the majority of the values in $\mathcal{H}_{\rho}$ are positive and the slope of the apparent elliptical shape is similar to that of the diagonal ($U = V$), as shown in (f). The absolute values of $\mathcal{H}_{\rho}$ represented in $\mathcal{H}_{\sigma}$ lead to a similar pattern in (g). In the normalized heatmap in (h) we can also perceive a strong increasing relationship, where the darker red colors stand out much more than in (f).

The relationship between the data variables in (i) is weaker ($\rho_{n}\approx -0.04$, $\sigma_{n} = 0.32$). In this case $\mathcal{H}_{\rho}$ exhibits pale shades of red and blue. This indicates that there are both increasing and decreasing trends in the data, but the relationship between the variables is only moderate. Since there is approximately the same amount of red as blue users could guess that $\rho_{C}$ should be close to $0$. In $\mathcal{H}_{\sigma}$ we see pale shades of purple, where the mean of the colors is $0.32$. However, this average is difficult to calculate visually. In $\mathcal{H}$, shown in (l), the normalization allows us to see the existence of increasing and decreasing trends in the data more clearly than in (j).

Finally, the four clusters in (m) were generated through two independent bimodal distributions. The majority of colors in the three heatmaps in (n), (o) and (p) are very close to white given their palettes. However, as a result of the normalization, the borders of $\mathcal{H}$ may present regions with values close to $1$ or $-1$.

We can also use $\mathcal{H}$ to assign colors to individual data pairs, which could be used in any visualization that shows these pairs such as scatter plots or parallel coordinate graphs. In particular, we propose using the coordinates of the pseudo-observations to recover colors of $\mathcal{H}$. Formally, given some data pair $(x_{k},y_{k})$, with corresponding pseudo-observations $(\hat{u}_{k},\hat{v}_{k})$, we associate it with the value (color) of $\mathcal{H}(i/n,j/n)$, where $(i/n,j/n)$, for $i,j\in\{1,\ldots,n-1\}$, is the closest point to $(\hat{u}_{k},\hat{v}_{k})$. The first column of Fig.~\ref{fig:coloredvisualizations} illustrates this association. In particular, we have used the four colored $\mathcal{H}$ heatmaps of Fig.~\ref{fig:heatmaps} and superimposed the pseudo-observations of the data. Essentially, each pair is assigned the color corresponding to the heatmap at the location of its pseudo-observation. 

The usefulness of this combination can be seen in scatter plots or other types of visualizations that may benefit from showing increasing and decreasing trends. The second, third and fourth columns of Fig.~\ref{fig:coloredvisualizations} shows colored pseudo-observations, scatter plots, and parallel coordinate plots, respectively. Specifically, red indicates increasing patterns, blue is related to decreasing trends, and white suggests independence. For the first data set the relationship between the variables is strictly decreasing and therefore all of the points receive the color (darkest blue) associated with $-1$ in (b), (c) and (d). This example illustrates the appropriateness of using a uniform colormap for a strictly monotonic relationship. 

For the second (row) data set $\mathcal{H}$ is predominantly red. Thus, even though the clusters have a negative correlation individually, they are colored red due to the overall increasing trend in the data. In this regard, it is important to note that the colors can not be interpreted individually. They must be analyzed globally and taking into account the colors of their neighbors. Observe that the increasing relationship is less clear in the two extreme clusters. Lastly, in the parallel coordinates plot the colors can also avoid misinterpretations. Note that the visualization shows five ``X'' patterns typical of decreasing relationships (such as the one in (d)). Again, the red colors help to perceive the overall increasing trend.

The figures in the third row illustrate how this approach allows us to visualize decreasing and/or increasing trends in the data. Lastly, the colors can also be used to detect independence effortlessly, as illustrated in the example of the last row.

\section{Experiments and results}
\label{sec:experiments}

\begin{figure*}[ht!]
    \centering

  \begin{minipage}{0.19\textwidth}
   \centering
     \includegraphics[clip=true,width=\textwidth]{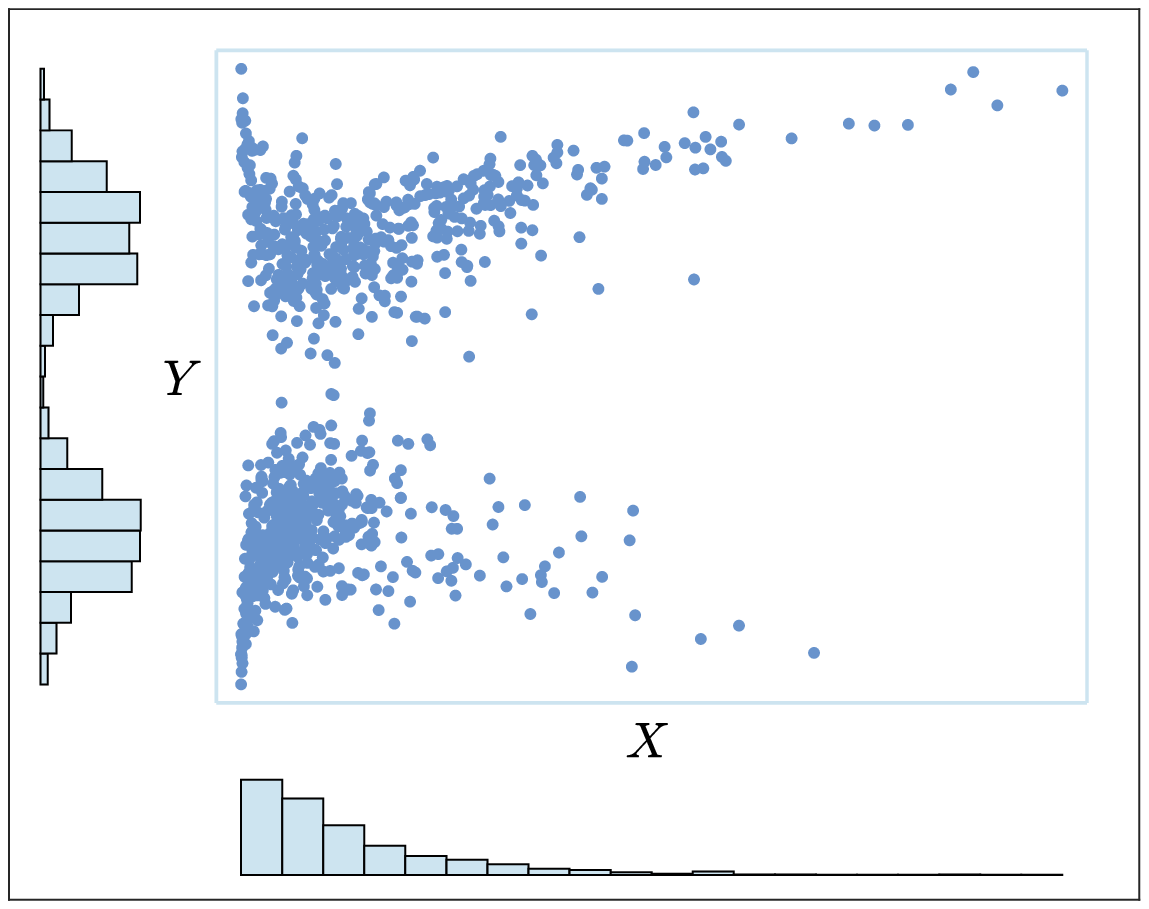}\\
     \footnotesize (a)
  \end{minipage}
  \hfill
  \begin{minipage}{0.19\textwidth}
   \centering
     \includegraphics[clip=true,width=\textwidth]{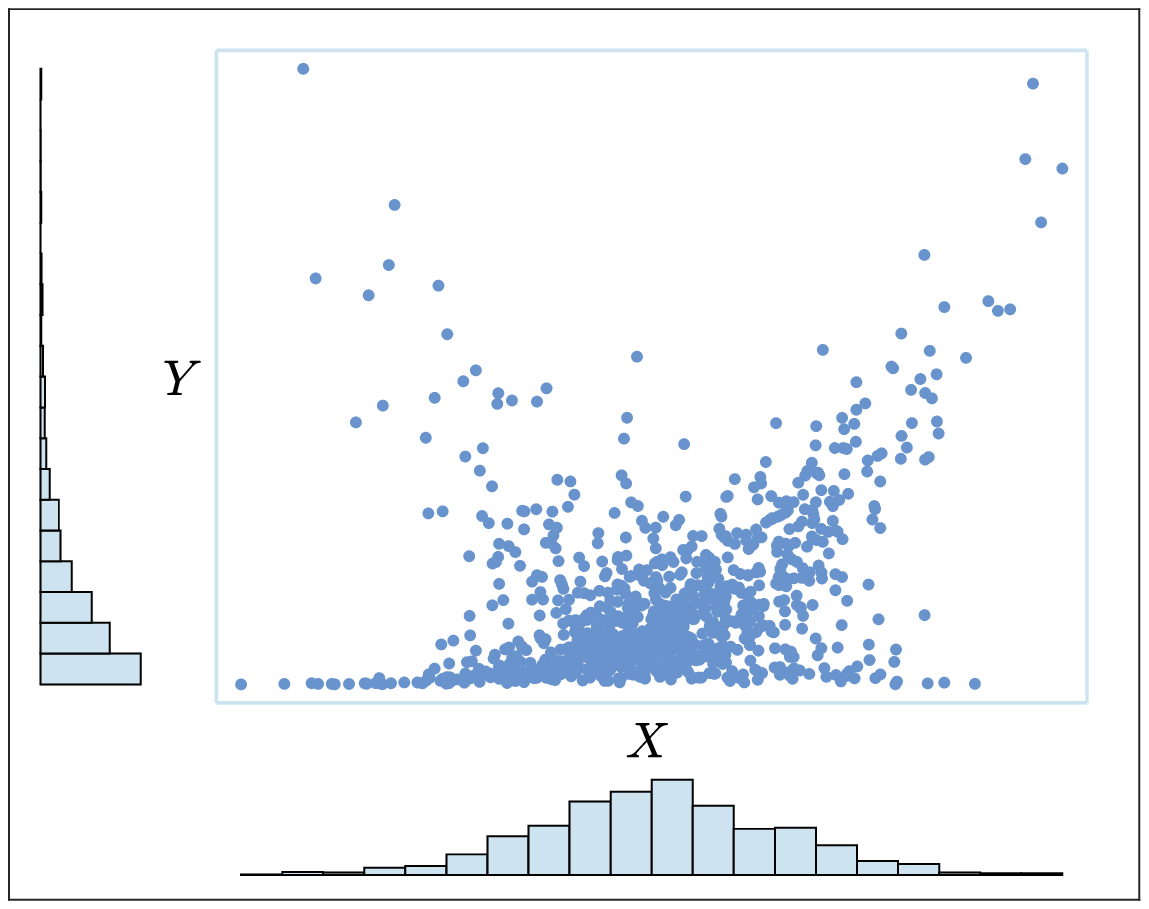}\\
     \footnotesize (b)
  \end{minipage}
  \hfill
  \begin{minipage}{0.19\textwidth}
   \centering
     \includegraphics[clip=true,width=\textwidth]{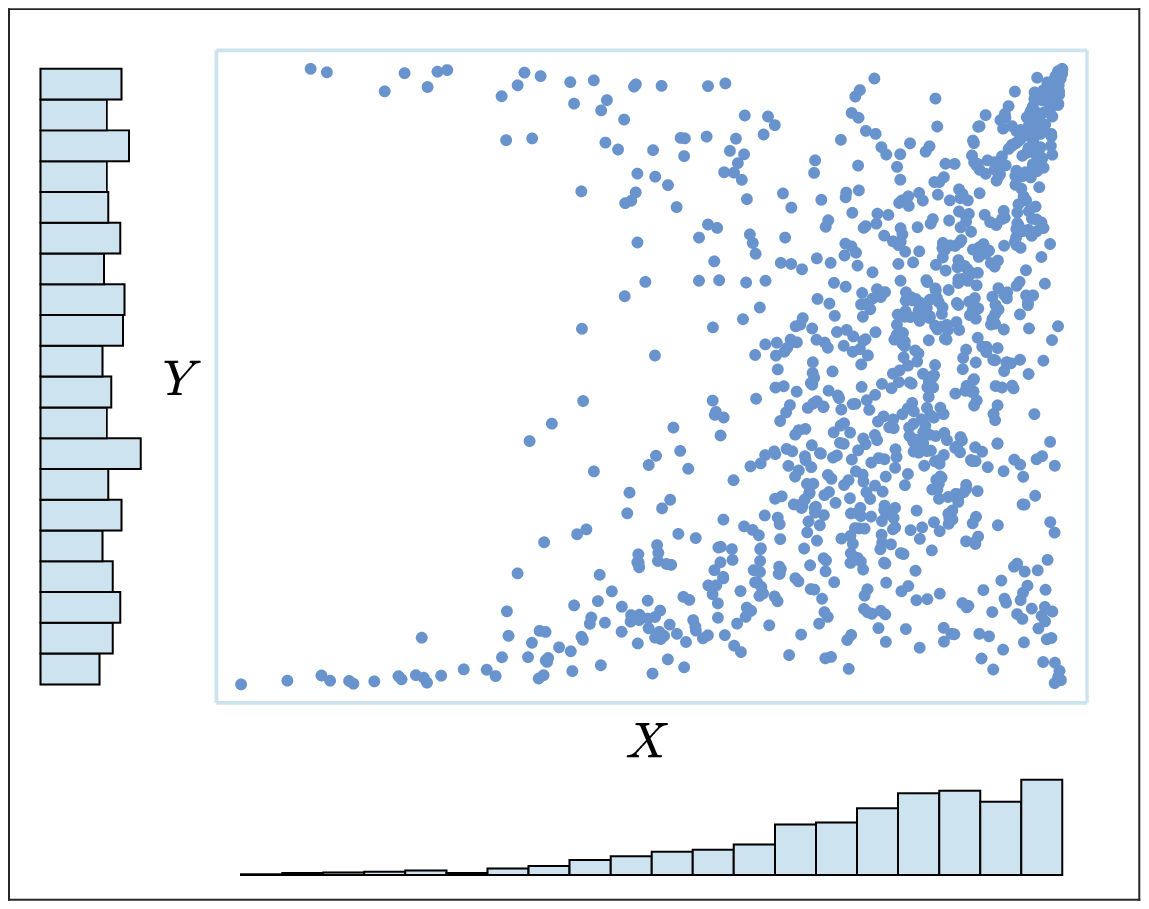}\\
     \footnotesize (c)
  \end{minipage}
  \hfill
  \begin{minipage}{0.19\textwidth}
   \centering
     \includegraphics[clip=true,width=\textwidth]{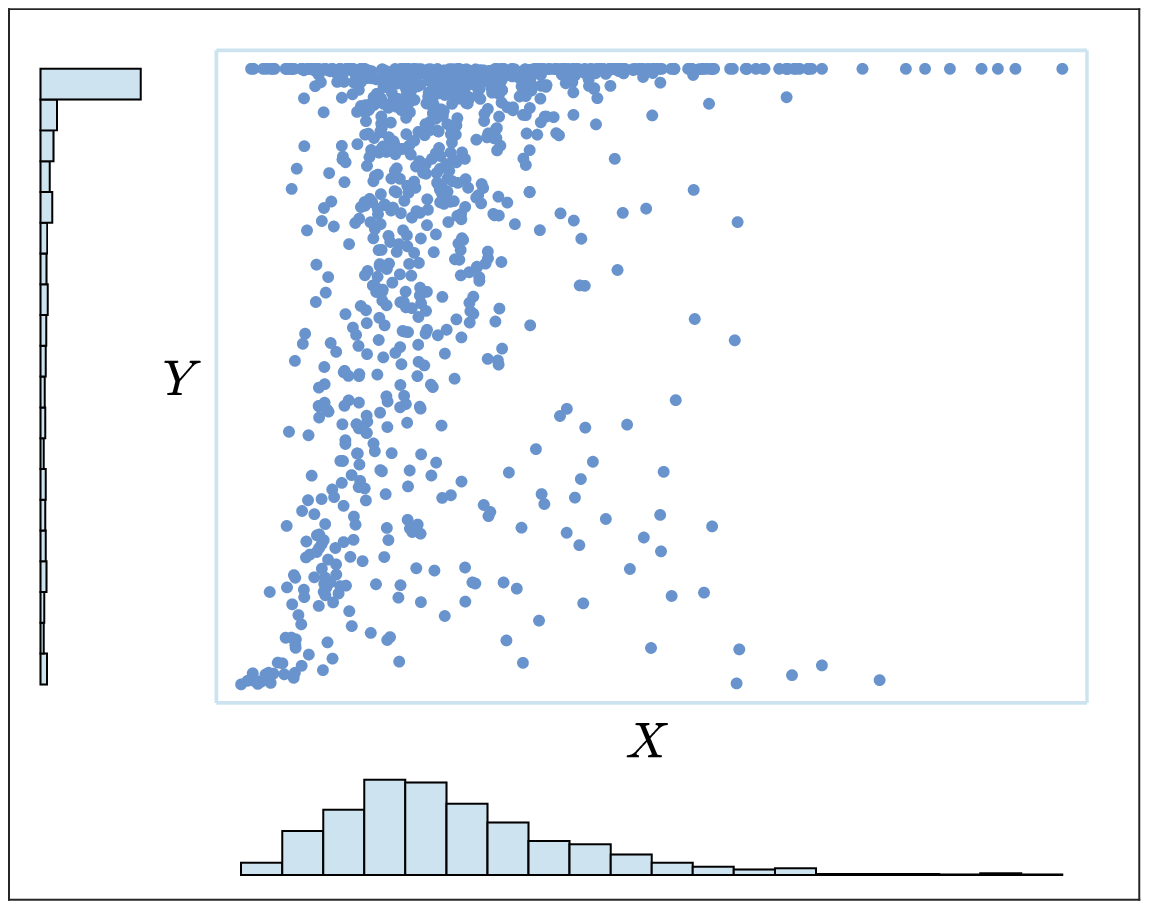}\\
     \footnotesize (d)
  \end{minipage}
  \hfill
  \begin{minipage}{0.19\textwidth}
   \centering
     \includegraphics[clip=true,width=\textwidth]{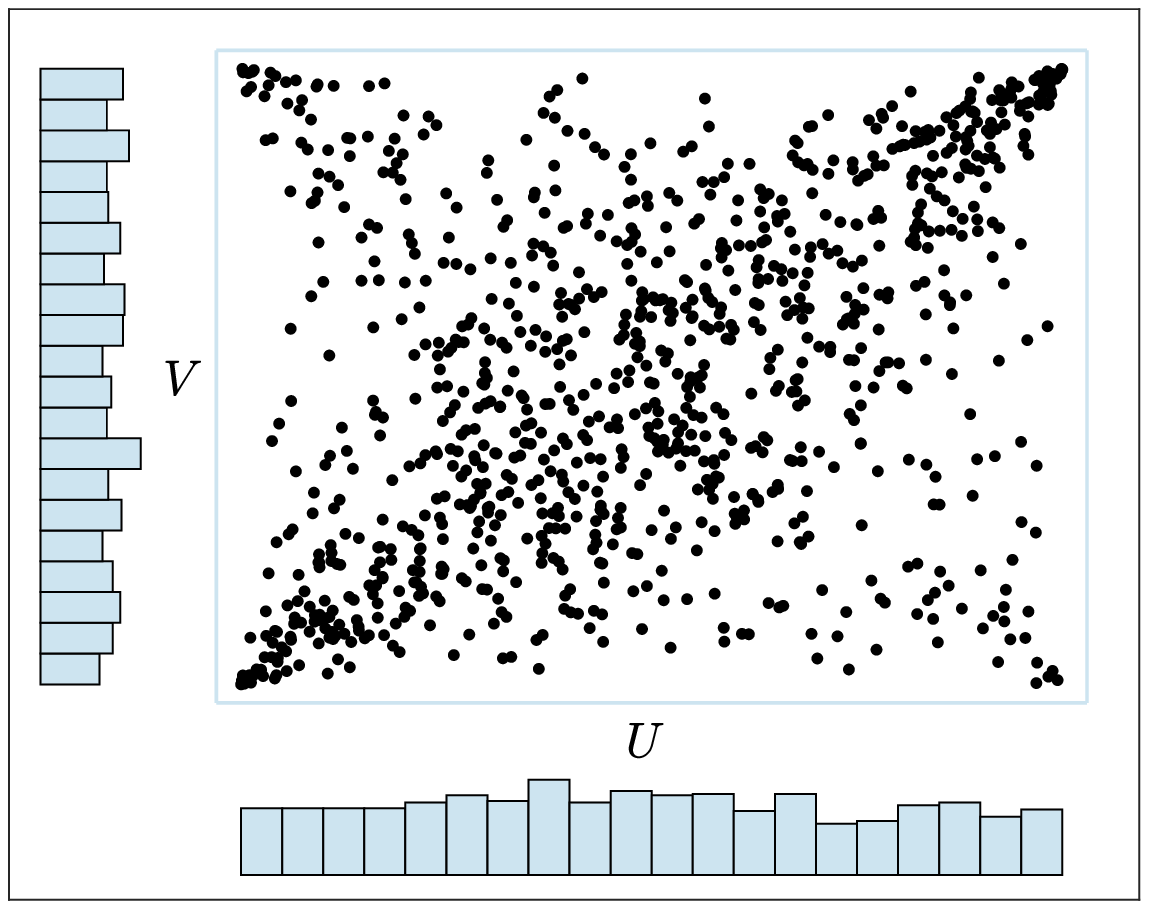}\\
     \footnotesize (e)
  \end{minipage}

    \mbox{} \bigskip

  \begin{minipage}{0.19\textwidth}
   \centering
     \includegraphics[clip=true,width=\textwidth]{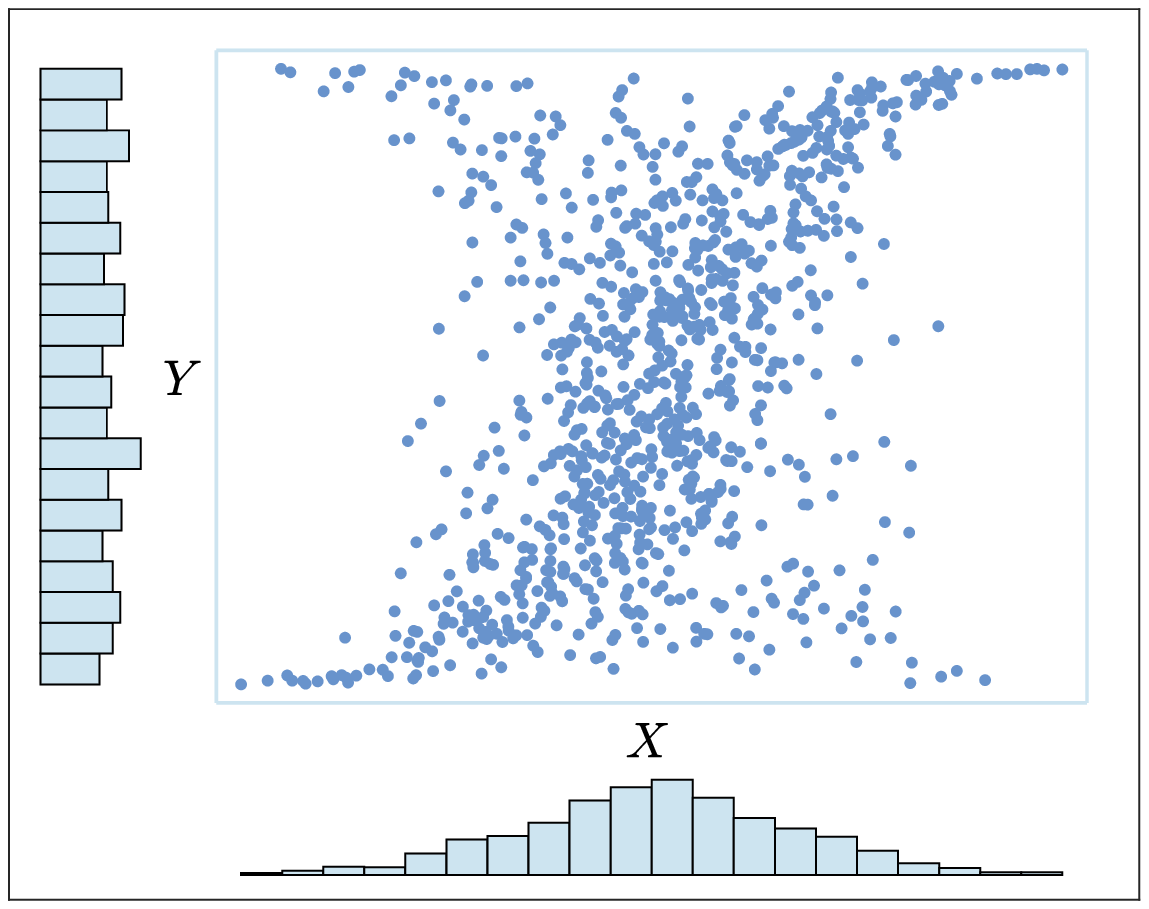}\\
     \footnotesize (f)
  \end{minipage}
  \hfill
  \begin{minipage}{0.19\textwidth}
   \centering
     \includegraphics[clip=true,width=\textwidth]{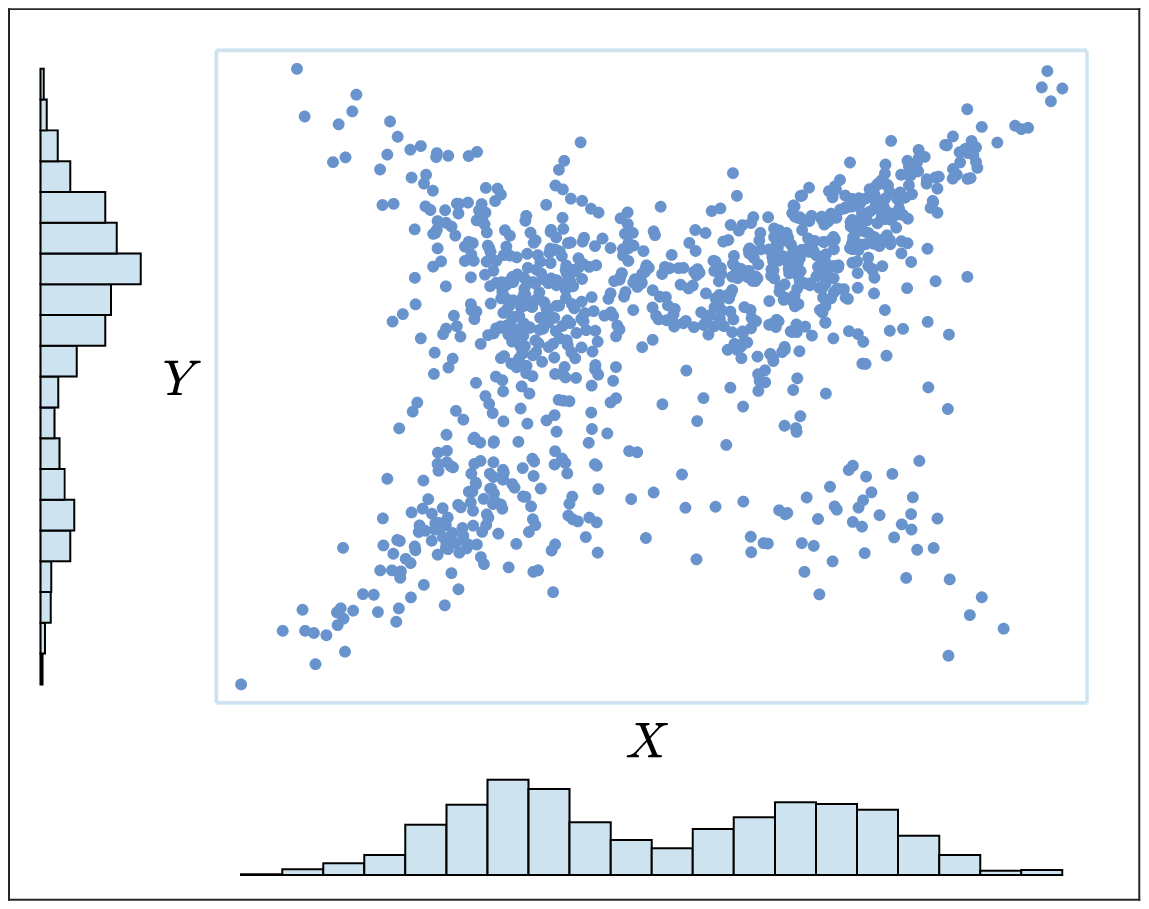}\\
     \footnotesize (g)
  \end{minipage}
  \hfill
  \begin{minipage}{0.19\textwidth}
   \centering
     \includegraphics[clip=true,width=\textwidth]{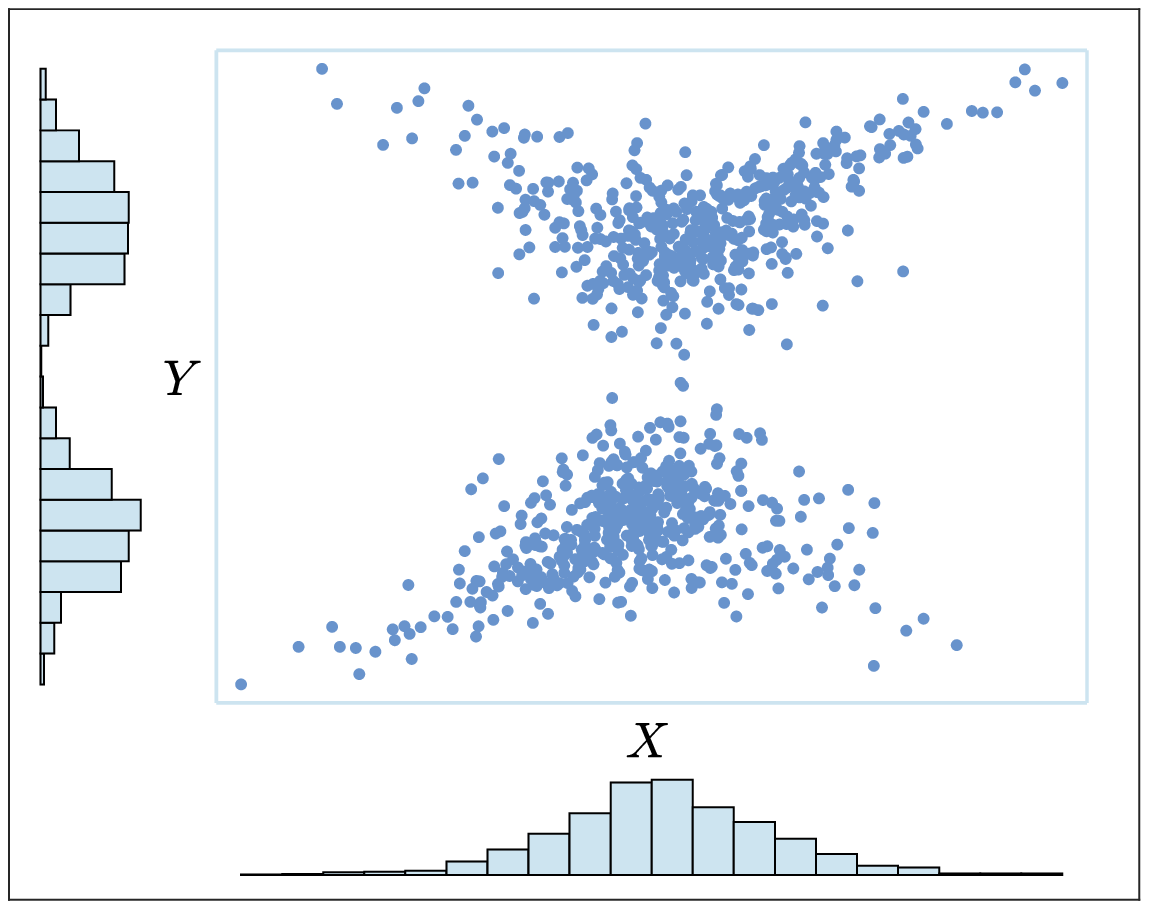}\\
     \footnotesize (h)
  \end{minipage}
  \hfill
  \begin{minipage}{0.19\textwidth}
   \centering
     \includegraphics[clip=true,width=\textwidth]{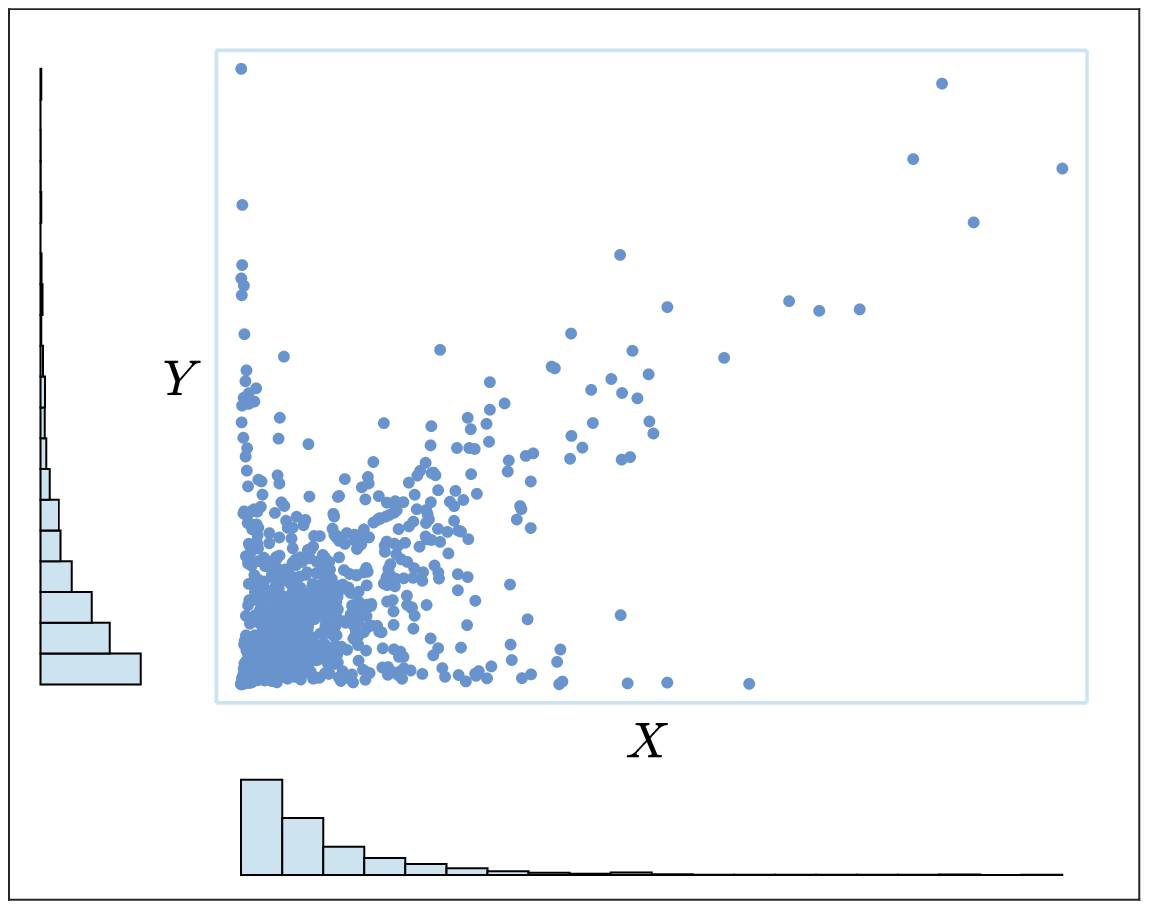}\\
     \footnotesize (i)
  \end{minipage}
  \hfill
  \begin{minipage}{0.19\textwidth}
   \centering
     \includegraphics[clip=true,width=\textwidth]{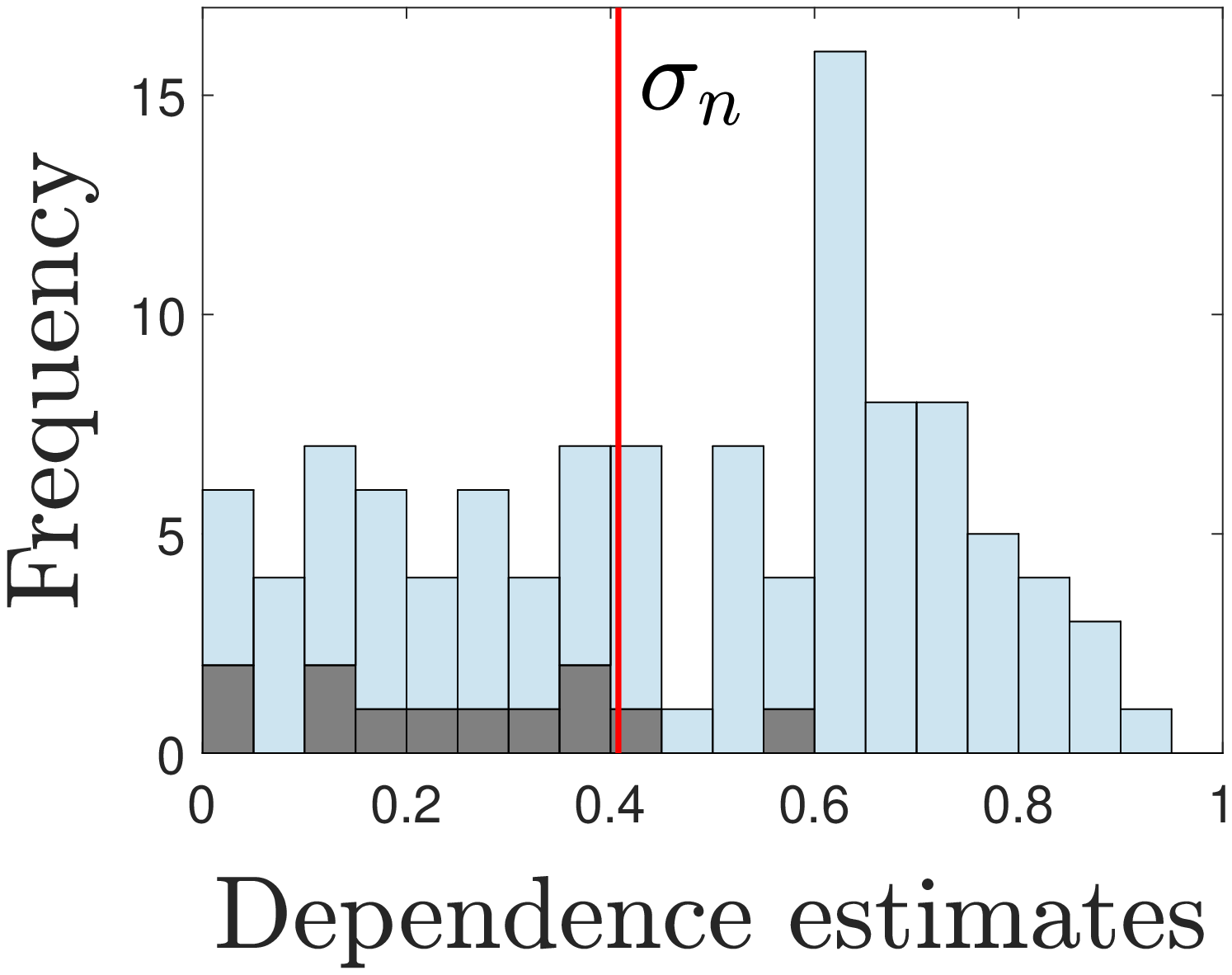}\\
     \footnotesize (j)
  \end{minipage}
  
  \caption{Nine scatter plots with the same dependence structure, where $\sigma_{n} = 0.407$ (and $\rho_{n} = 0.4$) for every distribution. The plot in (e) corresponds to pseudo-observations, which were used to generate the other scatter plots, through $X := F_{X}^{-1}(U)$ and $Y := F_{Y}^{-1}(V)$ for different marginal distributions, analogously as in Fig.~\ref{fig:pseudoobservations}. Note the variety of distributions that arise by changing the marginals. The stacked histogram in (j) shows the frequency of the dependence estimates provided by the 12 participants of our user study. The darker gray bars indicate the estimates associated with the plot of pseudo-observations, while the lighter blue bars are related to the estimates on the regular scatter plots. In this example the estimates are generally smaller for pseudo-observations. In addition, the large dispersion of the estimates related to the regular scatter plots is noteworthy, which indicates the high difficulty of the task.}
\label{fig:5x2}
\end{figure*}

\begin{figure*}[ht!]
    \centering

  \begin{minipage}{0.19\textwidth}
   \centering
     \includegraphics[clip=true,width=\textwidth]{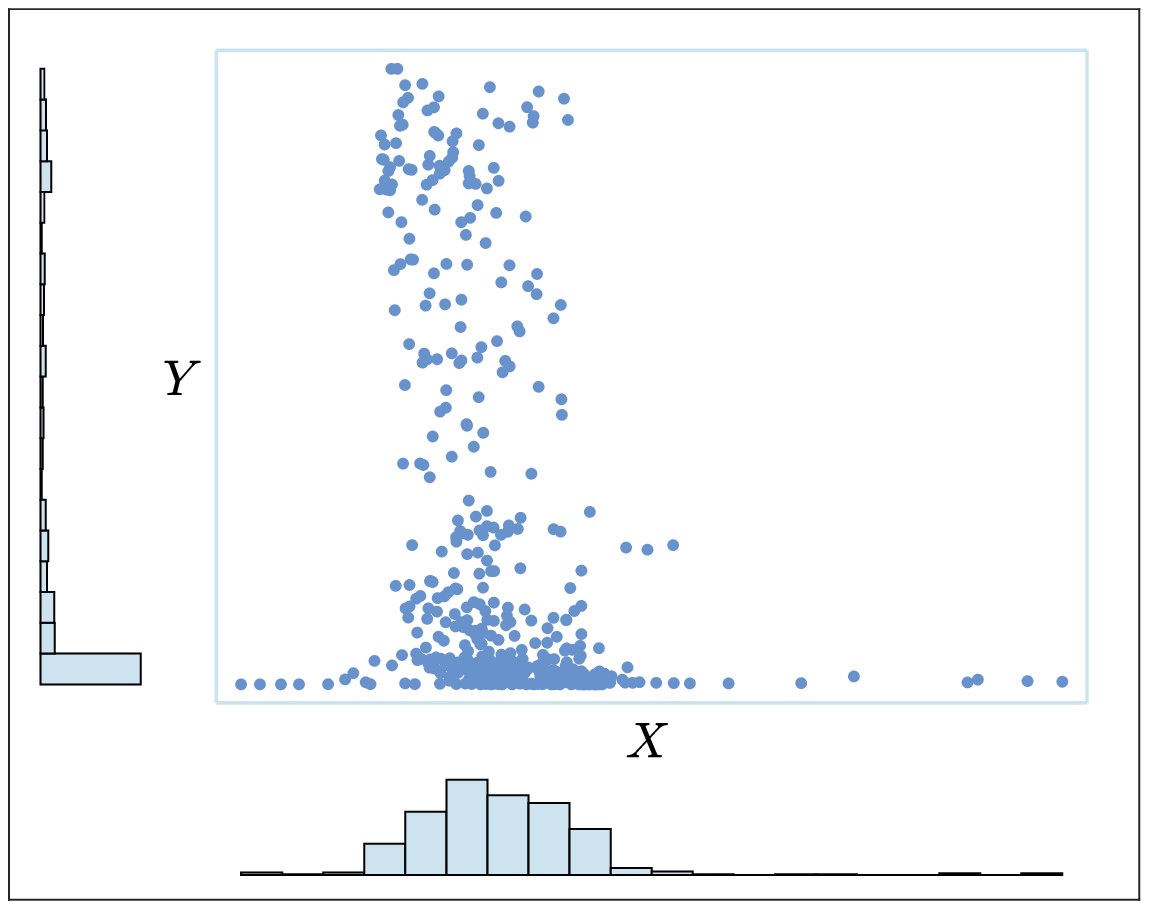}\\
     \footnotesize (1)
  \end{minipage}
  \hfill
  \begin{minipage}{0.19\textwidth}
   \centering
     \includegraphics[clip=true,width=\textwidth]{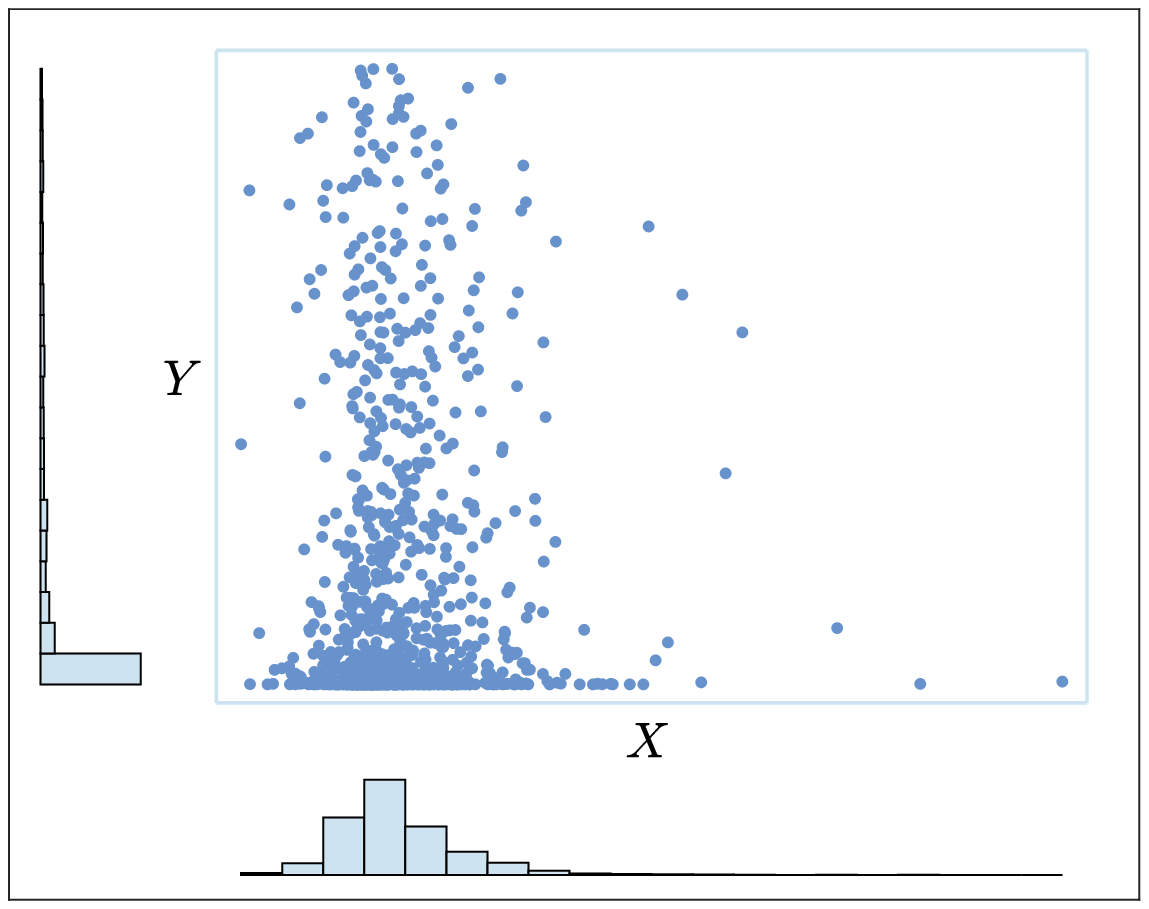}\\
     \footnotesize (2)
  \end{minipage}
  \hfill
  \begin{minipage}{0.19\textwidth}
   \centering
     \includegraphics[clip=true,width=\textwidth]{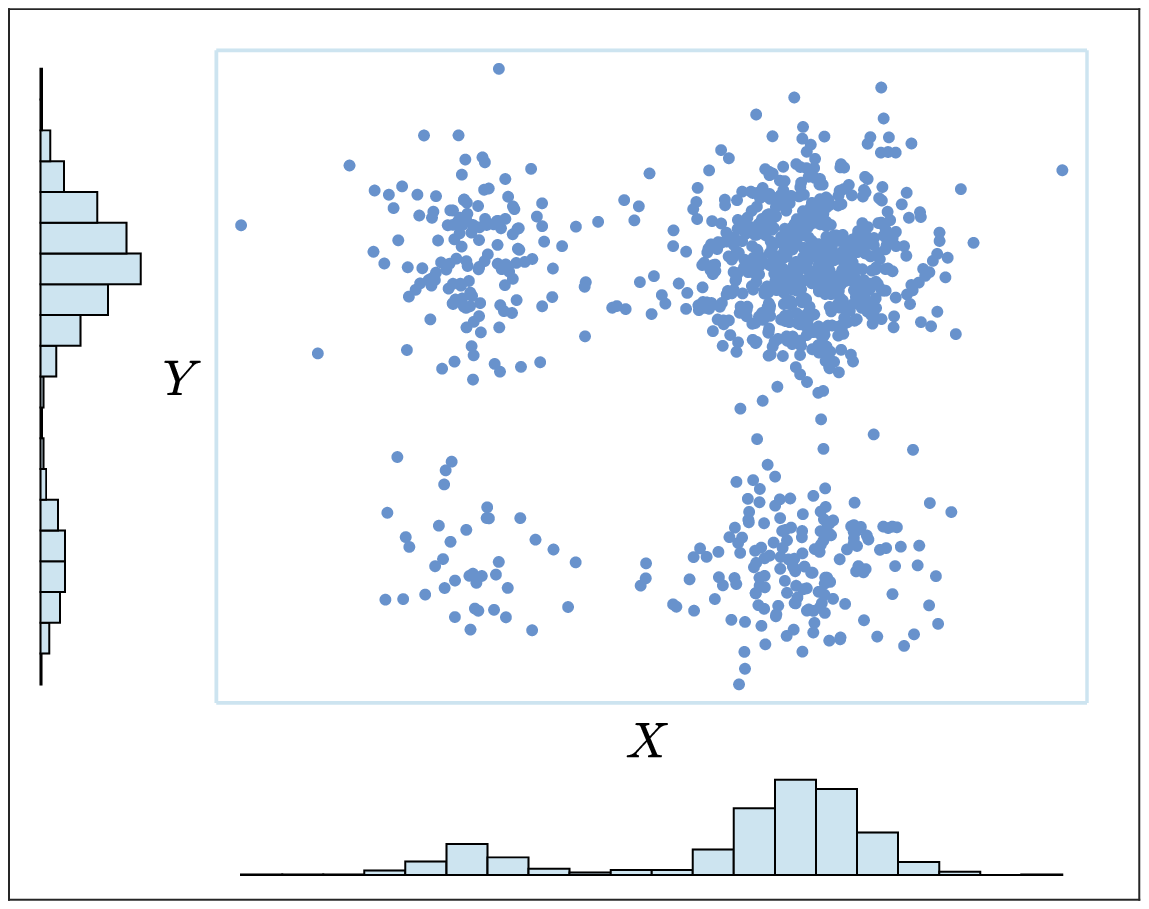}\\
     \footnotesize (3)
  \end{minipage}
  \hfill
  \begin{minipage}{0.19\textwidth}
   \centering
     \includegraphics[clip=true,width=\textwidth]{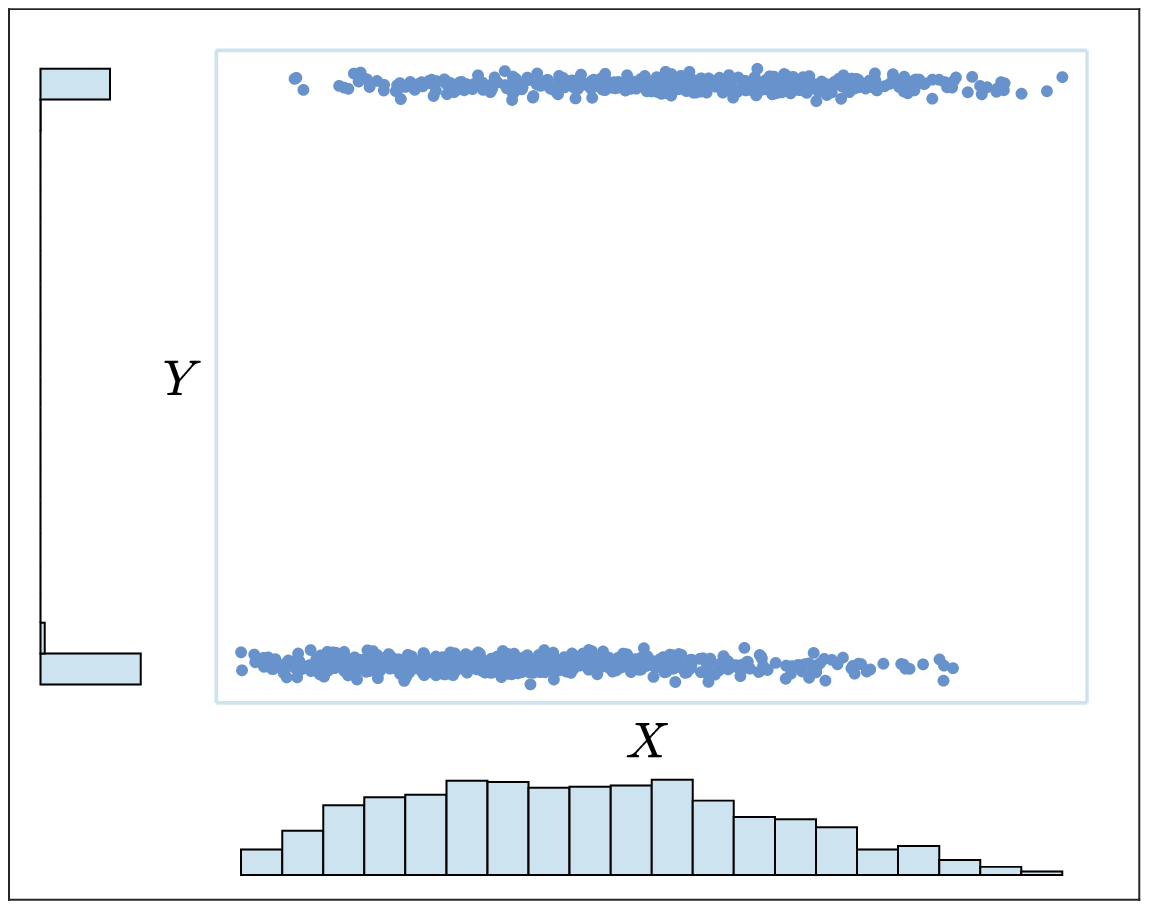}\\
     \footnotesize (4)
  \end{minipage}
  \hfill
  \begin{minipage}{0.19\textwidth}
   \centering
     \includegraphics[clip=true,width=\textwidth]{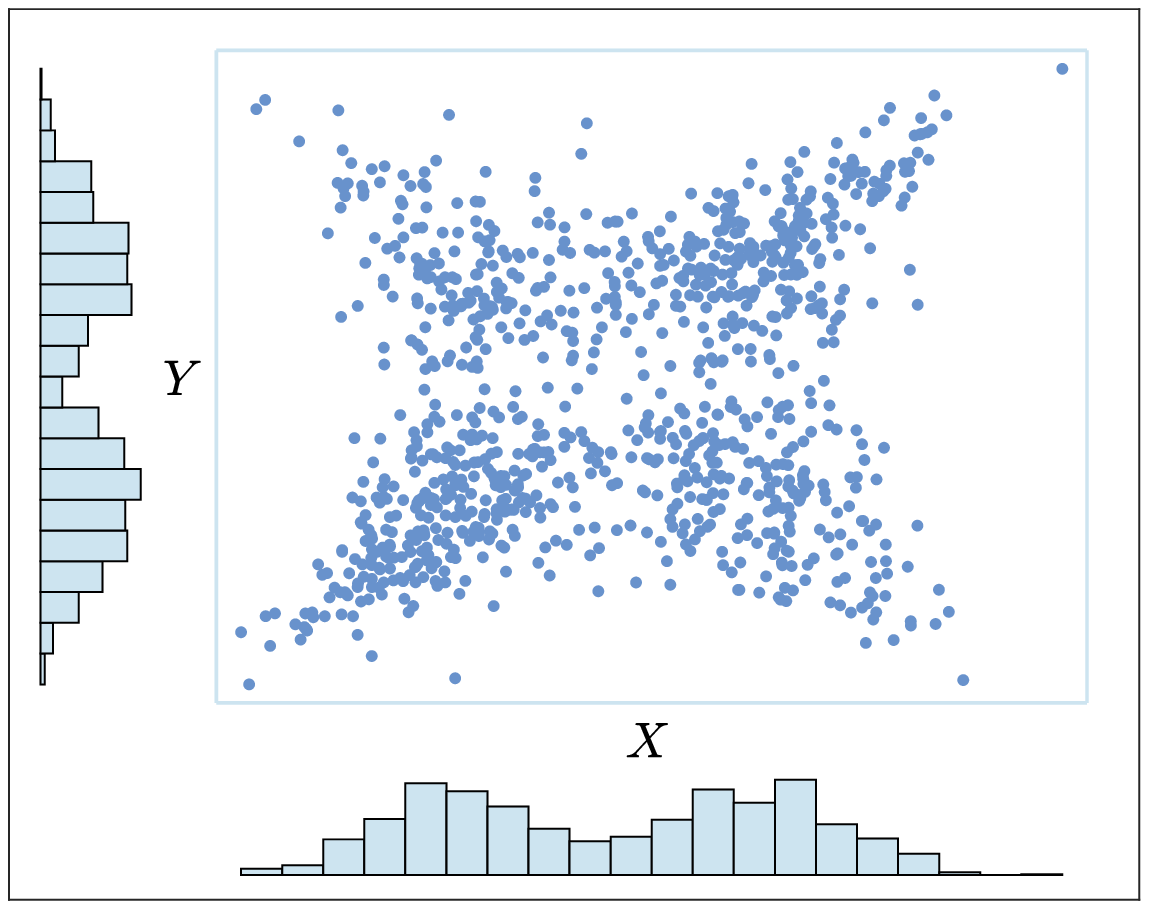}\\
     \footnotesize (5)
  \end{minipage}

    \mbox{} \\ \smallskip

  \begin{minipage}{0.19\textwidth}
   \centering
     \includegraphics[clip=true,width=\textwidth]{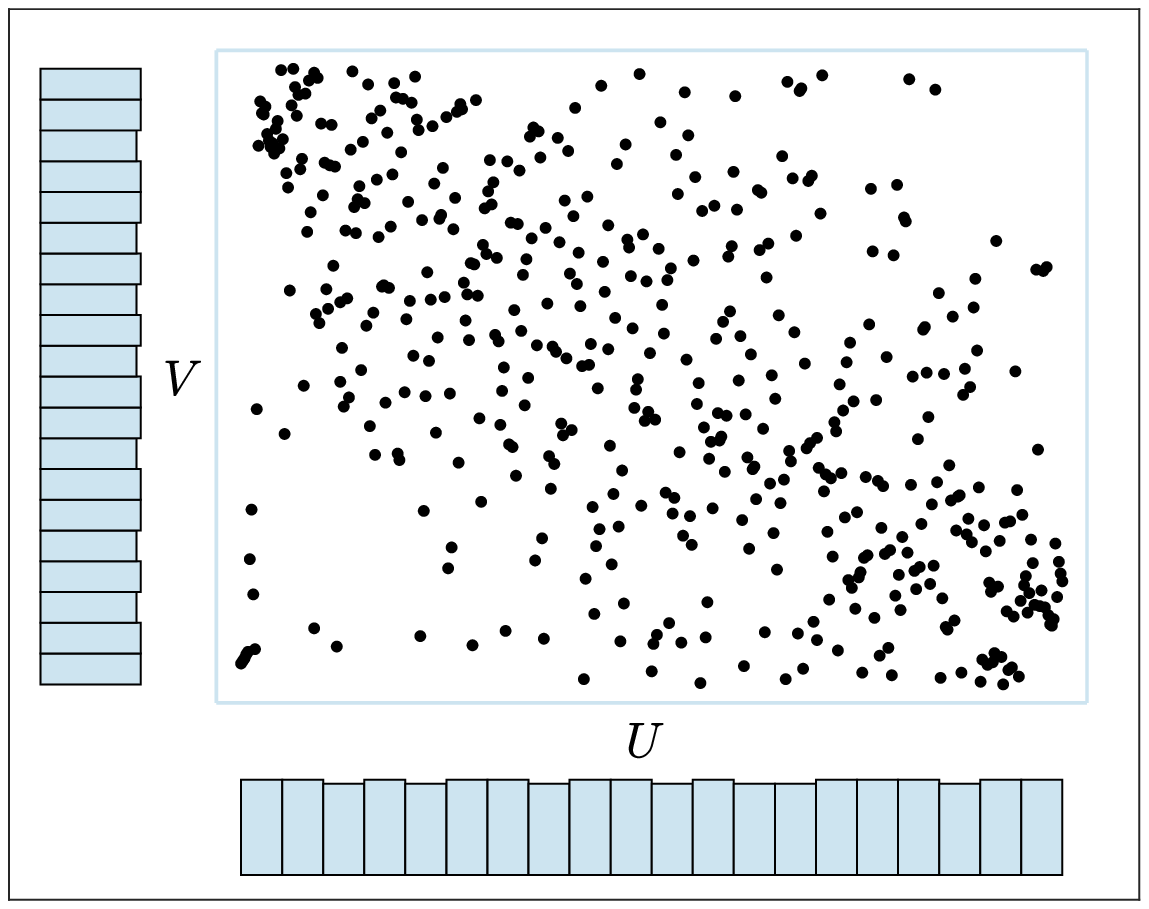}\\
     \footnotesize (16)
  \end{minipage}
  \hfill
  \begin{minipage}{0.19\textwidth}
   \centering
     \includegraphics[clip=true,width=\textwidth]{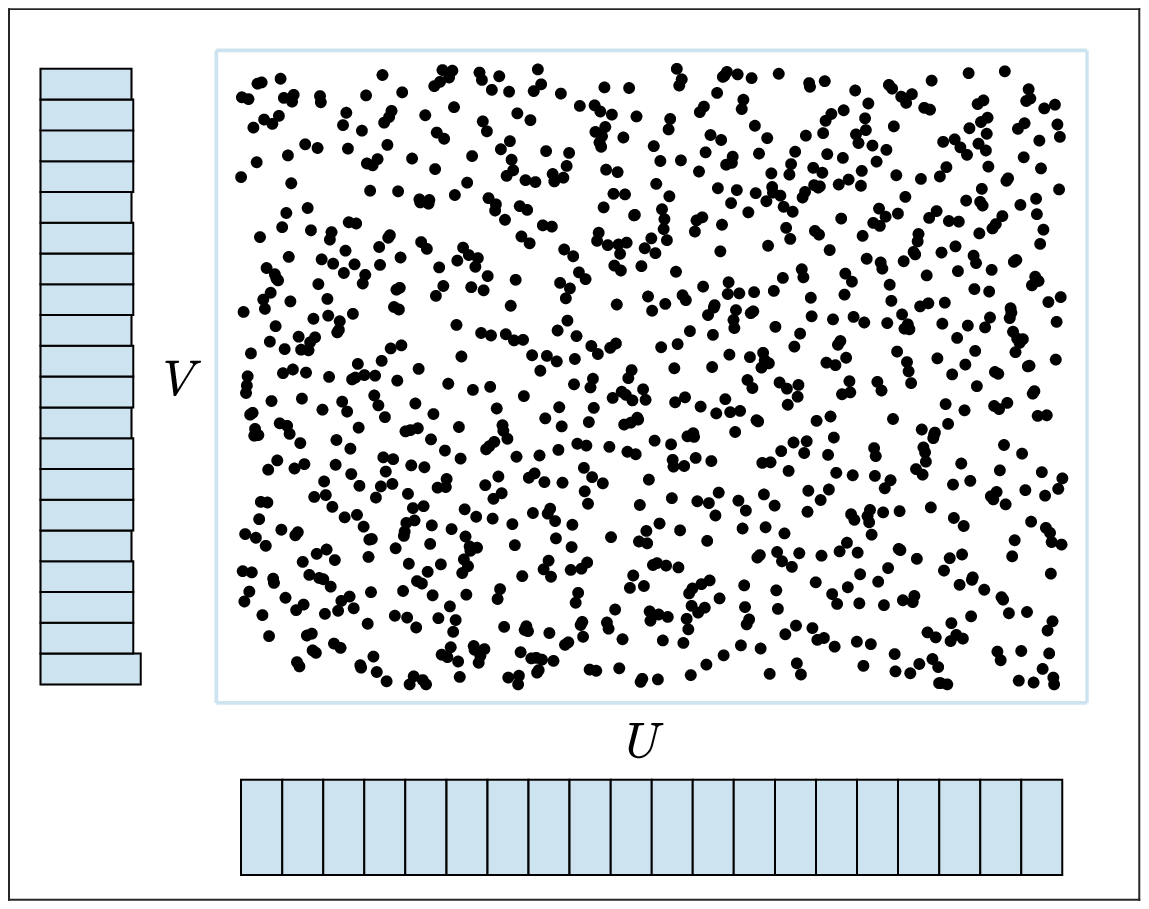}\\
     \footnotesize (17)
  \end{minipage}
  \hfill
  \begin{minipage}{0.19\textwidth}
   \centering
     \includegraphics[clip=true,width=\textwidth]{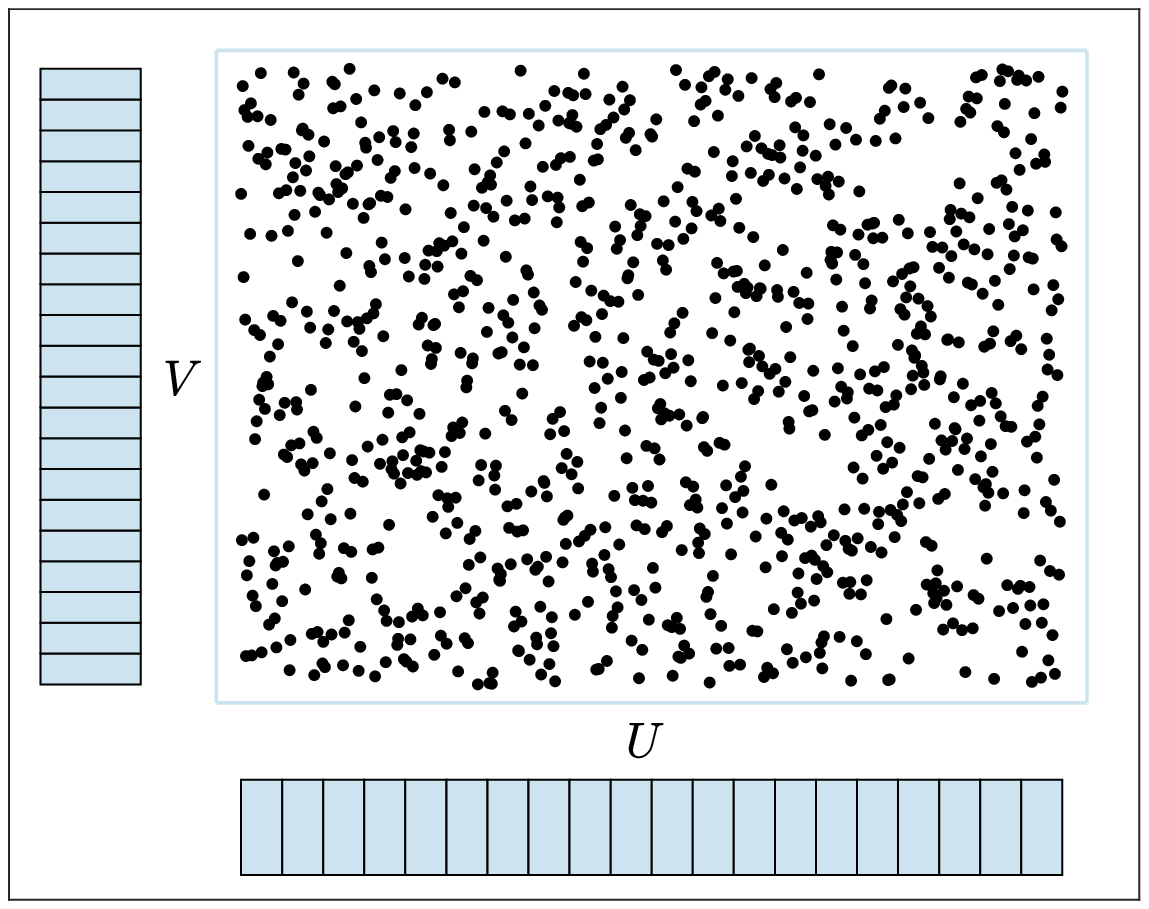}\\
     \footnotesize (18)
  \end{minipage}
  \hfill
  \begin{minipage}{0.19\textwidth}
   \centering
     \includegraphics[clip=true,width=\textwidth]{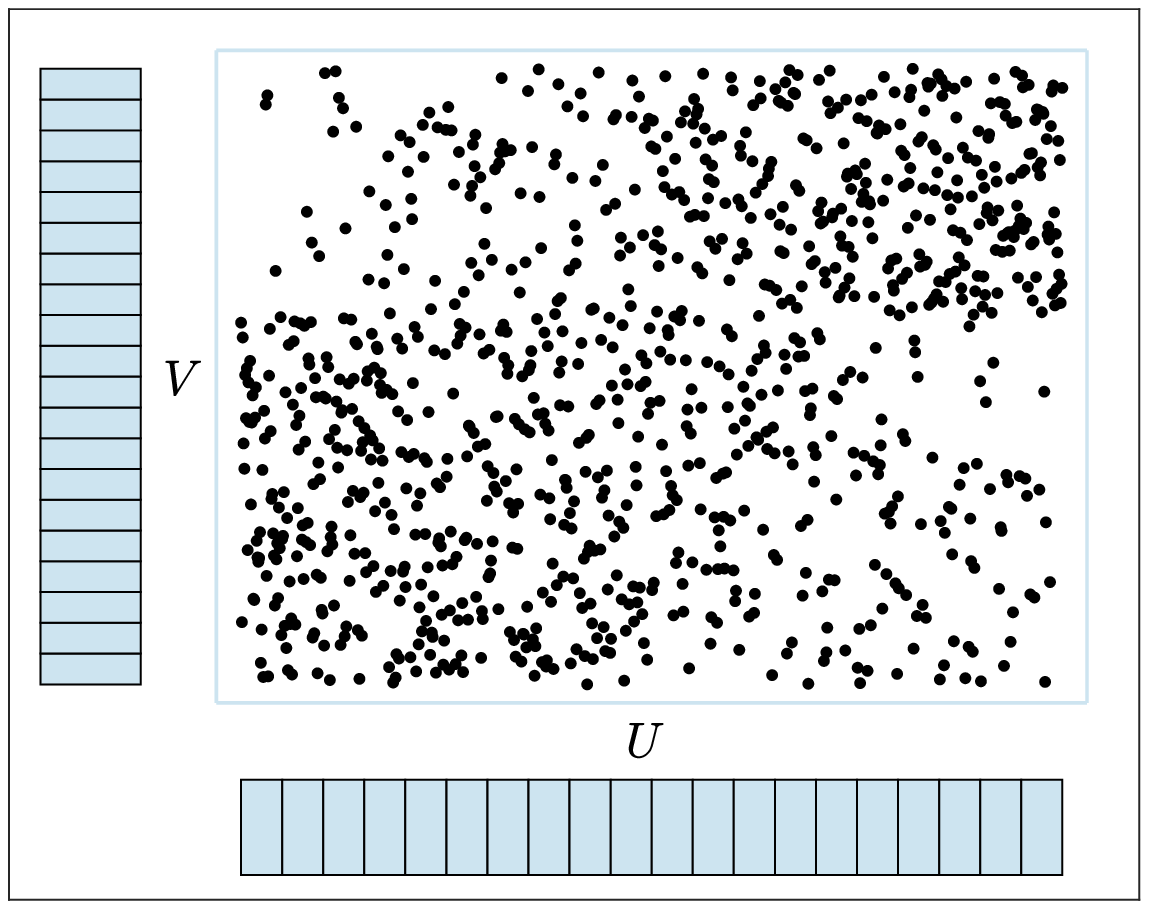}\\
     \footnotesize (19)
  \end{minipage}
  \hfill
  \begin{minipage}{0.19\textwidth}
   \centering
     \includegraphics[clip=true,width=\textwidth]{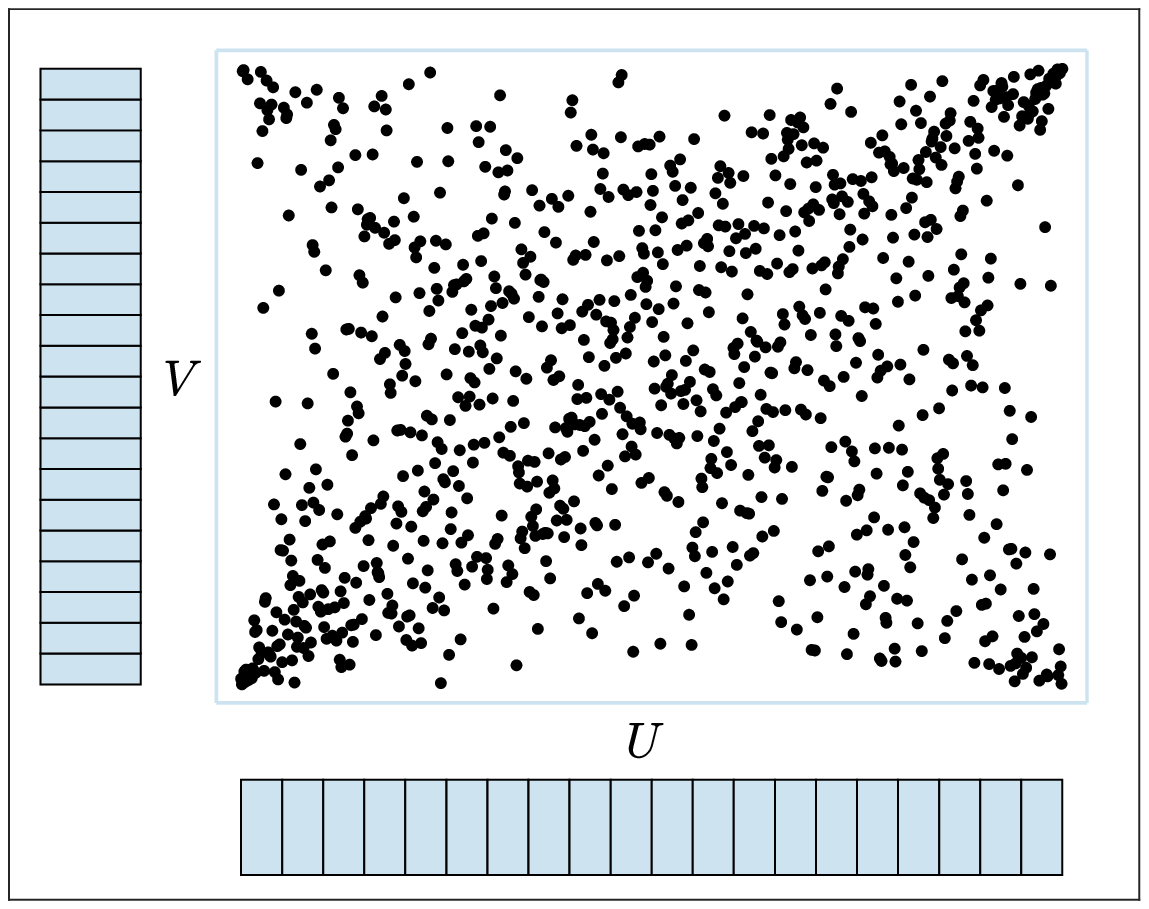}\\
     \footnotesize (20)
  \end{minipage}
  
    \mbox{} \\   \smallskip

  \begin{minipage}{0.19\textwidth}
   \centering
     \includegraphics[clip=true,width=\textwidth]{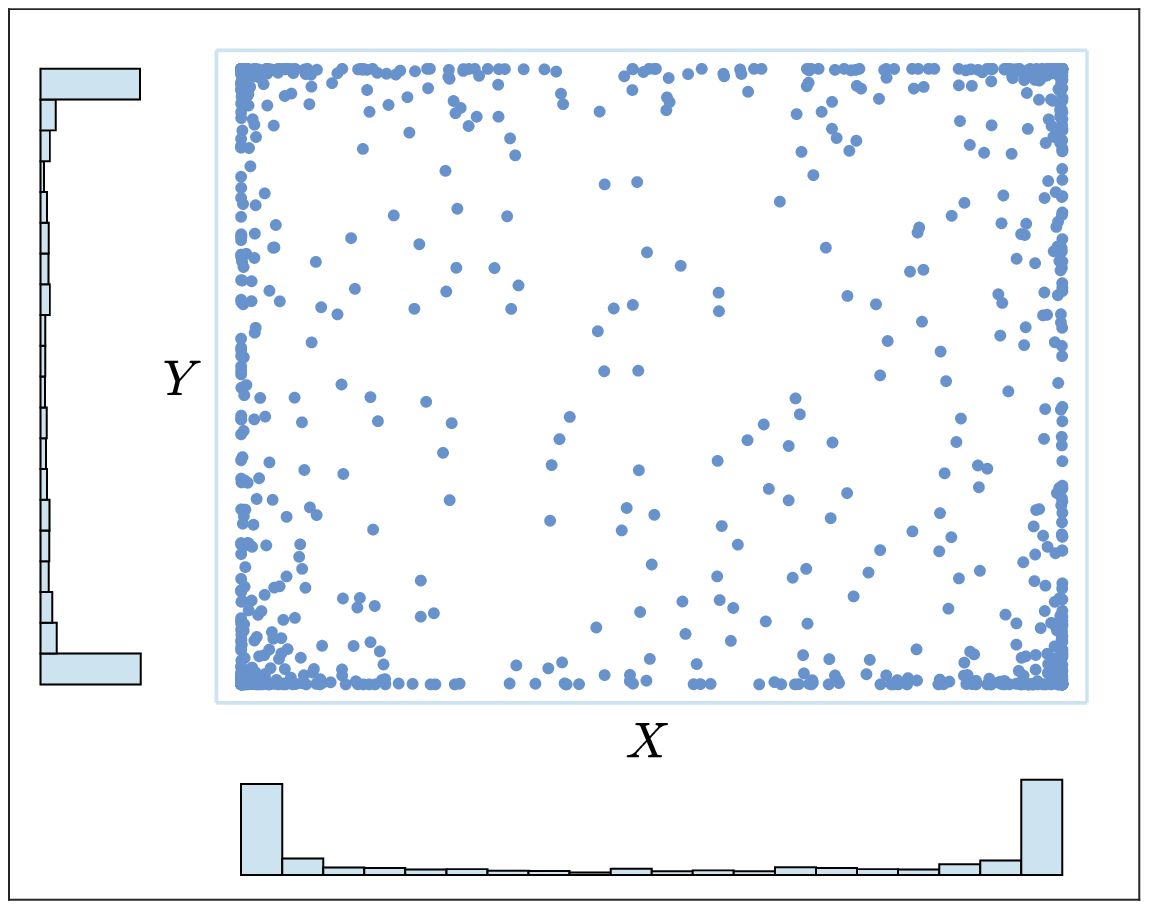}\\
     \footnotesize (6)
  \end{minipage}
  \hfill
  \begin{minipage}{0.19\textwidth}
   \centering
     \includegraphics[clip=true,width=\textwidth]{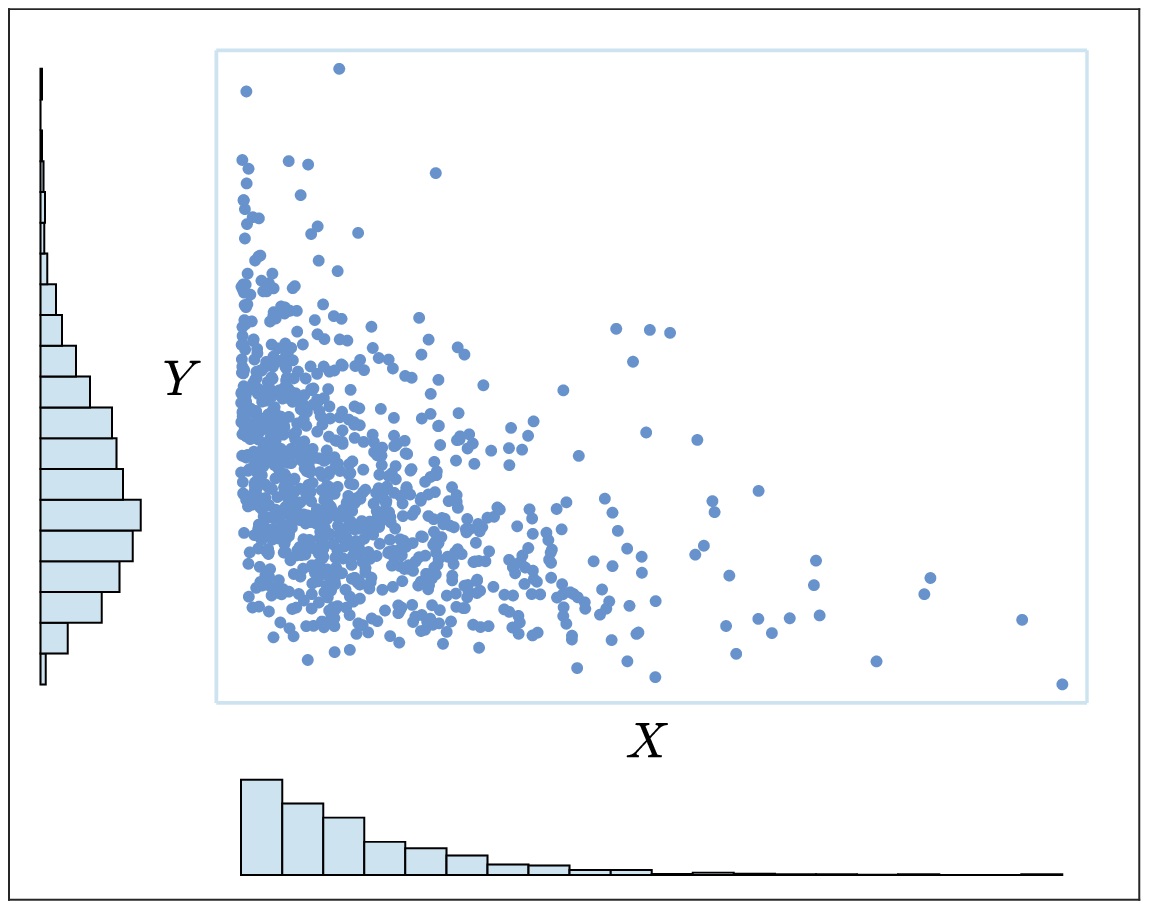}\\
     \footnotesize (7)
  \end{minipage}
  \hfill
  \begin{minipage}{0.19\textwidth}
   \centering
     \includegraphics[clip=true,width=\textwidth]{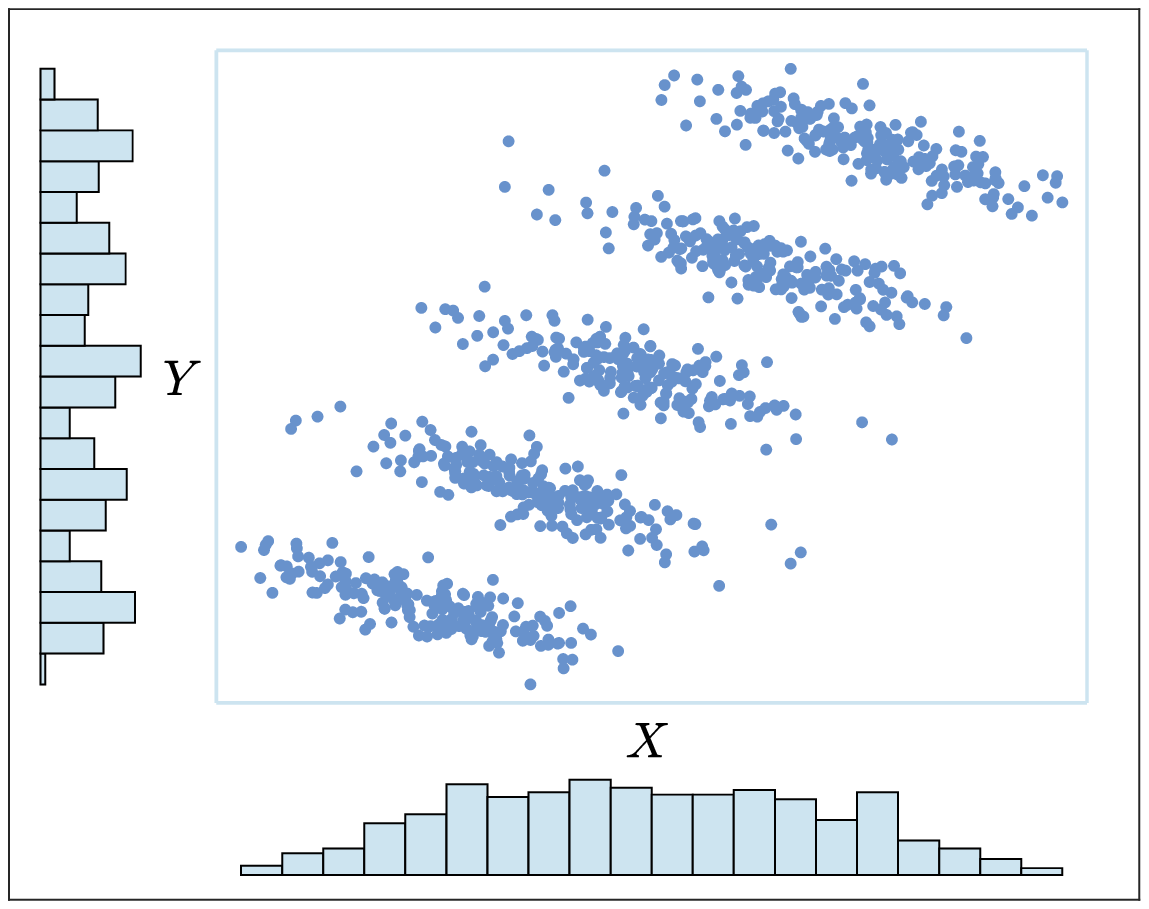}\\
     \footnotesize (8)
  \end{minipage}
  \hfill
  \begin{minipage}{0.19\textwidth}
   \centering
     \includegraphics[clip=true,width=\textwidth]{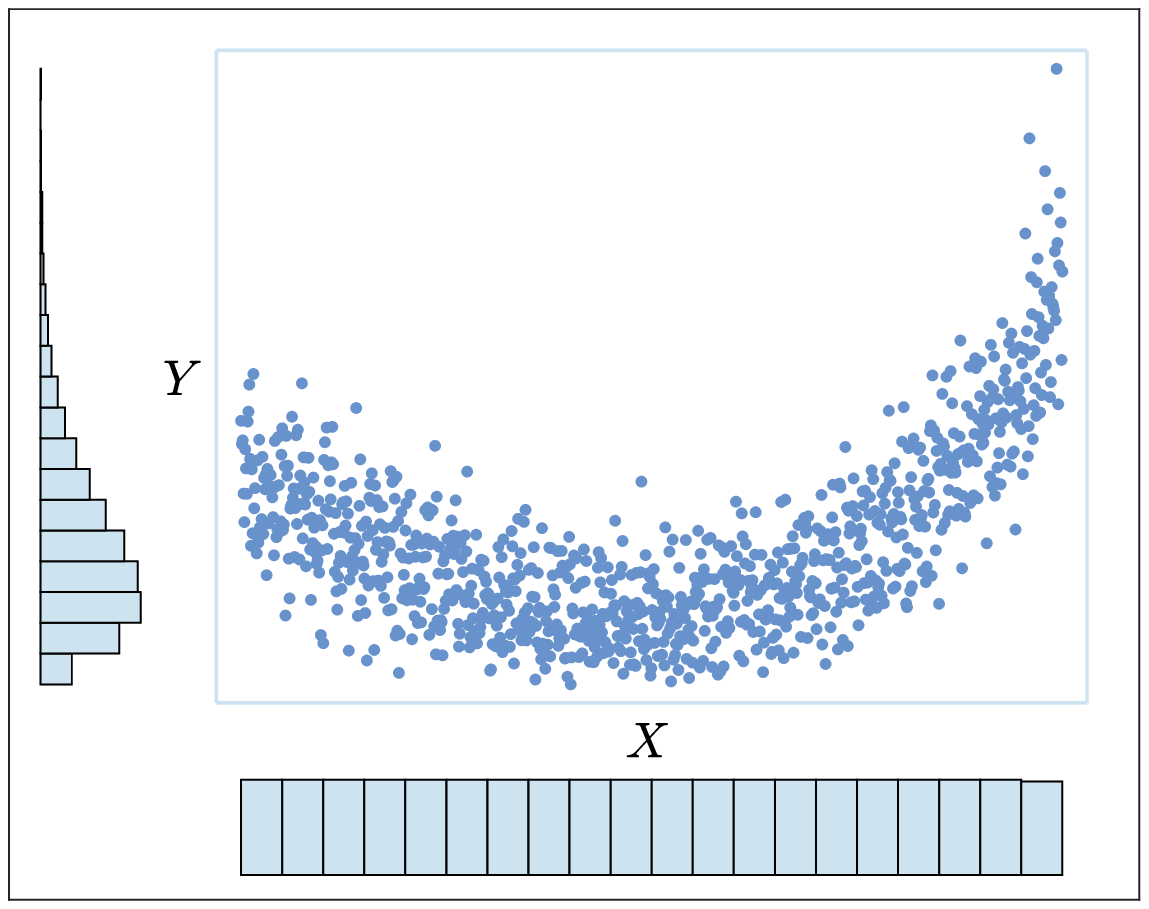}\\
     \footnotesize (9)
  \end{minipage}
  \hfill
  \begin{minipage}{0.19\textwidth}
   \centering
     \includegraphics[clip=true,width=\textwidth]{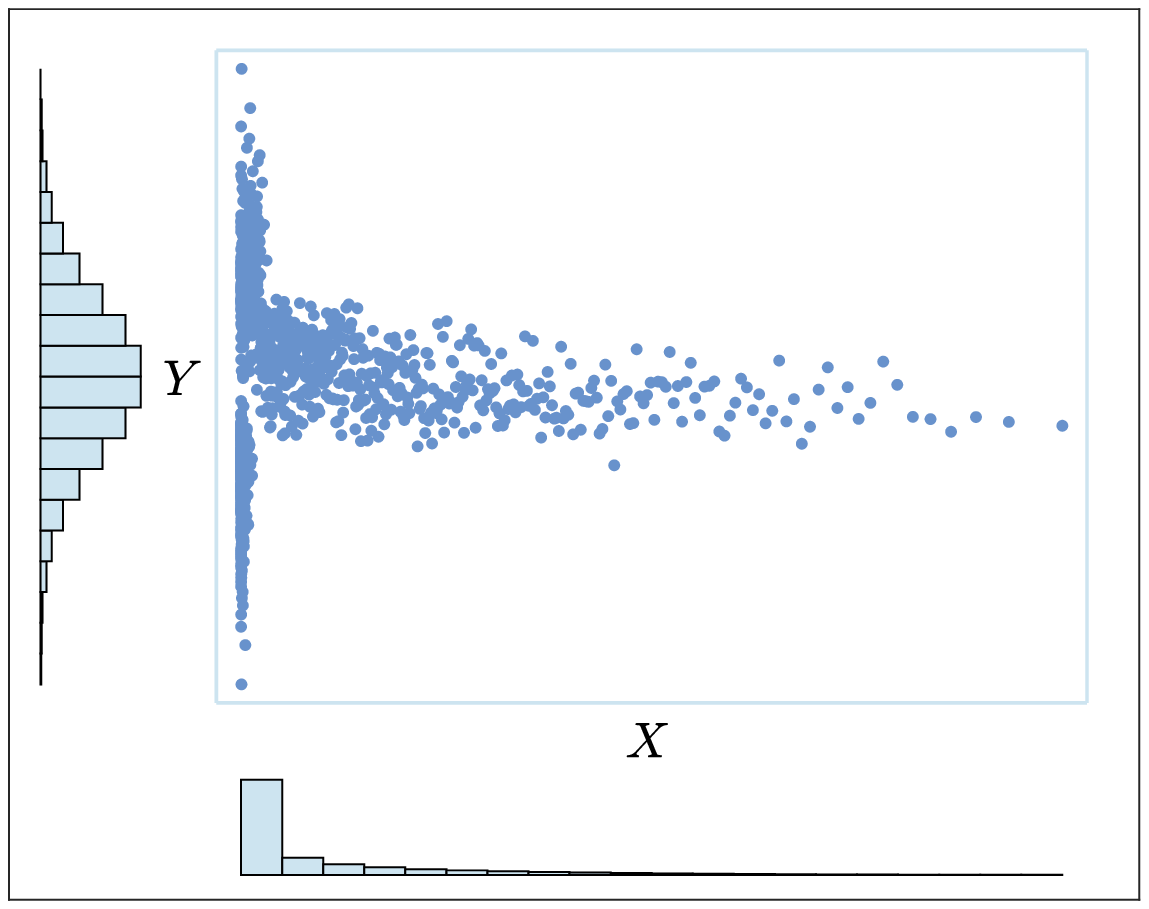}\\
     \footnotesize (10)
  \end{minipage}

    \mbox{}  \smallskip

  \begin{minipage}{0.19\textwidth}
   \centering
     \includegraphics[clip=true,width=\textwidth]{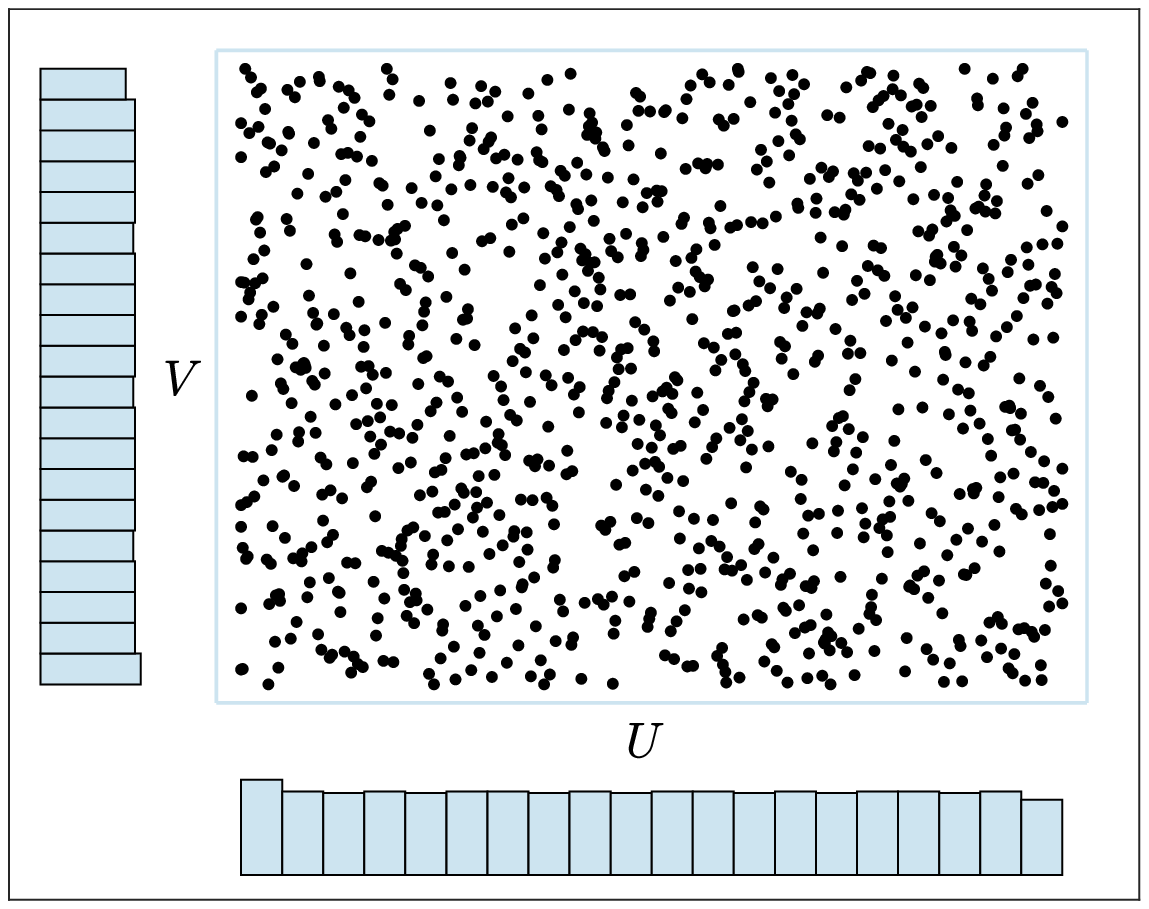}\\
     \footnotesize (21)
  \end{minipage}
  \hfill
  \begin{minipage}{0.19\textwidth}
   \centering
     \includegraphics[clip=true,width=\textwidth]{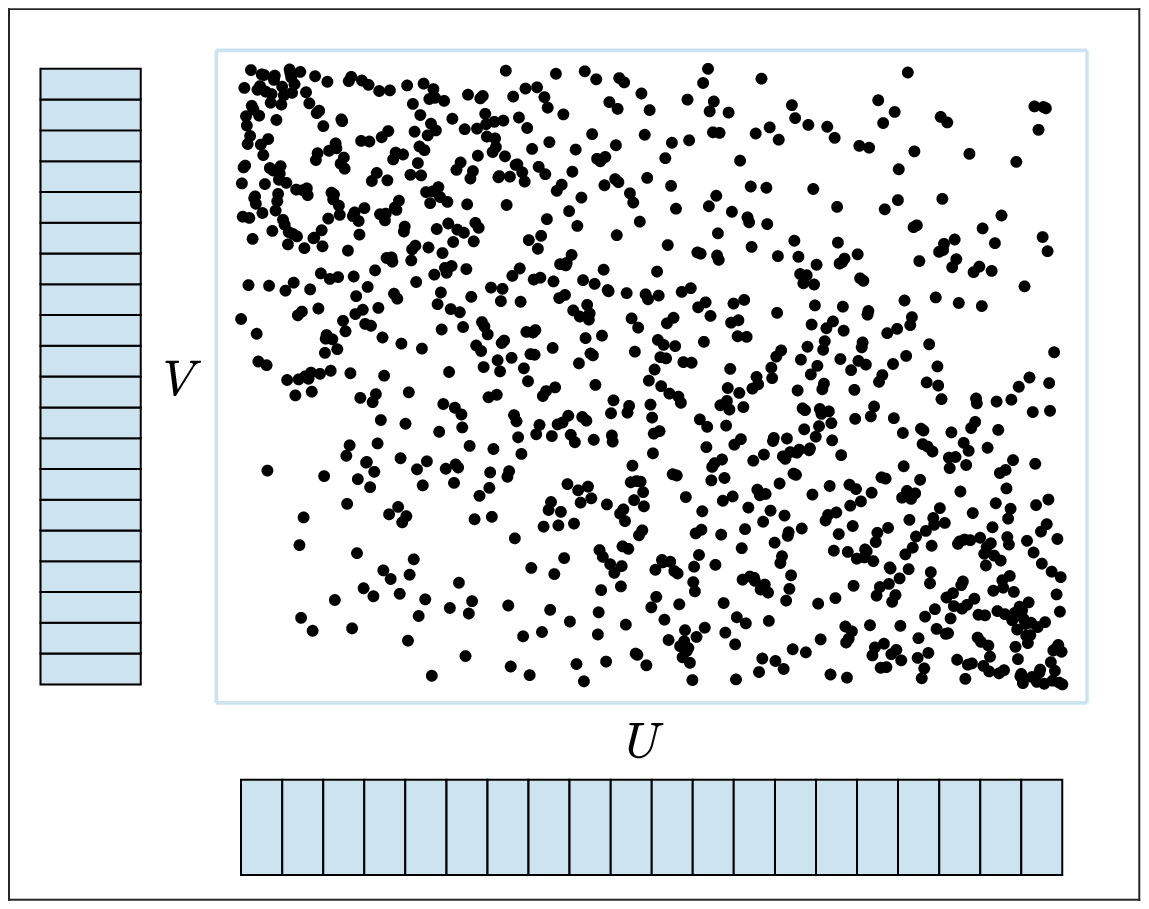}\\
     \footnotesize (22)
  \end{minipage}
  \hfill
  \begin{minipage}{0.19\textwidth}
   \centering
     \includegraphics[clip=true,width=\textwidth]{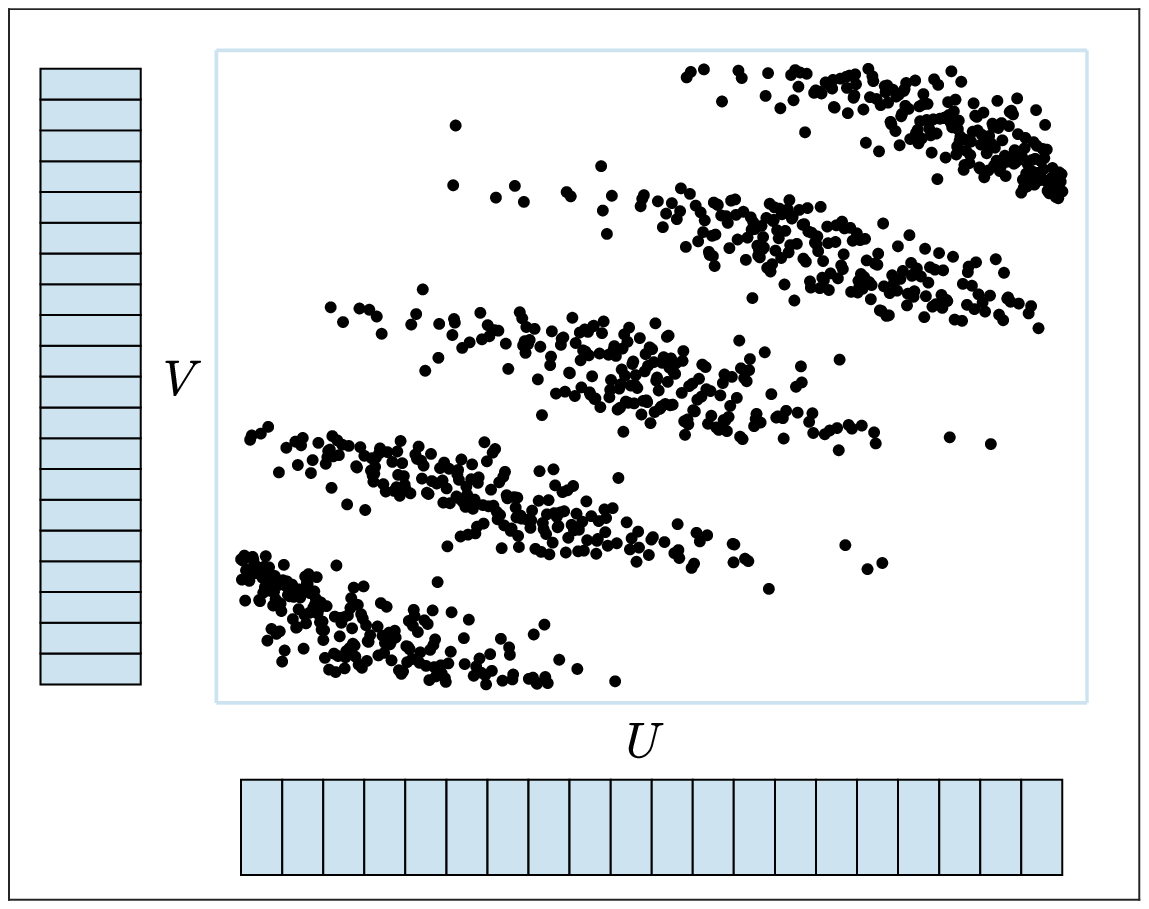}\\
     \footnotesize (23)
  \end{minipage}
  \hfill
  \begin{minipage}{0.19\textwidth}
   \centering
     \includegraphics[clip=true,width=\textwidth]{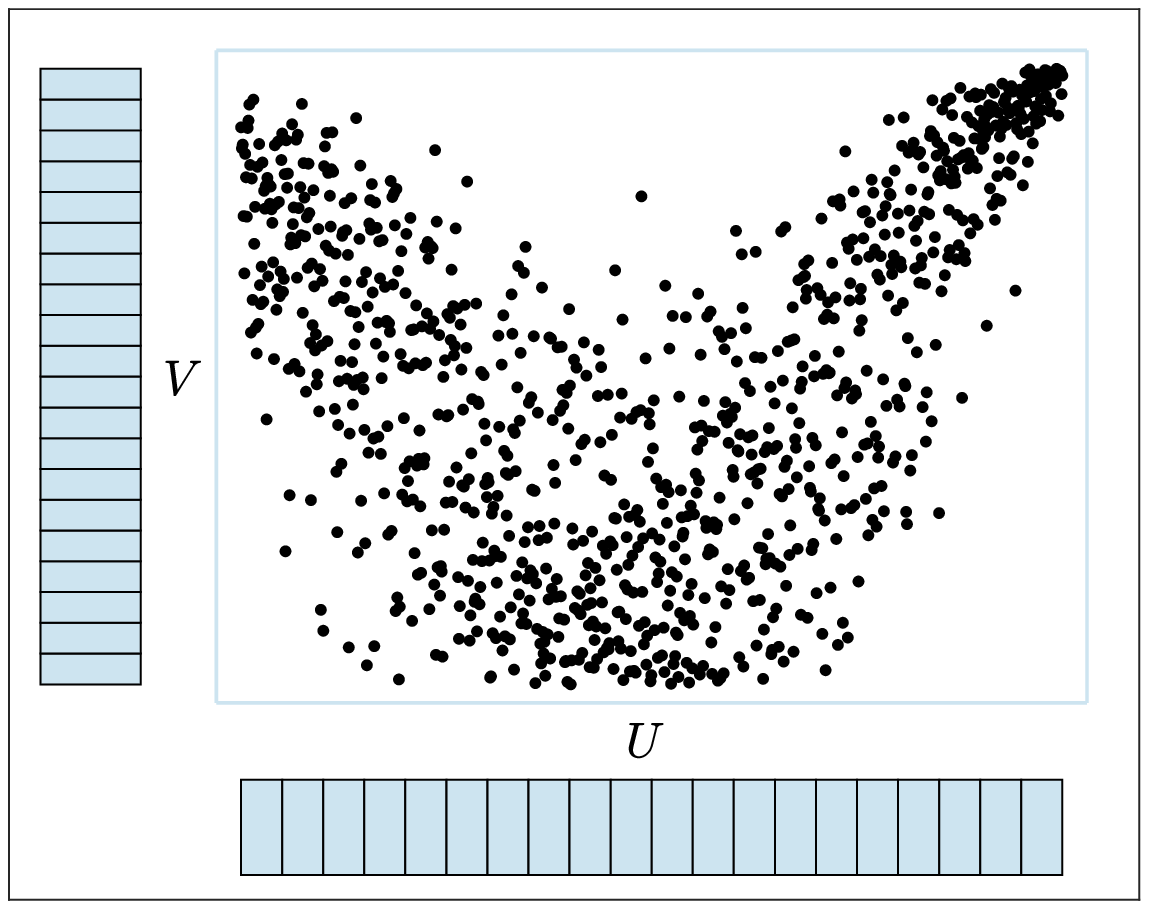}\\
     \footnotesize (24)
  \end{minipage}
  \hfill
  \begin{minipage}{0.19\textwidth}
   \centering
     \includegraphics[clip=true,width=\textwidth]{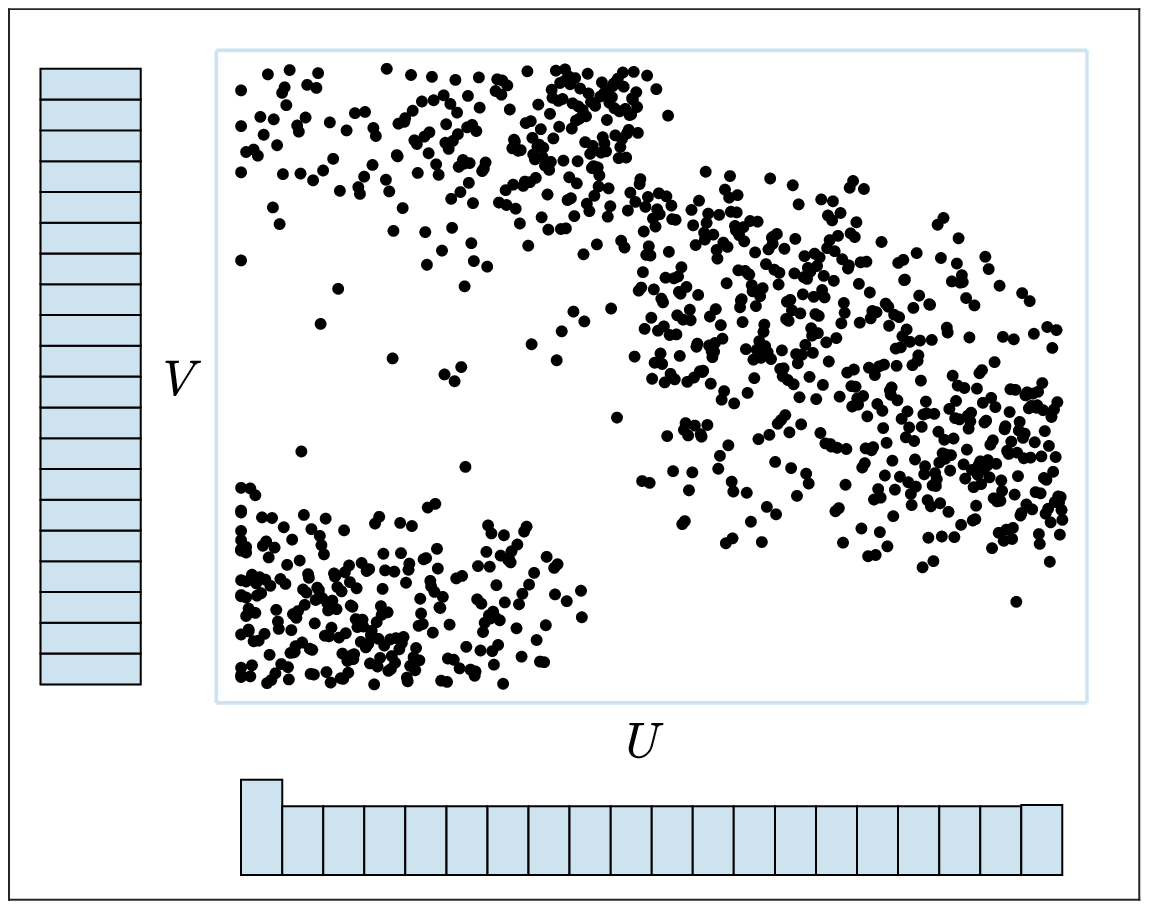}\\
     \footnotesize (25)
  \end{minipage}
  
    \mbox{} \\   \smallskip
  
  \begin{minipage}{0.19\textwidth}
   \centering
     \includegraphics[clip=true,width=\textwidth]{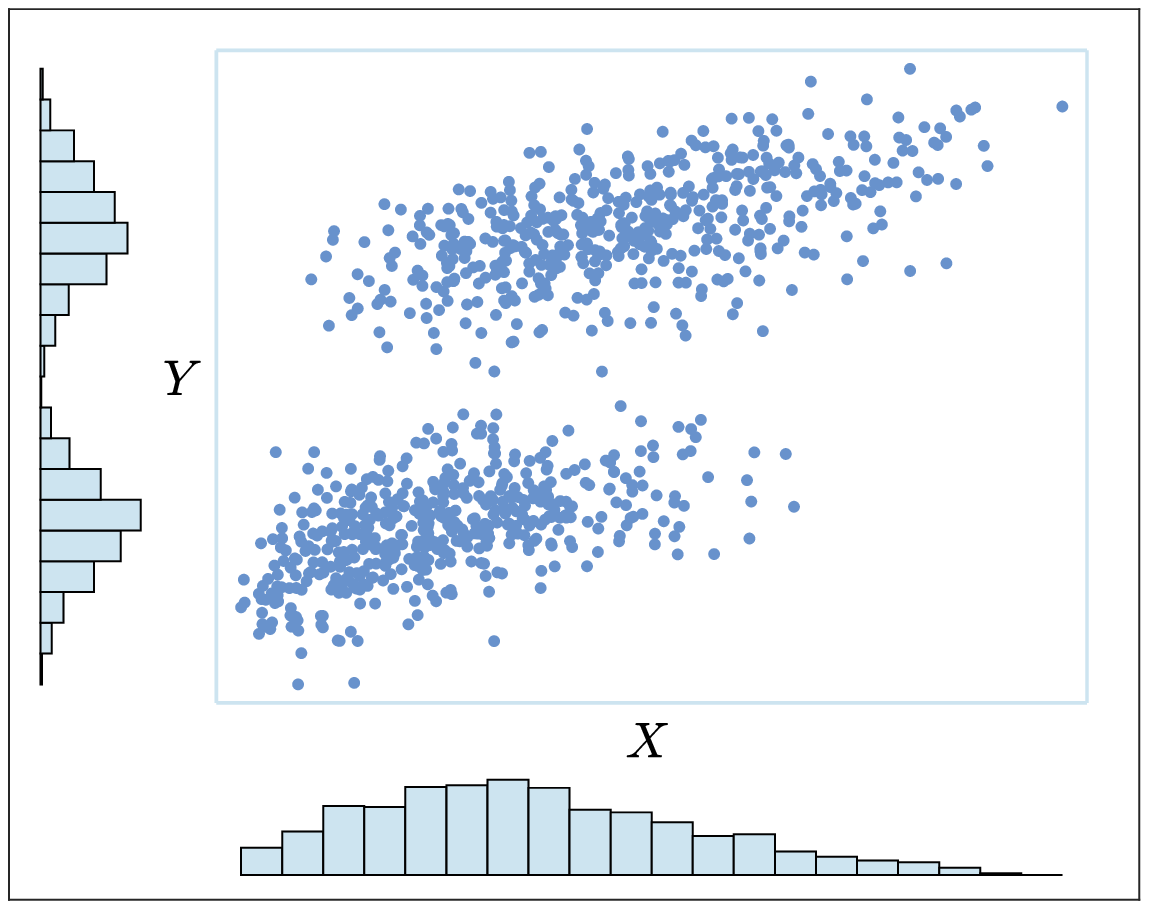}\\
     \footnotesize (11)
  \end{minipage}
  \hfill
  \begin{minipage}{0.19\textwidth}
   \centering
     \includegraphics[clip=true,width=\textwidth]{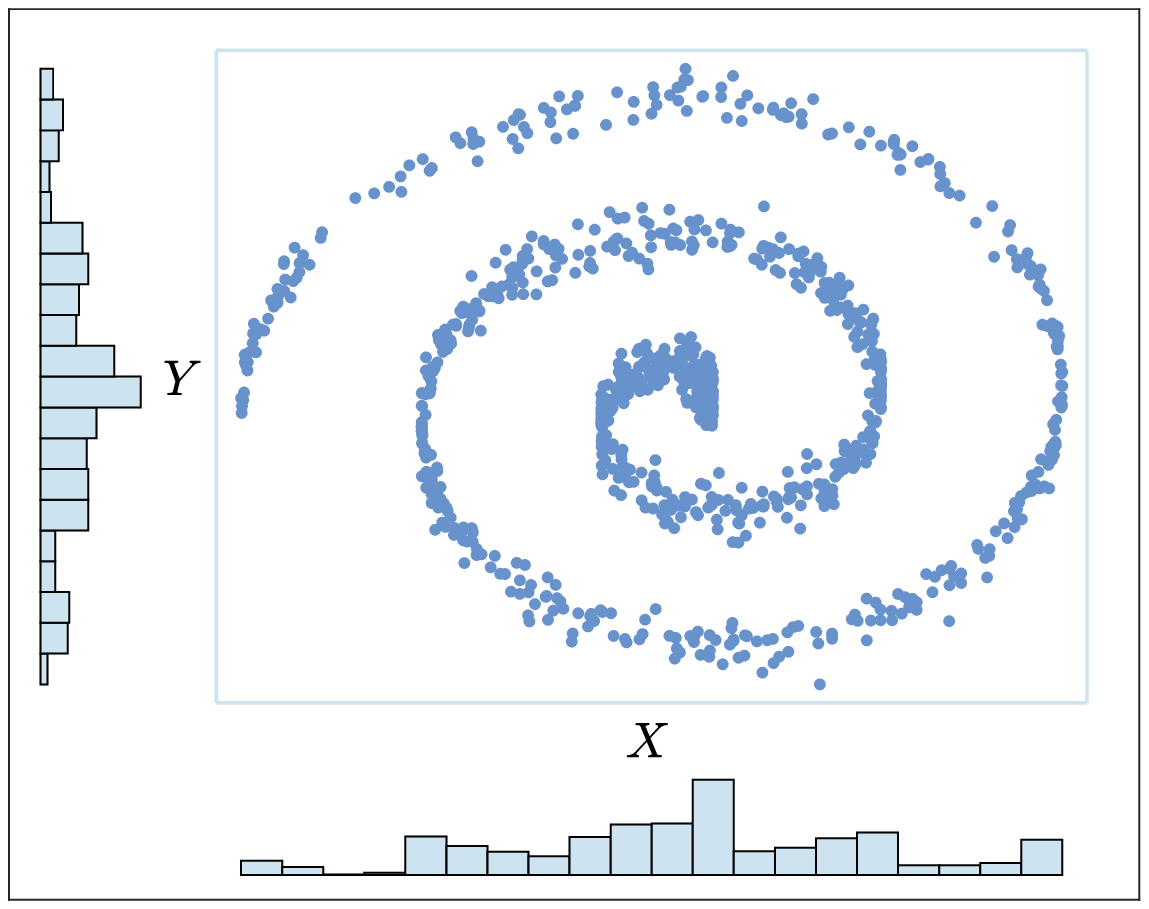}\\
     \footnotesize (12)
  \end{minipage}
  \hfill
  \begin{minipage}{0.19\textwidth}
   \centering
     \includegraphics[clip=true,width=\textwidth]{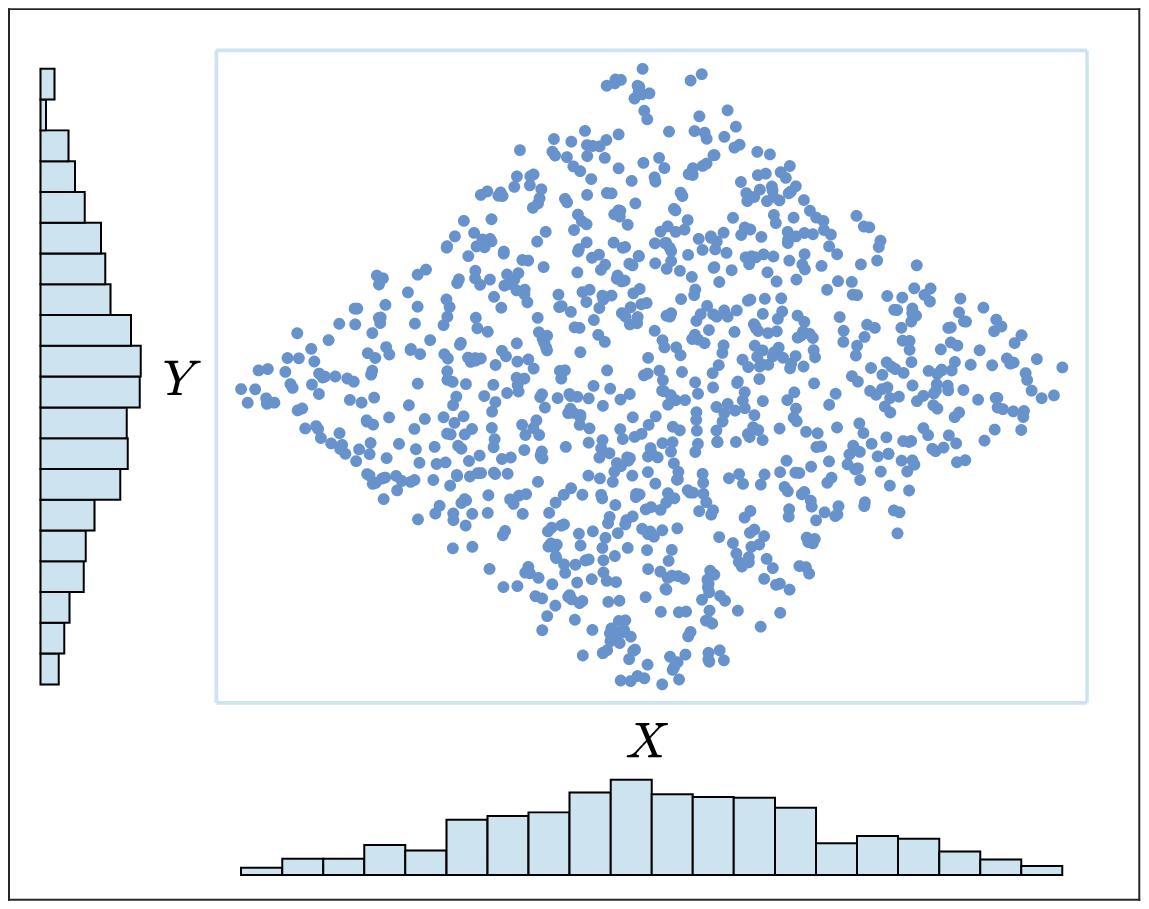}\\
     \footnotesize (13)
  \end{minipage}
  \hfill
  \begin{minipage}{0.19\textwidth}
   \centering
     \includegraphics[clip=true,width=\textwidth]{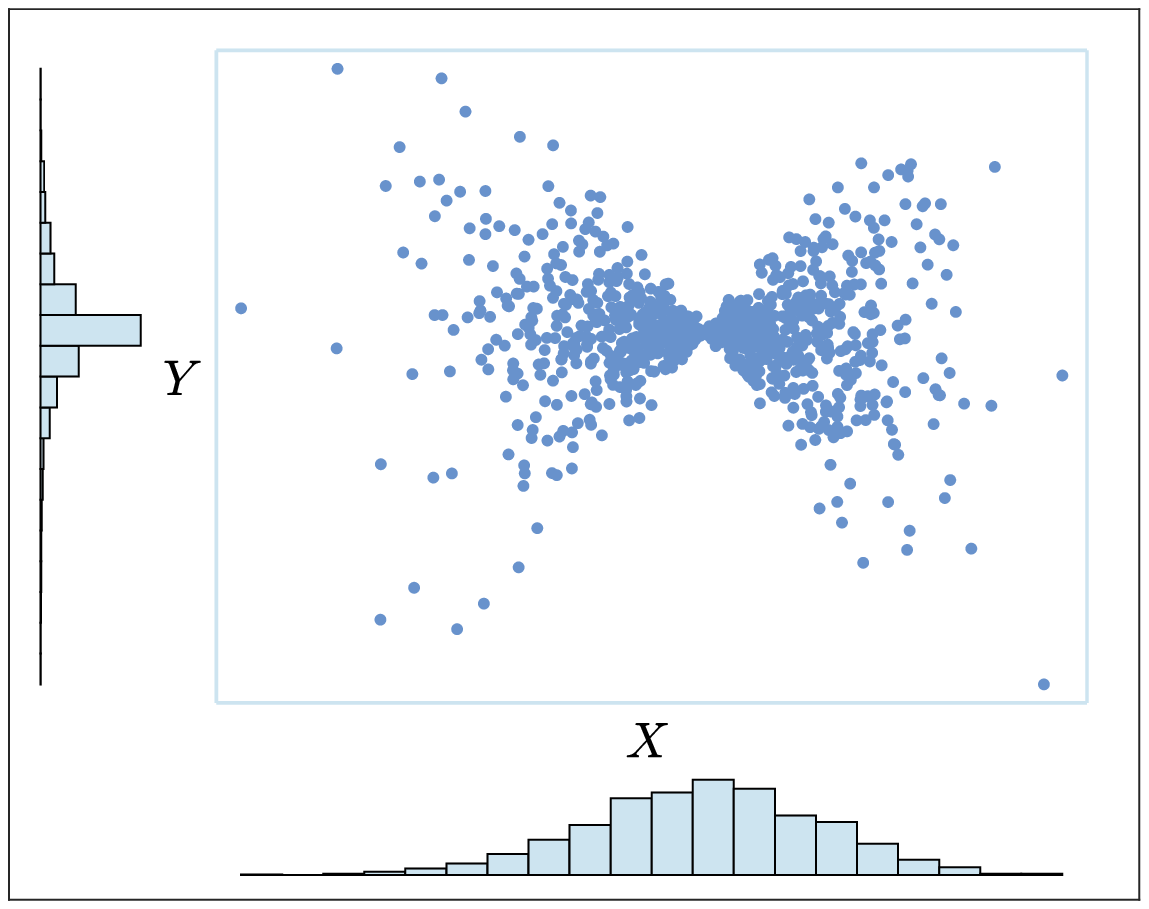}\\
     \footnotesize (14)
  \end{minipage}
  \hfill
  \begin{minipage}{0.19\textwidth}
   \centering
     \includegraphics[clip=true,width=\textwidth]{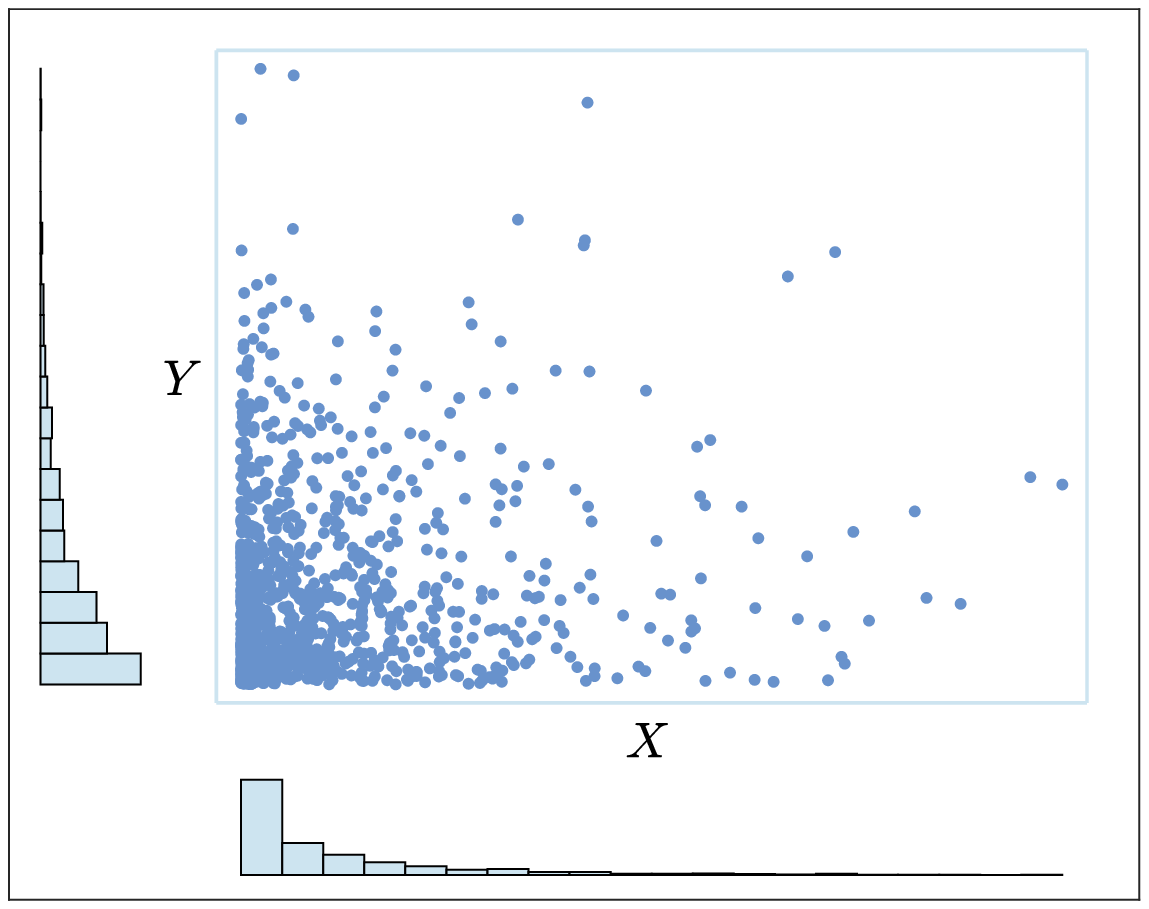}\\
     \footnotesize (15)
  \end{minipage}

    \mbox{}  \smallskip

  \begin{minipage}{0.19\textwidth}
   \centering
     \includegraphics[clip=true,width=\textwidth]{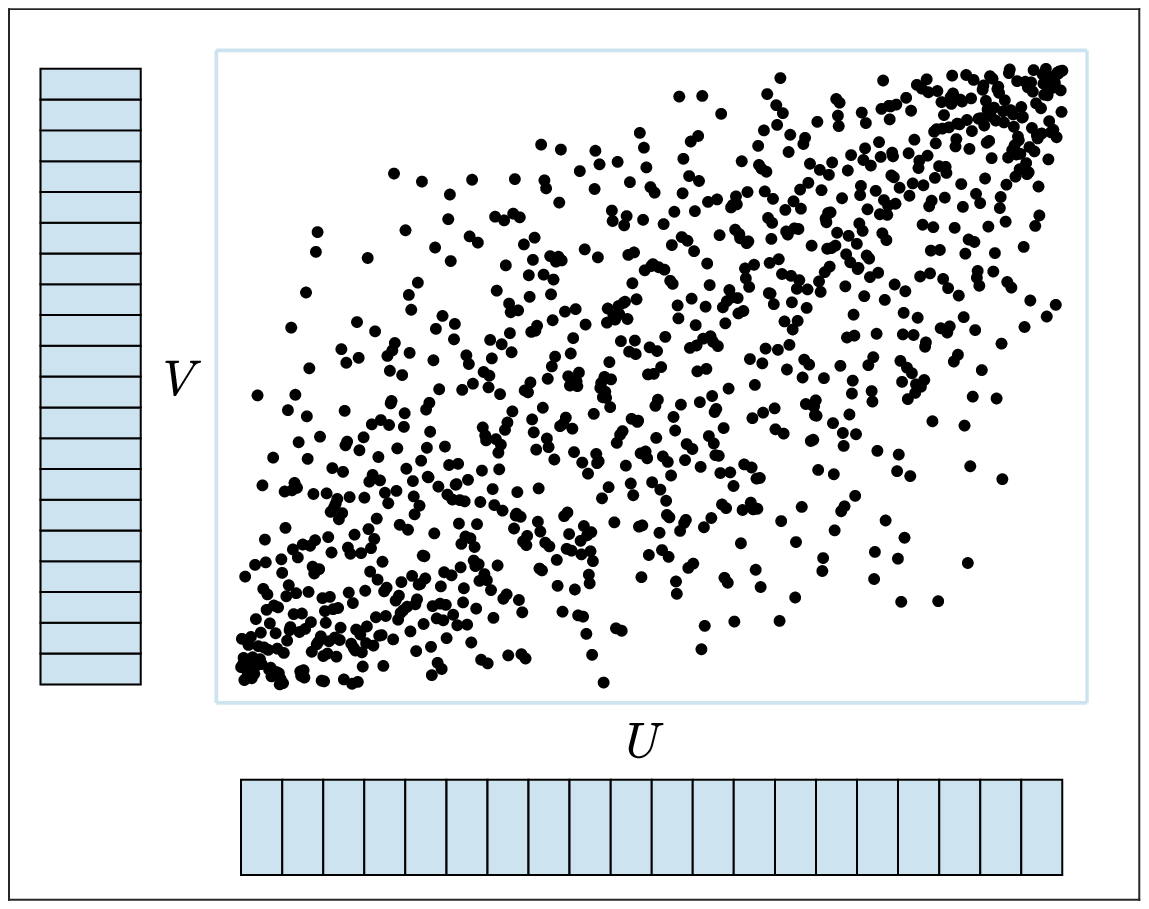}\\
     \footnotesize (26)
  \end{minipage}
  \hfill
  \begin{minipage}{0.19\textwidth}
   \centering
     \includegraphics[clip=true,width=\textwidth]{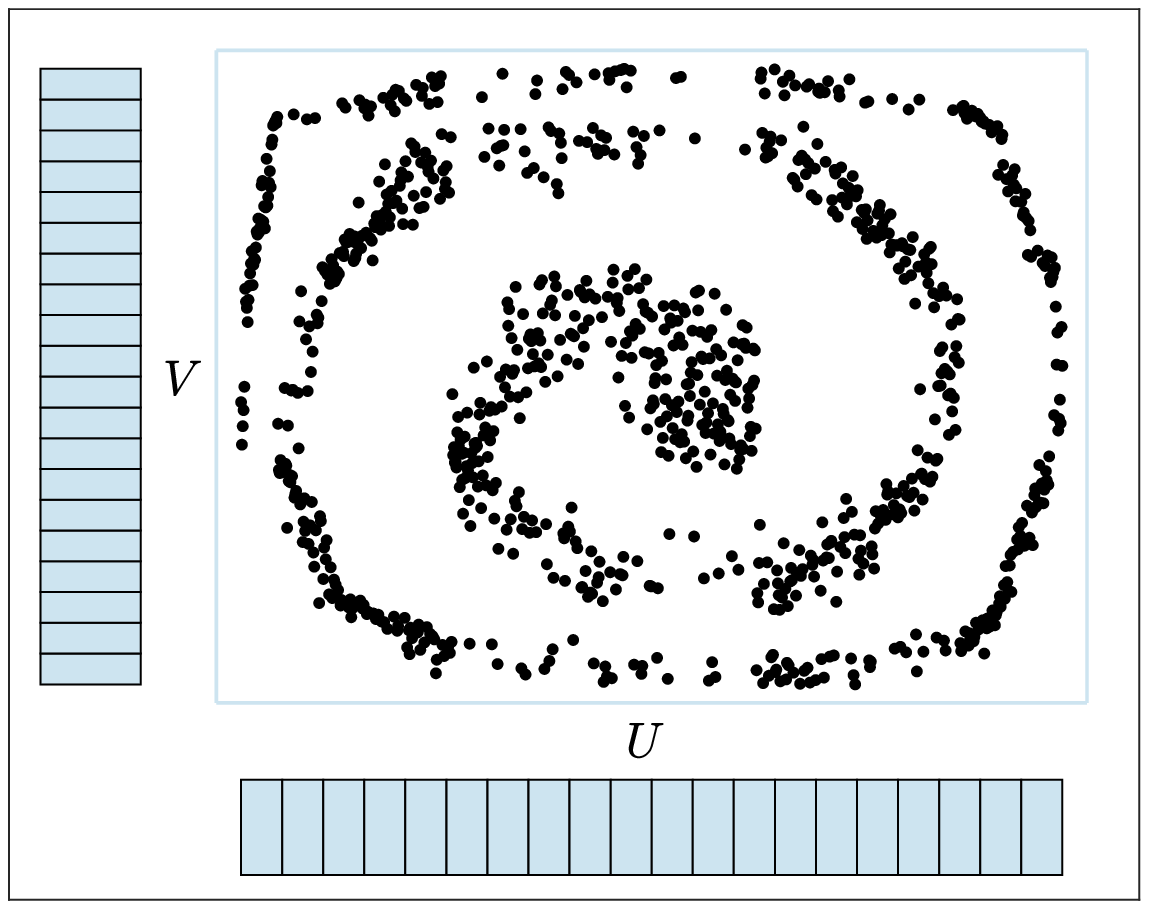}\\
     \footnotesize (27)
  \end{minipage}
  \hfill
  \begin{minipage}{0.19\textwidth}
   \centering
     \includegraphics[clip=true,width=\textwidth]{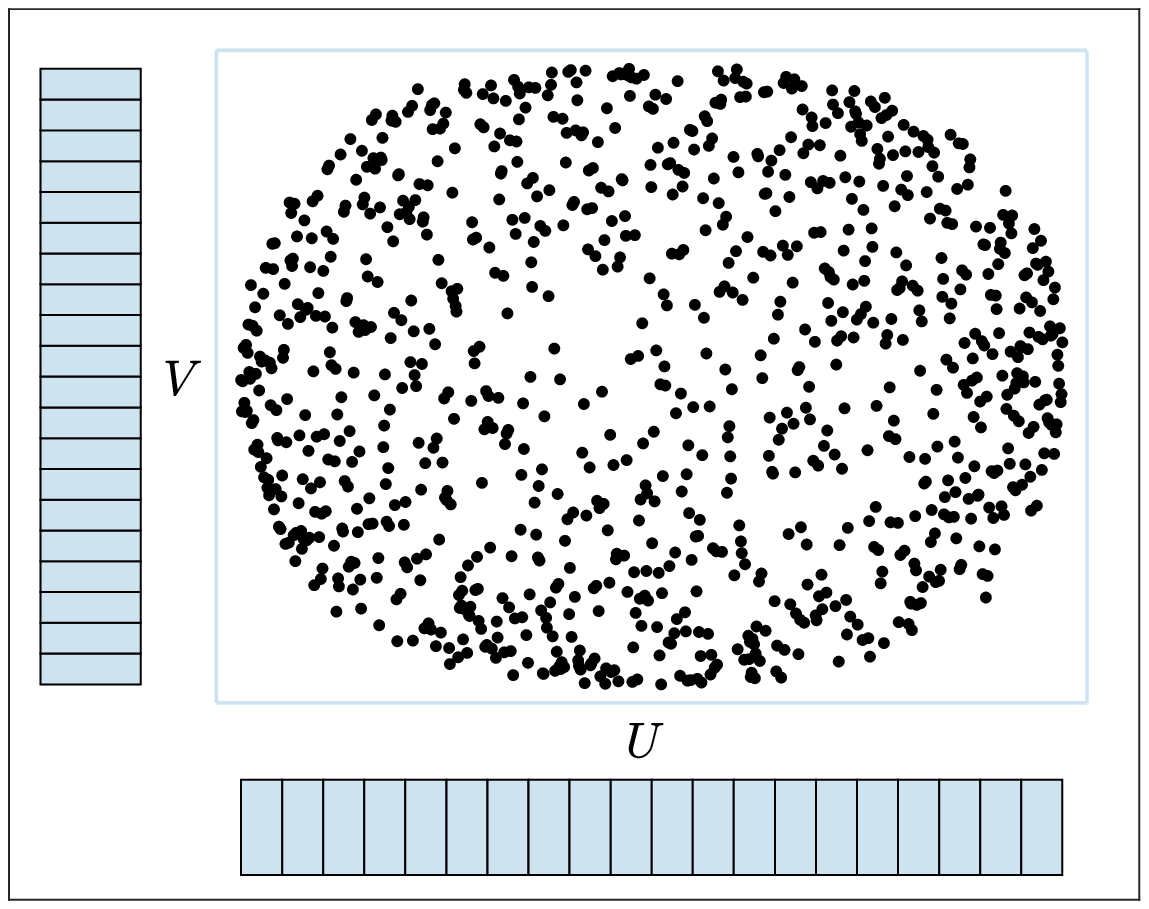}\\
     \footnotesize (28)
  \end{minipage}
  \hfill
  \begin{minipage}{0.19\textwidth}
   \centering
     \includegraphics[clip=true,width=\textwidth]{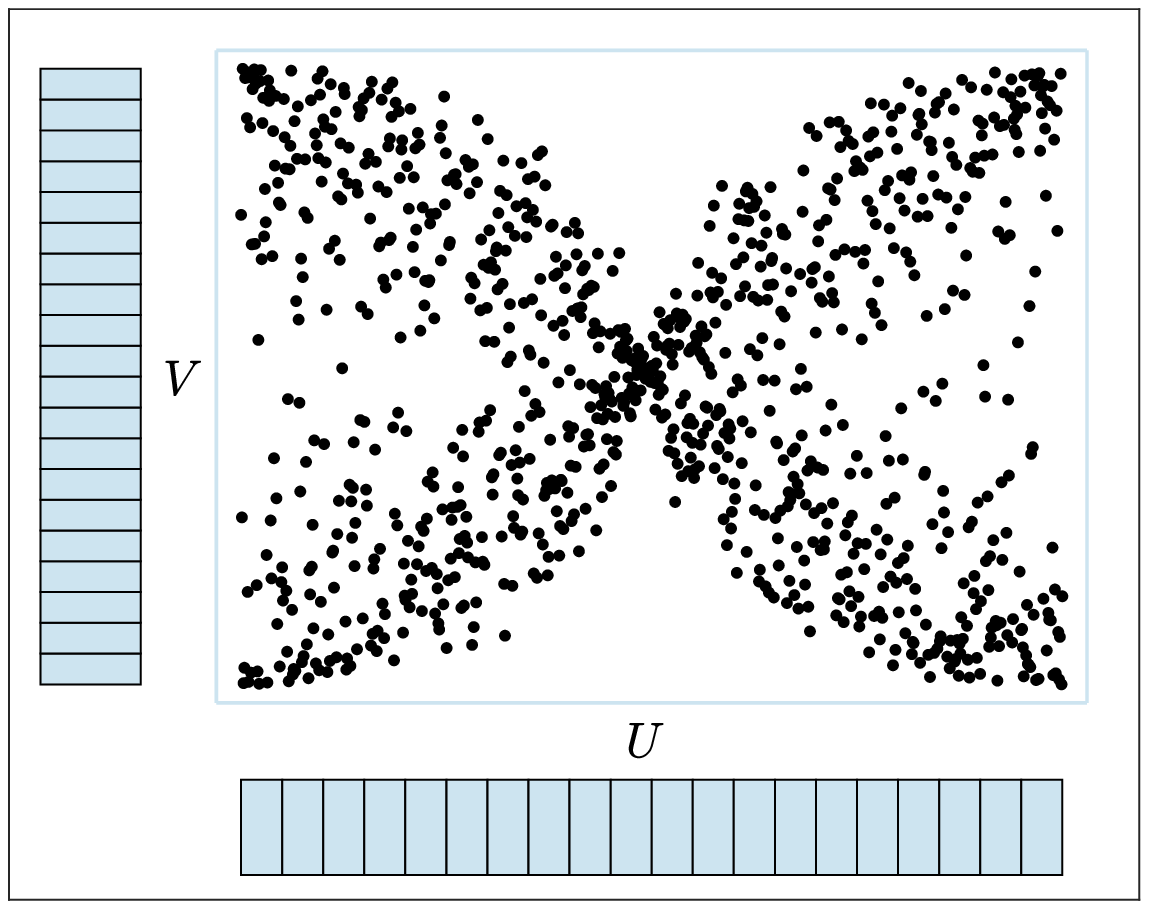}\\
     \footnotesize (29)
  \end{minipage}
  \hfill
  \begin{minipage}{0.19\textwidth}
   \centering
     \includegraphics[clip=true,width=\textwidth]{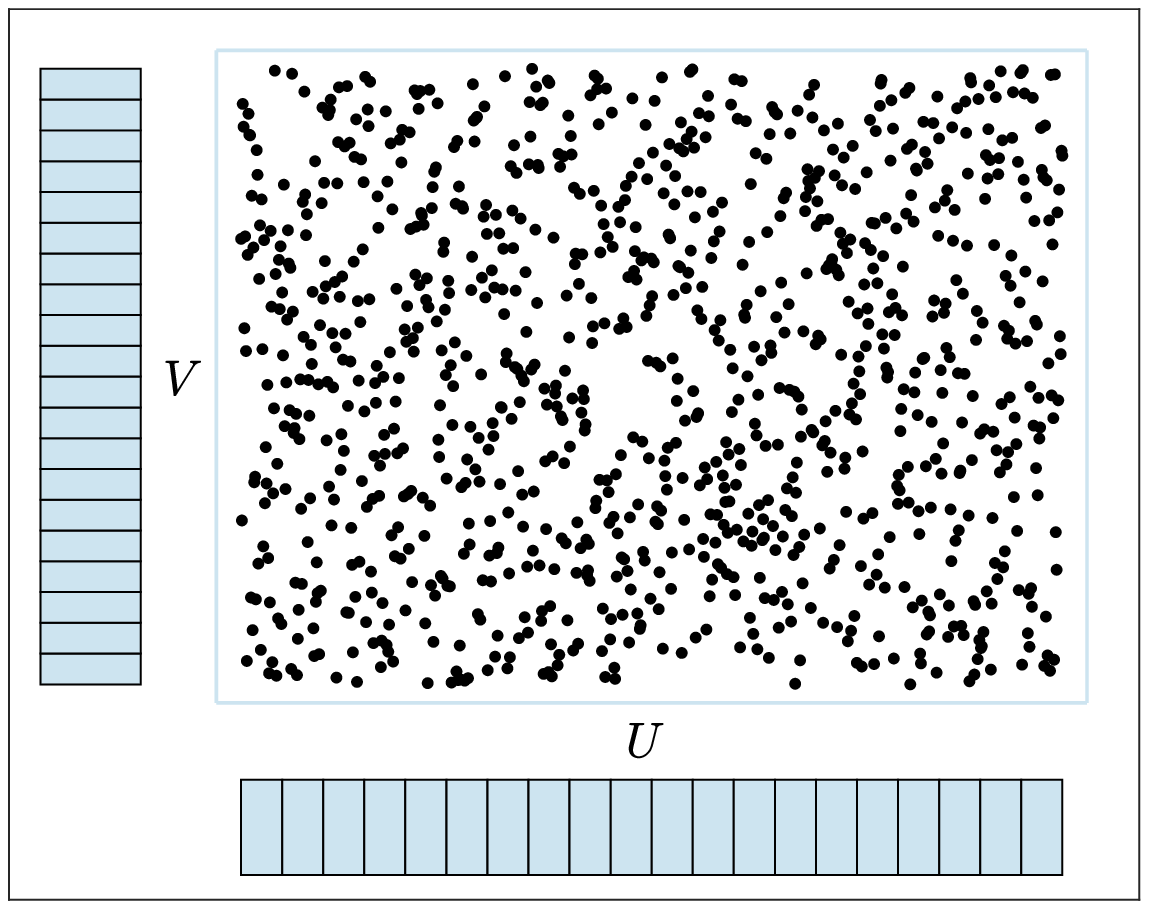}\\
     \footnotesize (30)
  \end{minipage}  
  \caption{Second set of scatter plots used in the experiments in section \ref{sec:experiments}. Each row of blue-dotted scatter plots is followed by a row of their corresponding black-dotted pseudo-observations. For example (16) are the pseudo-observations for (1) and therefore both have exactly the same dependence structure (copula).}
\label{fig:6x5}
\end{figure*}

We carried out an experiment to evaluate the ability of users to perceive dependence in regular scatter plots and ones based on pseudo-observations. The participants were eight electrical engineering professors familiar with statistics and data science methods, and four professors of statistics. Their ages ranged from 25 to 49. Each user was shown 39 scatter plots (see Fig.~\ref{fig:5x2} and Fig.~\ref{fig:6x5}) on a square figure, and they had to estimate a degree of statistical dependence between the two represented variables. The users controlled a slider in order to select values from 0 (independence) to 1 (``total'' dependence). We did not offer any explanation regarding the definition of independence, but clarified that total dependence would be a situation in which the value of one variable could be completely determined by the value of the other. The data sets used can be found in the supplemental material.

In order to analyze statistically significant differences between the users' dependence assessments based on scatter plots, in contrast to those based on the pseudo-observations corresponding to such scatter plots, we performed paired two-sample sign tests \cite{Mood1974}, but under a Bayesian approach \cite{Bernardo1994} since it is better suited for a small sample size ($m=12$ users). For each pair (scatter plot, pseudo-observations) we analyzed the pairs $(S_i,T_i),$ for $i\in\{1,\ldots,m\},$ where $S_i$ and $T_i$ are the perceived dependencies from the scatter plot and the pseudo-observations, respectively. We then define the indicator random variables $Z_i=\mathbb{1}_{\{S_i\,>\,T_i\}}$, which are independent and identically distributed Bernoulli random variables with unknown parameter $\theta=P(S>T).$ By using a conjugate family as a non-informative prior distribution we get a posterior distribution for $\theta=P(S>T)$, which is a Beta distribution with parameters $\frac{1}{2}+\sum z_i$ and $\frac{1}{2}+m-\sum z_i$. Therefore, as a point estimate (under the quadratic loss) the mean of such distribution is $\hat{\theta}=(\frac{1}{2}+\sum z_i)/(m+1).$ In addition, we built minimum length $100\gamma\%$ credible interval $[a,b]$ estimates for $\theta=P(S>T).$

The first set of scatter plots to evaluate are shown in Fig.~\ref{fig:5x2}. The plot in (e) corresponds to pseudo-observations, and we generated the other eight plots (a-d) and (f-i) by just choosing different marginal distributions (see Fig.~\ref{fig:pseudoobservations}). The pseudo-observations corresponding to each scatter plot are therefore the ones in (e). This implies that all of the plots have the same dependence structure (i.e the same underlying copula), and therefore the same value for Schweizer's dependence ($\sigma_{n} = 0.407$) and Spearman's concordance ($\rho_{n} = 0.4$). 

The stacked histogram in (j) shows the frequency of the dependence estimates provided by the users. The darker gray bars indicate the estimates associated with the plot of pseudo-observations, while the lighter blue bars are related to the estimates on the regular scatter plots. The large dispersion of the estimates related to the scatter plots stands out, which reveals the high difficulty of estimating dependence through these visualizations. Furthermore, in this example the estimates are generally smaller for pseudo-observations.

\begin{figure}[ht!]
    \centering

     \includegraphics[clip=true,width=0.75\columnwidth]{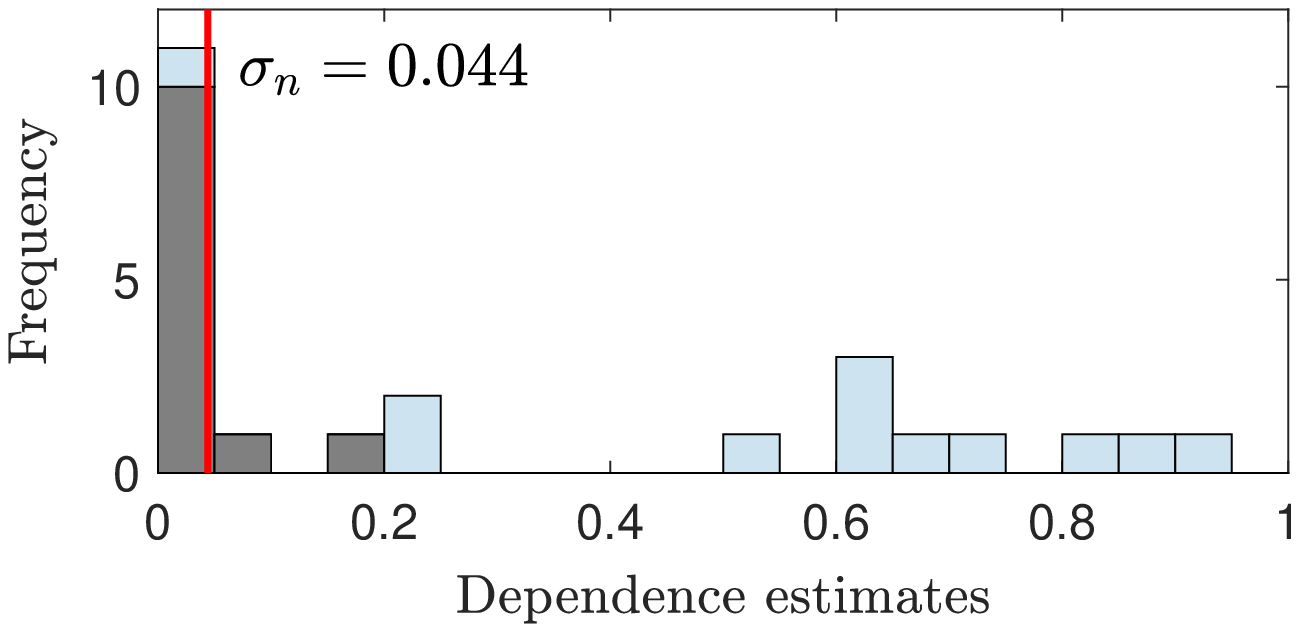}


  \caption{Histogram of dependence estimates for the data set in Fig.~\ref{fig:6x5}(3) containing four clusters. Ten out of the 12 participants identified independence (i.e., provided a 0 estimate) correctly when analyzing pseudo-observations, while the remaining two users provided low dependence values. However, the estimates were generally considerably larger when using scatter plots.}
\label{fig:histograms_independence_estimates}
\end{figure}

\begin{table}[h!]
    \centering
    \renewcommand{\arraystretch}{1.1}
    \begin{tabular}{|c|c|c|c|c|} \hline 
        plot & $a$ & $\hat{\theta}$ & $b$ & significant \\ \hline 
           a & 0.55 & 0.73 & 0.92 & Yes \\
           b & 0.89 & 0.96 & 1.00 & Yes \\
           c & 0.65 & 0.81 & 0.97 & Yes \\
           d & 0.45 & 0.65 & 0.86 & No \\
           f & 0.65 & 0.81 & 0.97 & Yes \\
           g & 0.76 & 0.88 & 1.00 & Yes \\
           h & 0.45 & 0.65 & 0.86 & No \\
           i & 0.55 & 0.73 & 0.92 & Yes \\ \hline
    \end{tabular} \\ \medskip
    
    \caption{Results of applying a paired two-sample sign Bayesian test to compare dependence assessments for each of the blue dotted plots in Fig.~\ref{fig:5x2} against their common black-dotted pseudo-observations, through a point estimate $\hat{\theta}$ and a minimum length $90\%$ credible interval $[a,b]$ estimate for $\theta=P(S>T).$ The test is significant when $0.5$ is not included in the interval estimate, and is not significant otherwise.
    }
    \label{tab:plots9}
\end{table}

 Table~\ref{tab:plots9} shows the results of applying the paired two-sample sign Bayesian test to contrast users' dependence assessments for the 8 plots against their dependence assessment based directly on the pseudo-observations. For each case if $\hat{\theta}$ is close to $0.5$ that means that there is not a big difference in the dependence assessments when visualizing the scatter plot or the pseudo-observations, while farther values of $\hat{\theta}$ from $0.5$ imply the opposite. The \textit{significant} column contains a \textit{Yes} if $0.5$ does not belong to the $[a,b]$ interval estimate for $\theta=P(S>T),$ and a \textit{No} when it does. In 6 out of 8 cases the difference is significant, and in the other 2 cases the $0.5$ value is closer to the lower bound of the interval rather than closer to the point estimate $\hat{\theta}.$ In all 8 cases we get point estimates for $\theta=P(S>T)$ above $0.5$. Lastly, the results support the claim that the estimates for pseudo-observations are different than those for regular scatter plots. In this example, the dependence estimates are greater for scatter plots even though the dependence is exactly the same for the pseudo-observations. Thus, using marginals that are not uniform seems to induce larger dependence estimates.

\begin{table}[ht!]
    \centering
    \renewcommand{\arraystretch}{1.1}
    \begin{tabular}{|c|c|c|c|c|c|} \hline 
        plots & $\sigma_n$ & $a$ & $\hat{\theta}$ & $b$ & significant \\ \hline 
         6, 21 &  0.035 & 0.54 & 0.73 & 0.92 & Yes \\
        15, 30 &  0.041 & 0.76 & 0.88 & 1.00 & Yes \\
         3, 18 &  0.044 & 0.76 & 0.88 & 1.00 & Yes \\
         2, 17 &  0.069 & 0.89 & 0.96 & 1.00 & Yes \\
        12, 27 &  0.122 & 0.36 & 0.58 & 0.80 & No \\
        13, 28 &  0.131 & 0.36 & 0.58 & 0.80 & No \\
        14, 29 &  0.180 & 0.36 & 0.58 & 0.80 & No \\
         5, 20 &  0.271 & 0.55 & 0.73 & 0.92 & Yes \\
         4, 19 &  0.356 & 0.55 & 0.73 & 0.92 & Yes \\
        10, 25 &  0.425 & 0.45 & 0.65 & 0.86 & No \\
         9, 24 &  0.438 & 0.65 & 0.81 & 0.97 & Yes \\
         7, 22 &  0.496 & 0.55 & 0.73 & 0.92 & Yes \\
         1, 16 &  0.513 & 0.36 & 0.58 & 0.80 & No \\
        11, 26 &  0.729 & 0.55 & 0.73 & 0.92 & Yes \\
         8, 23 &  0.805 & 0.14 & 0.35 & 0.55 & No \\ \hline
    \end{tabular} \\ \medskip
    
    \caption{Results of applying a paired two-sample sign Bayesian test to compare dependence assessments under 15 different Schweizer dependence values $\sigma_n$. The corresponding plots are shown in Fig.~\ref{fig:6x5}. For each case we calculated a point estimate $\hat{\theta}$ and a minimum length $90\%$ credible interval $[a,b]$ estimate for $\theta=P(S>T).$ The test is significant when $0.5$ is not included in the interval estimate, and is not significant otherwise.}
    \label{tab:plots15}
\end{table}

We also used a second set of 15 scatter plots and their associated pseudo-observations, but with different empirical values $\sigma_n$ for Schweizer's dependence. These plots are shown in Fig.~\ref{fig:6x5}. Table~\ref{tab:plots15} shows the results of the analysis where the differences were significant in 9 out of the 15 cases, with point estimates for $\theta=P(S>T)$ quite above $0.5$. This, again, might be interpreted as the effect of the marginals on dependence assessment, which can lead to larger estimates when analyzing scatter plots, despite sharing the same dependence with the pseudo-observations.

In this experiment we also observed the benefit of using pseudo-observations in cases of independence or weak dependence. Figure~\ref{fig:histograms_independence_estimates} shows a histogram of the estimates for the data set with $\sigma_{n} = 0.044$, which is the one shown in Fig.~\ref{fig:pseudoobservations}(b) containing four clusters. Clearly, most users identified independence by looking at the pseudo-observations, but provided quite large estimates when analyzing the scatter plot. We observed similar results in other data sets (see the supplemental material). Every participant identified independence on data sets with $\sigma_{n} = 0.041$ and $\sigma_{n} = 0.069$, while all but one participant estimated independence on the data set with $\sigma_{n} = 0.035$.

\section{Conclusions and discussion}
\label{sec:discussion}

In this paper we have presented several visualization approaches for helping users to understand and interpret the statistical dependence between continuous random variables. The present work has two main contributions. On the one hand, we show that regular scatter plots are not ideal tools for determining dependence. Instead, we suggest using copula pseudo-observations, which stem from transforming the data according to the (empirical) cumulative distribution functions. Pseudo-observations can be considered as marginal-free scatter plots, and are better suited for estimating dependence since the marginals, which do not contain information regarding the relationship between variables, may introduce noise and can hamper the interpretation of the dependence structure in the data.

Our experiments have shown that pseudo-observations are especially useful for detecting dependence when the two random variables are independent or present weak dependencies. However, it is important to note that the participants in the study did not receive any instructions or guidance (except for clarifying the maximum ``total'' dependence situation: when the value of one variable is completely determined by the value of the other). In the future we plan on studying the benefits of pseudo-observations over regular scatter plots of real and synthetic data, with users with prior training on how to interpret pseudo-observations.

Secondly, we have proposed alternative visualizations based on heatmaps for analyzing empirical copulas. Two are related to Spearman's concordance measure and Schweizer-Wolff's dependence measure, where users could approximate these values by averaging over the represented colors. However, it is difficult to obtain accurate estimates of these means visually. Thus, we have proposed a normalized measure of differences between an empirical copula and the copula that represents independence. With this measure we construct a heatmap that highlights normalized deviations from independence, i.e., increasing and/or decreasing trends in data. Moreover, these heatmaps can be used to assign colors to data pairs that can be exploited in other visualizations such as scatter plots or parallel coordinate plots.

In this work we have focused on the statistical dependence between continuous random variables, because in this case the underlying copula is unique and defined on the whole unit square ($[0,1]^2$). Future work will be devoted to the case when at least one of the random variables is not continuous. In such case the copula exists but it is not unique. What is unique is a \textit{subcopula} function \cite{Erdely2017} which has a similar definition but whose domain is a proper subset of the unit square. This brings different challenges in terms of measuring and visualizing dependence.





\bibliographystyle{abbrv-doi}

\bibliography{Infovis2022}
\end{document}